\documentclass[aps,prd,twocolumn,showpacs,superscriptaddress,groupedaddress,amsmath,floatfix,nofootinbib]{revtex4}  
\pdfoutput=1
\usepackage {amsmath}
\usepackage {amssymb}
\usepackage {amsfonts}
\usepackage {amsthm}
\usepackage {mathrsfs}
\usepackage {latexsym}
\usepackage {graphicx}
\usepackage {dsfont}
\usepackage {txfonts}
\usepackage {rotating}
\usepackage {wasysym}
\usepackage {multirow}
\usepackage {hhline}
\usepackage {graphicx}
\usepackage {hyperref}
\usepackage {color}
\usepackage {bm}
\usepackage{textcomp}
\usepackage{appendix}
\usepackage{acronym}
\usepackage{dcolumn}   
\usepackage{xspace}
\usepackage {url}

\newcommand{\dcc}{LIGO-P1300118}

\newcommand{\subtext}[1]{\text{\tiny{#1}}}
\newcommand{\submath}[1]{\text{\tiny{\emph{#1}}}}

\newcommand{\dg}{^\circ}
\newcommand{\Msol}{M_\odot}
\newcommand{\fstat}{\mathcal{F}}
\newcommand{\cstat}{\mathcal{C}}
\newcommand{\temp}{\mathcal{T}}
\newcommand{\flattemp}{\mathcal{T}_{\text{F}}}
\newcommand{\ftext}{$\mathcal{F}$\xspace}
\newcommand{\fstext}{$\mathcal{F}$-statistic\xspace}
\newcommand{\fstexts}{$\mathcal{F}$-statistics\xspace}
\newcommand{\ctext}{$\mathcal{C}$\xspace}
\newcommand{\cstext}{$\mathcal{C}$-statistic\xspace}

\newcommand{\stat}[1]{\mathcal{#1}}

\newcommand{\Na}{N_\subtext{a}}            	
\newcommand{\Ngw}{N_\subtext{GW}}      		
\newcommand{\Fx}{F_\subtext{X}}			
\newcommand{\Lx}{L_\subtext{X}}			
\newcommand{\dist}{d}                           

\newcommand{\tp}{t_\submath{p}}			
\newcommand{\ta}{t_\submath{a}}			
\renewcommand{\to}{t_\submath{0}}		

\newcommand{\Tspan}{T_\text{s}}			
\newcommand{\ro}{\rho_\submath{0}}		

\newcommand{\spinf}{\nu_{\text{s}}}        	
\newcommand{\spinomega}{\Omega_{\text{s}}}   	

\newcommand{\fn}{f_n}				

\newcommand{\ho}{h_\submath{0}}			
\newcommand{\mo}{m_\submath{0}}			%
\newcommand{\Mo}{M_\submath{0}}			%
\newcommand{\fo}{f_\submath{0}}			
\newcommand{\Po}{P_\submath{0}} 		
\newcommand{\atrue}{a_\submath{0}}		
\newcommand{\phio}{\phi_\submath{0}}		

\newcommand{\Porb}{P}				
\newcommand{\Pmeas}{P}				
\newcommand{\asini}{a}				
\newcommand{\ameas}{a}				

\newcommand{\atemp}{a'}				

\newcommand{\fres}{d_f}				
\newcommand{\lj}{l_{[j]}}			

\newcommand{\hUL}{h_\subtext{UL}}		
\newcommand{\cstar}{\cstat^*}			
\newcommand{\Pa}{P_\text{a}}			
\newcommand{\htorq}{h_{0}^{\text{EQ}}}          
\newcommand{\tspinw}{\tau_\text{s}}                  
\newcommand{\Tspansw}{T_{\text{s}}^{\text{spin}}} 
\newcommand{\nullH}{H_{\text{n}}}               
\newcommand{\sigH}{H_{\text{GW}}}               

\long\def\symbolfootnote[#1]#2{\begingroup\def\thefootnote{\fosymbol{footnote}}\footnote[#1]{#2}\endgroup}

\def\commitID{commitID: 1d41f0b94a6df042af7e99a97529058a81632422}
\def\commitDATE{Wed Nov 6 12:08:06 2013 +0000}

\begin{document}


\title{Implementation of the frequency-modulated sideband search method for gravitational waves from low mass X-ray binaries}
\author{L. Sammut}
\email{lsammut@student.unimelb.edu.au}
\affiliation{University Of Melbourne, Victoria 3052, Australia}
\affiliation{Albert Einstein Institute, Callinstra{\ss}e 38,D-30167
Hannover,Germany}
\author{C.~Messenger}
\affiliation{University of Glasgow, Glasgow, G12 8QQ, United Kingdom}
\affiliation{Albert Einstein Institute, Callinstra{\ss}e 38,D-30167
Hannover,Germany}
\author{A.~Melatos}
\affiliation{University Of Melbourne, Victoria 3052, Australia}
\author{B.J.~Owen}
\affiliation{The Pennsylvania State University, University Park, Pennsylvania
16802, USA}



\date{\today \\\mbox{\small \commitID} \\\mbox{\small commitDate: \commitDATE}\\\mbox{\dcc}}

\begin{abstract}
  We describe the practical implementation of the sideband search, a search for periodic gravitational waves from neutron stars in binary systems. The
  orbital motion of the source in its binary system causes frequency-modulation in the combination of matched filters known as the \fstext. The sideband search is based on the incoherent summation of these frequency-modulated \fstext sidebands. It provides a new detection statistic for sources in binary systems, called the \cstext. The search is well suited to low-mass X-ray binaries, the brightest of which, called Sco X-1, is an ideal target candidate. For sources like Sco X-1, with well constrained orbital parameters, a slight variation on the search is possible. The extra orbital information can be used to approximately demodulate the data from the binary orbital motion in the coherent stage, before incoherently summing the now reduced number of sidebands. We investigate this approach and show that it improves the sensitivity of the standard Sco X-1 directed sideband search. Prior information on the neutron star inclination and gravitational wave polarization can also be used to improve upper limit sensitivity. We estimate the sensitivity of a Sco X-1 directed sideband search on 10 days of LIGO data and show that it can beat previous upper limits in current LIGO data, with a possibility of constraining theoretical upper limits using future advanced instruments.
\end{abstract}

\maketitle


\acrodef{LIGO}{Laser Interferometer Gravitational-Wave Observatory}
\acrodef{LSC}{LIGO Scientific Collaboration}
\acrodef{LMXB}[LMXB]{low-mass X-ray binary}
\acrodef{QPO}[QPO]{quasi-periodic oscillations} 
\acrodef{ScoX1}[Sco X-1]{Scorpius X-1} 
\acrodef{BH}{black-hole} 
\acrodef{pdf}[PDF]{probability density function}
\acrodef{cdf}[CDF]{cumulative distribution function}
\acrodef{SNR}[SNR]{signal-to-noise ratio}
\acrodef{HJD}[HJD]{heliocentric Julian date} 
\acrodef{GPS}[GPS]{global positioning system} 
\acrodef{EM}[EM]{electromagnetic}
\acrodef{ROC}{receiver operator characteristic}
\acrodef{rms}[RMS]{root-mean-square} 
\acrodef{qpo}[QPO]{quasi-periodic oscillation} 
\acrodef{amxp}[AMXP]{accreting millisecond X-ray pulsar} 
\acrodef{msp}[MSP]{millisecond pulsar}
\acrodef{sft}[SFT]{short Fourier transform}
\acrodef{los}[LOS]{line-of-sight} 
\acrodef{SSB}[SSB]{Solar system barycenter} 
\acrodef{FFT}[FFT]{Fast-Fourier-Transform}
\acrodef{PDF}[PDF]{probability density function}

\section{Introduction}

Rotating neutron stars are the main candidates for sources of persistent, periodic
gravitational radiation detectable by ground-based, long-baseline
gravitational wave interferometers. Time varying quadrupole moments in (and
thus gravitational-wave emission from) these sources can result from
deformations of the solid crust (and possibly a solid core) supported by
elastic stresses \cite{Bildsten_1998, UCB_2000, Owen_2005, HaskellEA_2007,
Lin2007, JM_Owen_2013}, deformations of various parts of the star supported by
magnetic stresses \cite{B_G_1996, Cutler_2002, Melatos_Payne_2005,
Vigelius_Melatos_2008, HaskellEA_2008, Mastrano_Melatos_2012}, or free
precession \cite{Jones_Andersson_2002, Jones_2012} or long-lived oscillation
modes of the entire star \cite{OwenEA_1998, AKS_1999, Bondarescu2007,
Haskell2012, Bondarescu2013}.

Neutron stars in accreting binary systems are an important sub-class
of periodic gravitational wave sources.
Accretion may trigger or enhance the aforementioned gravitational-wave
emission mechanisms, creating or driving the quadrupole moment toward its maximum
value through thermal, magnetic, or other effects \cite{Bildsten_1998,
AKS_1999, Nayyar_Owen_2006, Melatos2007, vEysden_Melatos_2008,
Vigelius_Melatos_2009, AnderssonEA_2011}.
If a balance is assumed between the
gravitational radiation-reaction torque and the accretion
torque~\cite{Papaloizou_Pringle_1978, Wagoner_1984, Bildsten_1998}, then the
strongest emitters of continuous gravitational waves are predicted to be sources in low-mass X-ray binaries
(\acsp{LMXB}\acused{LMXB}), specifically those accreting at
the highest rate~\cite{Verbunt_1993, WattsEA_2008}.

Given the estimated ages (${\sim}10^{10}$ yrs) and observed accretion
rates of \acsp{LMXB} (reaching near the Eddington limit of
$\dot{M}_{\text{Edd}} = 2 \times 10^{- 8} M_{\odot} \text{yr}^{- 1}$),
accretion is expected to spin-up the neutron star beyond the breakup
frequency (${\sim}1.5$ kHz for standard neutron star equations-of-state
\cite{CST_1994,UBC_2000}). However, measured spin frequencies of LMXB
neutron stars (from X-ray pulsations or thermonuclear bursts) so far range
only from 95 to 619 Hz \cite{ChakrabartyEA_Nature_2003, Bhattacharyya2007, WattsEA_2008, GallowayEA_2010}. The spin
frequency cut-off lies well below breakup, and suggests the existence
of a spin-down torque to balance the spin-up from accretion. A possible explanation,
proposed by \citet{Papaloizou_Pringle_1978} and advanced by \citet{Wagoner_1984} and
\citet{Bildsten_1998}, is gravitational radiation. The torque-balance
scenario implies a relation between the X-ray flux, spin frequency and gravitational wave strain; the more luminous the X-ray source, the
greater the strain. \ac{ScoX1}, the brightest \ac{LMXB}, is therefore
a promising target for periodic gravitational wave searches.

The global network of kilometer-scale, Michelson-type laser
interferometers are sensitive to gravitational waves in the $O(10-1000)$ Hz
frequency band. The \ac{LIGO} detectors achieved design sensitivity
during the fifth science run (S5), between November 2005 and October
2007 \cite{LIGO_2009, LIGO_S5cal_Abadie2010}. \ac{LIGO} consists of
three Michelson interferometers (one with 4 km arms at Livingston,
Louisiana, and two co-located at Hanford, Washington, with 4 km and 2
km arms) separated by ${\sim}3000$ km. Together the \ac{LIGO}
\cite{LIGO}, Virgo \cite{VIRGO, Virgo2} and GEO600 \cite{GEO2,
  GEO600} detectors form a world-wide network of broad-band
interferometric gravitational wave observatories in an international effort to
directly detect gravitational wave emission for the first
time. 

The \ac{LIGO} Scientific Collaboration has so far published three main
types of searches for periodic or continuous gravitational wave
emission; targeted, directed and all-sky searches. Targeted searches
are the most sensitive since they have the most tightly constrained parameter
space. They target known sources, such as radio pulsars, with very well-constrained sky position, spin frequency and frequency evolution, and binary parameters (if any). These searches are fully-coherent, requiring
accurately known prior phase information, making them
computationally expensive to perform over large regions of parameter
space~\cite{BradyEA_1998,LIGO_2004_PRD69,LIGO_S5_KnownPulsars_2010}. Targeted
\ac{LIGO} and Virgo searches have already set astrophysically interesting upper
limits (e.g. beating theoretical indirect limits) on some pulsar
parameters such as the gravitational wave strain from the Crab
pulsar~\cite{LIGO_S5_Crab_2008,LIGO_S5_KnownPulsars_2010} and the Vela pulsar
\cite{LIGO_Vela_2011}. Directed searches aim at a particular sky location but search for unknown
frequency (and/or frequency evolution). In most cases so far, directed searches
have used a fully-coherent approach and approached the limits of
computational feasibility. The search directed at the (possible) neutron star in
the direction of the supernova remnant Cassiopeia A was able to beat indirect limits
\cite{LIGO_CasA_2010}. The third type of continuous gravitational wave
search, the wide parameter-space searches, are also
computationally intensive. They can involve searching over the entire
sky or any comparably large parameter space, and usually employ
semi-coherent approaches, combining short coherently analyzed segments
in an incoherent manner.  This process is tuned to balance the
trade-off between reduced computational load and reduced
sensitivity. The all-sky searches presented
in~\cite{LIGO_S4_AllSKy_2008,LIGO_S4_EaH_2009,LIGO_S5i_Powerflux_2009,
  LIGO_S5_EaH_2009,LIGO_S5full_Powerflux_2012} target isolated neutron star
sources (i.e. those not in binary systems). There is also an
all-sky search for neutron stars in binary systems currently being run
on \ac{LIGO}'s latest S6 data run~\cite{Goetz_Riles_TwoSpect_2011}.

Directed searches can also be made for known accreting neutron stars in
binaries, and \ac{LIGO} has previously conducted two of these searches for
\ac{ScoX1}. The first, a coherent analysis using data from the second
science run (S2), was computationally limited to the analysis of
six-hour data segments from the \ac{LIGO} interferometers, and placed
$95\%$ upper limits on the wave strain of $\ho^{95\%} \approx 2\times
10^{-22}$ for two different 20 Hz bands \cite{LIGO_S2_ScoX1_2007}. This
search utilized a maximum likelihood detection statistic based on matched
filtering called the \fstext~\cite{JKS1998}. The second, a directed version of
the all-sky, stochastic, cross-correlation analysis, known as the
``Radiometer'' search, was first conducted on all 20 days of data from
the S4 science run ~\cite{LIGO_S4_ScoX1_2007}, and later on the
${\sim}2$ year S5 data reporting a $90\%$ upper limit on the \ac{rms} strain
$h_\text{rms}^{90\%} > 5\times10^{-25}$ (over the range 40 -- 1500 Hz, with the minimum around 150 Hz) ~\cite{LIGO_S5stoch_ScoX1_2011}. 

Semi-coherent search methods provide a compromise between sensitivity and the computational cost of a fully coherent search. They should be the most sensitive at fixed computing cost \cite{Brady_Creighton_2000,Prix_Shaltev_2012}. A fast and robust search strategy for the detection of signals from binary
systems in gravitational wave data was proposed in~\cite{MW2007}. The signal from a source in a binary system is phase- (or frequency-) modulated due its
periodic orbital motion, forming ``sidebands'' in the gravitational wave frequency spectrum. In searching detector data, this technique, called the sideband search, uses the same coherent (\fstext) stage as the previous coherent (S2) search. It then combines the frequency-modulated sidebands arising in $\fstat$-statistic data in a (computationally inexpensive) incoherent stage, reducing the need for a large template bank.  This approach is based on a method that has successfully been employed in searches for binary pulsars in radio data~\cite{RansomEA_2003}.

Here we develop this sideband technique into a search pipeline and
present a detailed description of how it is applied to gravitational
wave detector data, as well as the expected sensitivity. The paper is
set out as follows. Section \ref{sec:LMXBs} briefly describes the
astrophysics of \acp{LMXB} and their predicted gravitational wave
signature. The search algorithm is described in detail in Section
\ref{sec:sideband}. Section \ref{sec:params} outlines the parameter
space of the search allowing primary sources to be identified in Section \ref{sec:sources}. The statistical
analysis of the results of the search is described in Section
\ref{sec:stats}, along with a definition of the upper limits of the search. 
The sensitivity of the search is discussed in Section \ref{sec:sensitivity}. A brief
summary is provided in Section \ref{sec:discussion}, with a discussion of the limitations and future prospects of the search.

\section{Low Mass X-ray Binaries}\label{sec:LMXBs}

\acp{LMXB} are stellar systems where a low-magnetic-field ($\lesssim
10^9$ G) compact object (primary) accretes matter from a lower-mass
(secondary) companion ($ < 1 M_\odot$) \cite{Verbunt_1993,Tauris_vandenHeuvel_ch16}. The compact objects in \ac{LMXB} systems can be black
holes, neutron stars or white dwarfs. For gravitational wave emission we are
interested in \acp{LMXB} with neutron stars as the primary mass
(typically $\sim{1.4}\Msol$), since neutron stars can sustain the largest
quadrupole moment.

Observations of thermal X-ray emission from the inner region of the
accretion disk provide a measurement of the accretion rates in
\acp{LMXB}. The range of observed accretions rates is broad, ranging
from $10^{- 11} M_{\odot} \text{yr}^{- 1}$ to the Eddington limit,
$\dot{M}_{\text{Edd}} = 2 \times 10^{- 8}M_{\odot} \text{yr}^{- 1}$
\cite{UBC_2000}. Some \acp{LMXB} also exhibit periodic pulsations or
burst type behavior, and so provide a means of measuring the spin
frequency $\spinf$ of the neutron star in the system. The measured
$\spinf$ of these systems lie in the range of $95 \leq \spinf \leq
619$ Hz \cite{ChakrabartyEA_Nature_2003, Bhattacharyya2007,
  GallowayEA_2010}. The broad range of accretion rates coupled with
the estimated age of these systems (${\sim}10^{10}$ years implied by
evolutionary models~\cite{vanParadijs_White_1995, UBC_2000}) would
suggest a greater upper limit on observed spin frequency since
accretion exerts substantial torque on the neutron star. However, none
of these systems have yet been observed to spin at or near the breakup
frequency $v_b{\sim}1.5$ kHz ($v_b \gtrsim$ 1 kHz for most equations
of state \cite{CST_1994, UBC_2000}). The maximum observed spin
frequency falls far below the theorized breakup frequency and suggests
a competing (damping) mechanism to the spin-up caused by
accretion. One explanation for the observed spin frequency
distribution of \acp{LMXB} is that the spin-up from the accretion
torque is balanced by a gravitational wave spin-down torque \cite{Wagoner_1984,
  Bildsten_1998}. Since the gravitational wave spin-down torque scales like
$\spinf^5$ (see Eq.~\ref{eq:N_gw} below), a wide range of accretion
rates then leads to a rather narrow range of equilibrium rotation
rates, as observed.

\subsection{Gravitational wave emission}\label{subsec:GWs}

Using the torque balancing argument from \citet{Wagoner_1984} and
\citet{Bildsten_1998}, we can estimate the gravitational wave strain
amplitude emitted from accreting binary systems from their observable
X-ray flux. This is a conservative upper limit as it assumes all
angular momentum gained from accretion is completely converted into
gravitational radiation.

The intrinsic strain amplitude $\ho$ for a system with angular spin frequency
$\spinomega=2\pi\spinf$ at a distance $d$ from an observer emitting gravitational waves via a
mass quadrupole can be expressed as
\begin{equation}
 \ho=\frac{4 G Q}{c^4 d}\spinomega^2\,,
\end{equation}
where $G$ is the gravitational constant, $c$ is the speed of light and
the quadrupole moment $Q=\epsilon I$ is a function of the ellipticity
$\epsilon$ and moment of inertia $I$ \cite{JKS1998}. This can be
expressed in terms of the spin-down (damping) torque $\Ngw$ due to
gravitational radiation giving
\begin{equation}
 \ho^2 = \frac{5 G}{8c^3 d^2\spinomega^2} \Ngw, \label{eq:h02}
\end{equation}
where 
\begin{equation}
 \Ngw = \frac{32GQ^2}{5c^5}\spinomega^5. \label{eq:N_gw}
\end{equation}
The accretion torque $\Na$ applied to a neutron star of mass $M$ and radius
$R$ accreting at a rate $\dot{M}$ is given by
\begin{equation}
\Na=\dot{M}\sqrt{GMR} \label{eq:Na}.
\end{equation}
Assuming that the X-ray luminosity can be written as
$\Lx=GM\dot{M}/R$, the accretion rate $\dot{M}$ can be expressed as a
function of the X-ray flux $\Fx$, such that
\begin{equation}
 \dot{M}=\frac{4\pi R d^2}{GM} \Fx \label{eq:Mdot},
\end{equation}
since $\Lx=4\pi d^2 \Fx$. In equilibrium, where gravitational
radiation balances accretion torque, $\Ngw=\Na$. The square of the
gravitational wave strain from Eq.~\ref{eq:h02} can then be expressed
in terms of the observable X-ray flux such that
\begin{equation} \label{eq:h02}
 \ho^2 = \frac{5}{3}\frac{\Fx}{\spinf} {\left(\frac{GR^3}{M}\right)}^{1/2}.
\end{equation}
Selecting fiducial values for the neutron star mass, radius, X-ray flux,
and spin frequency (around the middle of the observed range), we can
express the equilibrium strain upper limit $\htorq$ in terms of $\spinf$
and $\Fx$ via
\begin{eqnarray}
  \htorq  &=&  5.5\times 10^{- 27}
\left(\frac{F_{\text{x}}}{F_*}\right)^{1/2}\left(\frac{R}{10\text{km}}\right)^{
3/4} \left(\frac{1.4M_{\odot}}{M}\right)^{1/4} 
\left(\frac{300\text{Hz}}{\spinf}\right)^{1/2}, 
\nonumber \\ \label{eq:hc_Fx_300}
\end{eqnarray}
where $F_*=10^{-8}\text{ erg cm}^{-2}\text{ s}^{-1}$. If the system emits gravitational waves via current quadrupole radiation instead, as is the
case with $r$-mode oscillations, the relation between gravitational wave frequency and spin frequency differs. In this case the preceding equations are modified slightly, requiring roughly $\spinf \rightarrow (2/3) \spinf$ \cite{Owen2010}. However these expressions, and the rest of the analysis except where otherwise noted, do not change if expressed in terms of the gravitational-wave frequency.

The resulting relation in Eq. \ref{eq:hc_Fx_300} implies that \acp{LMXB} that accrete close to the
Eddington limit are potentially strong gravitational wave emitters. Of these
potentially strong sources, \ac{ScoX1} is the most promising due to its observed X-ray flux ~\cite{WattsEA_2008}.

\section{Search Algorithm}\label{sec:sideband}

In this section we define our detection statistic and show how it
exploits the characteristic frequency-modulation pattern inherent to
sources in binary systems.

Fully-coherent, matched-filter searches for continuous gravitational
waves can be described as a procedure that maximizes the likelihood
function over a parameter space. The amplitude parameters
(gravitational wave strain amplitude $\ho$, inclination $\iota$,
polarization $\psi$ and reference phase $\phio$) can, in general, be
analytically maximized, reducing the dimensions of this parameter
space ~\cite{JKS1998}. These parameters define the signal amplitudes
in our signal model. Analytic maximization leaves the phase-evolution
parameters (gravitational wave frequency $\fo$ and its derivatives
$f^{(k)}$ and sky position $[\alpha,\delta]$) to be numerically
maximized over. Numerical maximization is accomplished through a
scheme of repeated matched-filtering performed over a template bank of
trial waveforms defined by specific locations in the phase parameter
space, which is typically highly computationally expensive
\cite{BradyEA_1998,LIGO_S2_ScoX1_2007,WattsEA_2008}.

The method we outline here makes use of the fact that we know the
sky position of our potential sources and, hence, the phase evolution
due to the motion of the detector can be accurately accounted for.  We
also know that the phase evolution due to the binary motion of the
source will result in a specific distribution of signal power in
the frequency domain.  This distribution has the characteristics that
signal power is divided amongst a finite set of frequency-modulation
sidebands.  The number of sidebands and their relative frequency spacing
can be predicted with some knowledge of the binary
orbital parameters. 

In order to avoid the computational limits imposed by a fully-coherent
parameter space search, we propose a single fully-coherent
analysis stage, that accounts for detector motion only, is followed by
a single incoherent stage in which the signal power contained within
the frequency-modulated sidebands in summed to form a new detection
statistic. This summing procedure is accomplished via the convolution
of an approximate frequency domain power template with the output of
the coherent stage.  

\begin{figure*}
\includegraphics[width=2\columnwidth]{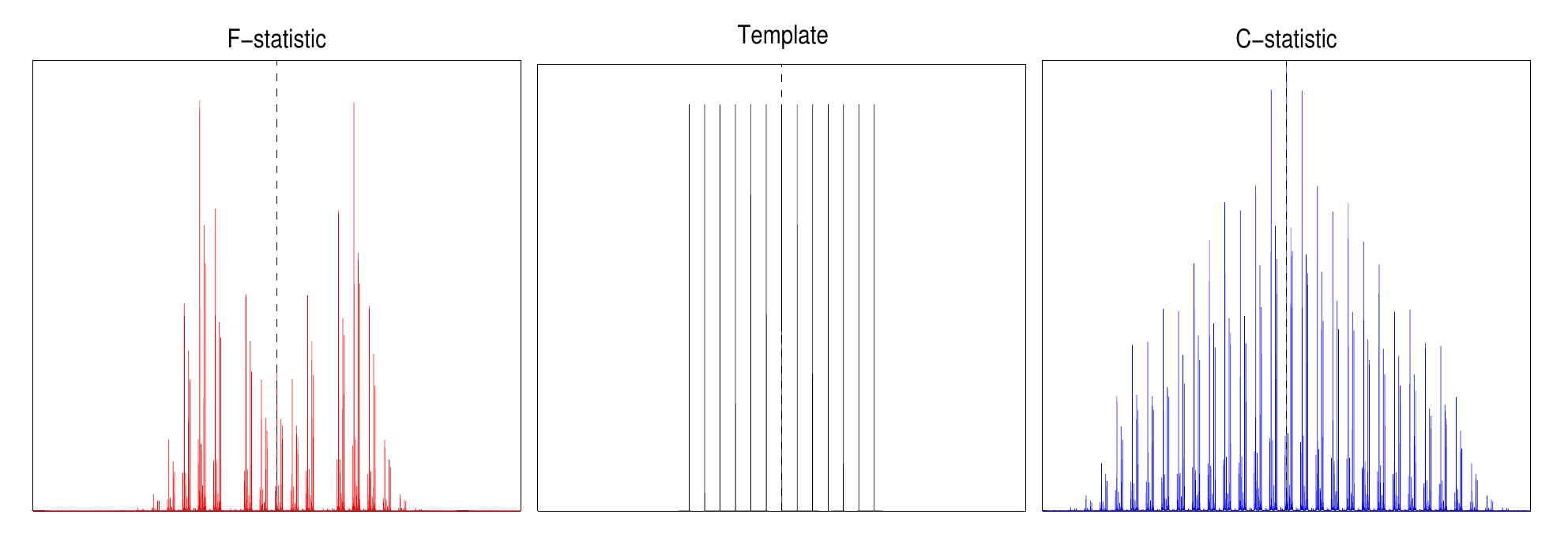}
\caption{\label{fig:search}Graphical illustration of the sideband search
pipeline, showing the frequency-modulated \fstext (left, red), the sideband
template (middle, black), and their convolution, known as the \cstext{} (right,
blue). In this noise free case, a signal of $\fo = 200$ Hz with amplitude
$\ho=1$ in a system with $\asini = 0.005$ s, $\Porb = 7912.7$ s, sky position
$\alpha = 2.0$ rad, $\delta=1.0$ rad, and phase parameters $\psi = 0.2 \text{
rad},\, \cos\iota = 0.5,\, \phi_0 = 1$ rad, was simulated for a 10 day
observation span. Frequency increases along the horizontal axis, which ranges
from 199.998 to 200.002 Hz on each plot. In each case the location of the
injected signal at 200 Hz is indicated by the vertical dashed black line.}
\end{figure*}

The three main stages of the search, the \fstext, sideband template,
and \cstext, are graphically illustrated in Figure~\ref{fig:search}.
In this noise-free example, the frequency-modulation sidebands are
clearly visible. The \ftext-statistic is also amplitude-modulated due
to the daily variation of the detector antenna response, resulting in the
amplitude-modulation applied to each frequency-modulation sideband.
The second panel represents the approximate frequency domain template,
a flat comb function with unit amplitude teeth (the spikes or delta functions).  When
convolved with the \ftext-statistic in the frequency domain we obtain
the \cstext shown in the right-hand panel.  The maximum power is
clearly recovered at the simulated source frequency.

The following sub-sections discuss each of the search components in
more detail. Section~\ref{subsec:phase} presents the phase model used
to characterize the gravitational wave signal from a source in a
binary system. The signal model is introduced in
Sec.~\ref{subsec:signal}. The \fstext{} is introduced in
Section~\ref{subsec:fstat} and its behavior as a function of search
frequency is described in Sec. \ref{subsec:fstatmm}. Section
\ref{subsec:fmod} then goes on to describe how matching a filter for
an isolated neutron star system to a signal from a source in a binary
system results in frequency-modulated sidebands appearing in the
output of the \fstext. The detection statistic for gravitational wave
sources in binary orbits is fully described as the incoherent sum of
frequency-modulated \fstexts in Sec.~\ref{subsec:Cstat}. The simplest,
unit amplitude, sideband template, and its justification over a more
realistic template, are discussed in Sec.~\ref{subsec:template}.  A
more sensitive implementation, incorporating an approximate binary
phase model in the calculation of the \fstext and reducing the width
of the frequency-modulated sideband pattern by the fractional errors
on the semi-major axis parameters, is discussed in
Sec.~\ref{subsec:demod}. Beginning in this section and continuing in
the following sections, we have made a distinction between the
intrinsic values of a search parameter $\bf{\theta_\submath{0}}$
(denoted with a subscript zero) and the observed values $\bf{\theta}$
(no subscripts).

\subsection{Phase model}\label{subsec:phase}

The phase of the signal at the source can be modeled by a Taylor
series such that
\begin{equation}\label{eq:phitaylor}
  \Phi(\tau) = 2\pi\sum^s_{k=0}\frac{f^{(k)}}{(k+1)!}(\tau-\tau_{0})^{k+1},
\end{equation}
where $f^{(k)}$ represents the $k^\text{th}$ time derivative of the
gravitational wave frequency evaluated at reference time $\tau_0$. For
the purposes of this work we restrict ourselves to a monochromatic
signal and hence set $f^{(k)}=0$ for all $k>0$ and define
$f_0=f^{(0)}|_{\tau_0}=f(\tau_0)$ as the intrinsic frequency. We
discuss this choice in Sec.~\ref{subsec:spinf}. The phase received at
the detector is $\Phi[\tau(t(t'))]$, where we define the retarded
times measured at the \ac{SSB} and detector as $t$ and $t'$
respectively. The relation between $t$ and $t'$ is a function of
source sky position relative to detector location and, since we only
concern ourselves with sources of known sky position, we assume that
the effects of phase contributions from detector motion can be exactly
accounted for.  For this reason we work directly within the \ac{SSB}
frame.  The relationship between the source and retarded times for a
non-relativistic eccentric binary orbit is given
by~\cite{Taylor_Weisberg_1989}
\begin{equation}\label{eq:bintime}
  t = \tau +\atrue\left[\sqrt{1-e^{2}}\cos{\omega}\sin{E}+\sin{\omega}\left(\cos{E}-e\right)\right],
\end{equation}
where $\atrue$ is the light crossing time of the semi-major axis
projected along the line of sight. The orbital eccentricity is defined
by $e$ and the argument of periapse, given by $\omega$, is the angle
in the orbital plane from the ascending node to the direction of
periapse.  The variable $E$ is the eccentric anomaly defined by
$2\pi(\tau-\tp)/P=E-e\sin{E}$, where $\tp$ is the time of passage through
periapse (the point in the orbit where the two bodies are at their
closest) and $P$ is the orbital period.

It is expected that dissipative processes within \acp{LMXB} drive the
orbits to near circularity. In the low eccentricity limit $e\ll1$, we
obtain the following expression
\begin{eqnarray}\label{eq:bintime2}
  t &=& \tau + \atrue\left\{\sin\left[\Omega(\tau-\ta)\right]+\frac{e\cos\omega}{2}\sin\left[2\Omega(\tau-\tp)\right]\right.\nonumber\\
  &&\left.\left.+\frac{e\sin\omega}{2}\cos\left[2\Omega(\tau-\tp)\right]\right\}+\mathcal{O}(e^{2})\right\}, 
\end{eqnarray}
where $\Omega=2\pi/P$ and we have used the time of passage through the
ascending node $\ta=\tp-\omega/\Omega$ as our reference phase in the
first term.  For circular orbits, where $e=0$, the expression reduces
to only this first term. Using $\ta$ is sensible in this case since
$\tp$ and $\omega$, are not defined in a circular orbit. Note that the
additional eccentric terms are periodic at multiples of twice the
orbital frequency. Using Eq. \ref{eq:bintime2}, we would expect to
accumulate timing errors of order $\mu$s for the most eccentric known
\ac{LMXB} systems. We shall return to this feature in
Sec.~\ref{subsec:ecc}.

To write the gravitational wave phase as a function of \ac{SSB} time, we invert
Eq.~\ref{eq:bintime2} to obtain $\tau(t)$. The binary phase can be
corrected for in a standard matched filter approach, where
Eq.~\ref{eq:bintime} is solved numerically.  In our method we instead
approximate the inversion as
\begin{equation}\label{eq:binphase}
  \Phi(t)\simeq 2\pi\fo(t - \to) - 2\pi\fo\atrue\sin\left[\Omega(t-\ta)\right]
\end{equation}
for circular orbits.\footnote{Phase errors caused by this inversion approximation amount to maximum phase offsets of $\sim 2\pi\fo \atrue^{2}\Omega$.} Under this approximation the signal phase can
be represented as a linear combination of the phase contributions from
the spin of the neutron star $\phi_\text{spin}$ and from the binary
orbital motion of the source $\phi_\text{bin}$, such that
\begin{subequations}
\begin{eqnarray}
\phi_\text{spin} &=&  2\pi\fo (t-\to) \label{eq:phi_spin}, \\
\phi_\text{bin}  &=& -2\pi\fo \atrue\sin \left[\Omega(t-\ta)\right]. \label{eq:phi_bin}
\end{eqnarray}
\end{subequations}
%

\subsection{Signal model} \label{subsec:signal}

We model the data $\bm{x}(t)$ collected by a detector located at the
\ac{SSB} as the signal $\bm{s}(t)$ plus stationary Gaussian noise $\bm{n}(t)$ so that
\begin{equation}
 \bm{x}(t) = \bm{s}(t) + \bm{n}(t) \label{eq:tseries}
\end{equation}
with
\begin{equation}
\bm{s}(t) = \mathcal{A}_\mu \bm{h}_\mu(t)\label{eq:s}, 
\end{equation}
where we employ the Einstein summation convention for
$\mu=1,2,3,4$. The coefficients $\mathcal{A}_\mu$ are independent of
time, detector location and orientation. They depend only on the
signal amplitude parameters $\bm{\lambda} =
\{\ho,\,\psi,\,\iota,\,\phio\}$, where $\ho$ is the dimensionless
gravitational wave strain amplitude, $\psi$ is the gravitational wave
polarization angle, $\iota$ is the source inclination angle and
$\phio$ is the signal phase at a fiducial reference time. The
coefficients $\mathcal{A}_\mu$ are defined as
\begin{subequations} \label{eq:A1234}
\begin{align}
 \mathcal{A}_1 &= A_+\cos\phio\cos 2\psi - A_\times \sin\phio\sin2\psi\\
 \mathcal{A}_2 &= A_+\cos\phio\sin 2\psi - A_\times \sin\phio\cos2\psi\\
 \mathcal{A}_3 &= A_+\sin\phio\cos 2\psi - A_\times \cos\phio\sin2\psi\\
 \mathcal{A}_4 &= A_+\sin\phio\sin 2\psi - A_\times \cos\phio\cos2\psi
\end{align}
\end{subequations}
where
\begin{eqnarray} \label{eq:Apc}
 A_+ = \tfrac{1}{2} \ho(1+\cos^2\iota) &,& A_\times = \ho\cos\iota
\end{eqnarray}
are the polarization amplitudes. The time dependent signal components
$\bm{h}_\mu(t)$ are defined as
\begin{equation}
\begin{split}
  h_1=a(t)\cos\Phi(t), &\hspace{0.5cm} h_2=b(t)\cos\Phi(t), \\
  h_3=a(t)\sin\Phi(t), &\hspace{0.5cm} h_4=b(t)\sin\Phi(t), \label{eq:hs}
\end{split}
\end{equation}
where $\Phi(t)$ is the signal phase at the detector (which we model as
located at the \ac{SSB}) given by Eq.\ref{eq:binphase} and the antenna
pattern functions $a(t)$ and $b(t)$ are described by Eqs.~(12) and
(13) in~\cite{JKS1998}.

\subsection{\fstext}\label{subsec:fstat}

The \fstext is a matched-filter based detection statistic derived via
analytic maximization of the likelihood over unknown amplitude
parameters~\cite{JKS1998}.  Let us first introduce the multi-detector inner product
\begin{equation}
 (\bm{x}|\bm{y})=\sum_{X}(x_X|y_X)=\sum_{X}\frac{2}{S_X(f)}\int\limits_{-\infty}^{\infty}w_{X}(t)\,x_X(t)y_X(t)\,dt, \label{eq:ipx}
\end{equation}
where $X$ indexes each detector and $S_{X}(f)$ is the detector
single-sided noise spectral density.  We modify the definitions
of~\cite{JKS1998} and~\cite{Prix_2007} to explicitly include gaps in
the time-series by introducing the function $w_{X}(t)$ which has value
$1$ when data is present and $0$ otherwise.  This also allows us to
extend the limits of our time integration to $(-\infty,\infty)$ since
the window function will naturally account for the volume and span of
data for each detector. 

The \fstext itself is defined as
\begin{equation}
  2\fstat = x_{\mu}\mathcal{M}^{\mu\nu}x_{\nu}, \label{eq:fstat}
\end{equation}
where $\mathcal{M}^{\mu\nu}$ form the matrix inverse of $\mathcal{M}_{\mu\nu}$ and we follow the shorthand notation of~\cite{Prix_2007} defining
$x_\mu \equiv (\bm{x}|\bm{h}_\mu)$ and $\mathcal{M}_{\mu\nu} \equiv (\bm{h}_\mu|\bm{h}_\nu)$. Evaluation of $\mathcal{M}$ leads to a
matrix of the form
\begin{equation}
  \mathcal{M}= \frac{1}{2}
  \left( \begin{array}{cc}
      \mathcal{C} & 0 \\
      0 & \mathcal{C}

    \end{array} \right),\hspace{1cm}
  \text{where }\mathcal{C}=
  \left( \begin{array}{cc}
      A & C \\
      C & B
    \end{array} \right)
\end{equation}
where the components 
\begin{equation}
  A=(a|a),\, B=(b|b),\, C=(a|b),
\end{equation}
are antenna pattern integrals. For a waveform with exactly known phase
evolution $\Phi(t)$ in Gaussian noise, the \fstext is a random
variable distributed according to a non-central
$\chi^{2}$-distribution with 4 degrees of freedom. The non-centrality
parameter is equal to the optimal \ac{SNR}
\begin{equation}
  \ro^{2}=\frac{1}{2}\left[A (\mathcal{A}_1^2 + \mathcal{A}_3^2) + B(\mathcal{A}_2^2 + \mathcal{A}_4^2) + 2C
    (\mathcal{A}_1 \mathcal{A}_2 + \mathcal{A}_3 \mathcal{A}_4)\right]\label{eq:rho2}
\end{equation}
such that the expectation value and variance of $2\fstat$ are given by
\begin{subequations}
\begin{eqnarray}
  \text{E}[2\fstat]&=&4+\ro^{2}\\
  \text{Var}[2\fstat]&=&8+4\ro^{2},
\end{eqnarray}
\end{subequations}
respectively. In the case where no signal is present in the data, the
distribution becomes a central $\chi^{2}$-distribution with 4 degrees
of freedom.

\subsection{\fstext and mismatched frequency}\label{subsec:fstatmm}

In this section we describe the behavior of the \fstext as a function
of search frequency $f$ for a fixed source frequency $\fo$.  In this
case the inner product that defines $x_{\mu}$ becomes
\begin{equation}\label{eq:xmu2}
  x_{\mu}= \mathcal{A}_{\nu}(\bm{h}_{\nu}|{\bm{h}_{\mu}}') + (\bm{n}|{\bm{h}_{\mu}}'), 
\end{equation}
where $\bm{h}_{\nu}$ are the components of a signal with frequency
$\fo$, and $\bm{h}'_{\mu}$ is a function of search frequency $f$.  If
we focus on the $\mu=\nu=1$ component as an example we find that
\begin{align}\label{eq:hp1}
  (\bm{h}_{1}|{\bm{h}_{1}}')&\cong\sum_{X}\frac{2}{S_X(f)}\int\limits_{-\infty}^{\infty}w_{X}(t)\,a^{2}_X(t)\cos\left(2\pi
    ft\right)\cos\left(2\pi\fo t\right)dt\nonumber\\
\end{align}
where we note that the product of cosine functions results in an
integrand that contains frequencies at $f-\fo$ and $f+\fo$.  Since
both $a_{X}(t)$ and $w_{X}(t)$ are functions that evolve on timescales
of hours--days we approximate the contribution from the $f+\fo$
component as averaging to zero.  We are left with 
%
\begin{align}
   (\bm{h}_{1}|{\bm{h}_{1}}')&\cong\text{Re}\left[\frac{1}{2}\sum_{X}\frac{2}{S_X(f)}\int\limits_{0}^{\infty}w_{X}(t)\,a^{2}_X(t)e^{-2\pi i
    (f-\fo)t}dt\right] \nonumber \\
&\cong\frac{1}{2}\sum_{X}\text{Re}\left[A_{X}(f-\fo)\right] \label{eq:hp2}
\end{align}
%
where we have defined the result of the complex integral as
$A_{X}(f)$.  This is the Fourier transform of the antenna pattern
functions weighted by the window function and evaluated at $f-\fo$.
It is equal to the quantity $(a_{X}|a_{X})$ when its argument is zero.

The quantity $a^{2}_{X}(t)$ (and similarly $b_{X}^{2}(t)$ and
$a_{X}(t)b_{X}(t)$) are periodic quantities with periods of 12 and 24
hours plus a non-oscillating component. When in a product with a
sinusoidal function and integrated over time they will result in
discrete amplitude-modulated sidebands with frequencies at $0,\pm
1/P_{\oplus},\pm 2/P_{\oplus},\pm3/P_{\oplus},\pm4/P_{\oplus}$ Hz
where $P_{\oplus}$ represents the orbital period of the Earth (1
sidereal day). We will ignore all but the zero-frequency components of
these functions for the remainder of this paper. We do note that
complications regarding the overlap of amplitude-modulated and
frequency-modulated sidebands (discussed in the next section) will
only arise for sources in binary orbits with periods equal to those
present in the antenna pattern functions.

In addition, the window function describing the gaps in the data will
influence $A_{X}(f)$.  For a gap-free observation the window function
serves to localize signal power to within a frequency range $\sim 1/T$
where $T$ is the typical observation length.  When gaps are present
this range is broadened and has a deterministic shape given by the
squared modulus of the Fourier transform of the window function.  We
can therefore use a further approximation that
\begin{equation}\label{eq:AXf}
  A_{X}(f-\fo)\approx (a_{X}|a_{X})\frac{\tilde{w}_{X}(f-\fo)}{\tilde{w}_{X}(0)}
\end{equation}
where the Fourier transform of the window function is normalized by
$\tilde{w}_{X}(0)\equiv \int dt w_{X}(t)$ such that it has a value of unity at
the true signal frequency.

We now define the antenna-pattern weighted window function as
\begin{subequations}\label{eq:Wf}
\begin{align}
  \tilde{W}(f-\fo)&\cong\sum_{X}\frac{(a_{X}|a_{X})}{A}\frac{\tilde{w}_{X}(f-\fo)}{\tilde{w}_{X}(0)}\\
  &\cong\sum_{X}\frac{(b_{X}|b_{X})}{B}\frac{\tilde{w}_{X}(f-\fo)}{\tilde{w}_{X}(0)}\\
  &\cong\sum_{X}\frac{(a_{X}|b_{X})}{C}\frac{\tilde{w}_{X}(f-\fo)}{\tilde{w}_{X}(0)},
\end{align}
\end{subequations}
which is true for observation times $T_{X}\gg$ days.  This complex
window function has the property that $\tilde{W}(0)$ is a real
quantity with maximum absolute value of unity when the template
frequency matches the true signal frequency. 

Finally we are able to combine Eqs.~\ref{eq:hp2},~\ref{eq:AXf},
and~\ref{eq:Wf} to obtain
\begin{equation}
  (\bm{h}_{1}|{\bm{h}_{1}}')\cong\frac{A}{2}\text{Re}\left[\tilde{W}(f-\fo)\right]
\end{equation}
which together with similar calculations for the additional components
in Eq.~\ref{eq:xmu2} give us 
%
%
\begin{eqnarray} \label{eq:hpmn}
  (\bm{h}_{\nu}|{\bm{h}_{\mu}}')\cong\frac{\mathcal{M}}{2}\left\{\left( \begin{array}{cc}\text{I}&0\\0&\text{I}\end{array}\right)
    \text{Re}\left[\tilde{W}(f-\fo)\right]+\left( \begin{array}{cc}0&\text{I}\\-\text{I}&0\end{array}\right)
  \text{Im}\left[\tilde{W}(f-\fo)\right]\right\} \hspace{-1.2cm} \nonumber \\ 
\end{eqnarray}
as the complete set of inner products between frequency-mismatched
signal components where $\text{I}$ is the $2\times 2$ identity matrix.
Note that when $f=\fo$ this expression reduces to Eq.~\ref{eq:fstat}.

If we now form the expectation value of the \fstext
(Eq.~\ref{eq:fstat}) for mismatched frequencies we find that
\begin{equation}
  \text{E}[2\fstat(f)]=4+\ro^{2}|\tilde{W}(f-\fo)|^{2}.
\end{equation}
Here we see that the fraction of the optimal \ac{SNR} that contributes
to the non-centrality parameter of the \fstext $\chi^{2}$ distribution
is reduced by evaluation of the mod-squared of the antenna-pattern
weighted window function with a non-zero argument.

\subsection{Frequency-modulation and the \fstext}\label{subsec:fmod}

We now consider the computation of the \fstext in the case where the
data contains a signal from a source in a circular binary orbit but
the phase model used in the \fstext template is that of a
monochromatic signal of frequency $f$. We again expand $x_{\mu}$ as
done in Eq.~\ref{eq:xmu2} where no prime indicates the signal and the
prime represents the monochromatic template.  We again focus on the
mismatched signal inner-product $(\bm{h}_{1}|{\bm{h}_{1}}')$ as an
example.  Starting with Eq.~\ref{eq:hp1} we discard the rapidly
oscillating terms inside the integral that will average to zero.  We
are then left with
\begin{align}
  (\bm{h}_{1}|{\bm{h}_{1}}')\cong\frac{1}{2}\sum_{X}\text{Re}\Bigg\{&\frac{2}{S_{X}(f)}\int\limits_{0}^{\infty}w_{x}(t)a^{2}_{X}(t)e^{-2\pi
      i t(f-\fo)}\nonumber\\
    &\exp{\left(-2\pi \fo \atrue\sin\left(\Omega(t-t_{a})\right)\right)}\Bigg\}
\end{align}
where the final term involving the exponential of a sinusoidal
function can be represented using the Jacobi-Anger expansion
\begin{equation}\label{eq:eiz}
e^{iz\sin\theta} = \sum_{n=-\infty}^{\infty} J_{n}(z)e^{in\theta}, 
\end{equation} 
where $J_{n}(z)$ is the $n^\text{th}$ order Bessel function of the first
kind.  This expansion allows us to transform the binary phase term
into an infinite sum of harmonics such that we can now write
%
\begin{align}
  (\bm{h}_{1}|{\bm{h}_{1}}')\cong\sum_{n=-\infty}^{\infty}&J_{n}(2\pi
    \fo\atrue)\nonumber \\
\quad \times\text{Re}\Bigg\{&\frac{e^{i\phi_{n}}}{2}\sum_{X}\frac{2}{S_{X}(f)}\int\limits_{0}^{\infty}w_{X}(t)a^{2}_{X}(t)e^{-2\pi it(f-\fn)}dt\Bigg\}\nonumber\\
\cong\sum_{n=-\infty}^{\infty}&J_{n}(2\pi
    \fo\atrue)\frac{A}{2}\text{Re}\left\{\tilde{W}(f-f_{n})e^{i\phi_{n}}\right\} \label{eq:hpmn2}
  \end{align}
%
It follows that all of the signal components can be expanded in the
same way giving us
\begin{align} \label{eq:hpmn3}
  (\bm{h}_{\nu}|{\bm{h}_{\mu}}')\cong\frac{\mathcal{M}}{2}&\sum_{n=-\mo}^{\mo}J_{n}(2\pi\fo\atrue)\left\{\left( \begin{array}{cc}\text{I}&0\\0&\text{I}\end{array}\right)
    \text{Re}\left[\tilde{W}(f-f_{n})e^{i\phi_{n}}\right]\right.\nonumber\\
    & \quad +\left.\left( \begin{array}{cc}0&\text{I}\\-\text{I}&0\end{array}\right)
  \text{Im}\left[\tilde{W}(f-f_{n})e^{i\phi_{n}}\right]\right\} \hspace{-1cm}
\end{align}
where we have truncated the infinite summation (explained below) and
defined the monochromatic modulated sideband frequencies and their
respective phases as
\begin{subequations}
 \begin{align}
   f_{n}&=\fo - n/\Po\,,\\
   \phi_{n}&=n\Omega t_{a}\,.
\end{align}
\end{subequations}
The Jacobi-Anger expansion has allowed us to represent the complex phase of
a frequency-modulated signal as an infinite sum of discrete signal
harmonics, or sidebands, each separated in frequency by $1/\Po$ Hz.
Each is weighted by the Bessel function of order $n$ where $n$ indexes
the harmonics and has a complex phase factor determined by the orbital
reference time $t_{a}$.  In the limit where the order exceeds the
argument, $n\gg z$, the Bessel function rapidly approaches zero
allowing approximation of the infinite sum in Eqs.~\ref{eq:eiz} and~\ref{eq:hpmn2}
as a finite sum over the finite range $[-\mo,\mo]$ where
$\mo=\texttt{ceil}[2\pi \fo \atrue]$. The summation format of Eq.~\ref{eq:hpmn3}
highlights the effects of the binary phase modulation. The signal can
be represented as the sum of $\Mo=2\mo+1$ discrete harmonics at
frequencies $\fn$ centered on the intrinsic frequency $\fo$, where
each harmonic peak is separated from the next by $1/\Po$
Hz.

Combining Eqs.~\ref{eq:fstat},~\ref{eq:xmu2}, and~\ref{eq:hpmn3}, we
can express the expectation value of the \fstext for a binary signal
as a function of search frequency $f$ as
\begin{equation}
E[2\fstat(f)]=4+\ro^{2}\sum_{n=-\mo}^{\mo}J^{2}_{n}(2\pi \fo\atrue)|\tilde{W}(f-f_{n})|^{2}.\label{eq:Fstat3}
\end{equation}
This expression should be interpreted in the following manner.  For a
given search frequency $f$ the contribution to the non-centrality
parameter (the \ac{SNR} dependent term) is equal to the sum of all
sideband contributions at that frequency.  Each sideband will
contribute a fraction of the total optimal \ac{SNR} weighted by the
$n^\text{th}$ order Bessel function squared, but will also be strongly
weighted by the window function.  The window function will only
contribute significantly if the search frequency is close to the
sideband frequency.  Hence, at a given search frequency close to a
sideband, for observation times $\gg \Porb$, the sidebands will be far
enough separated in frequency such that only one sideband will
contribute to the \fstext. 

\subsection{\cstext} \label{subsec:Cstat}

The \fstext is numerically maximized over the phase parameters of
the signal on a discrete grids.  For this search the search frequency
$f$ is such a parameter and consequently the \fstext is computed over
a uniformly spaced set of frequency values $f_{j}$ spanning the region
of interest. In this section we describe how this \fstext
frequency-series can be used to approximate a search template that is
then used to generate a new statistic sensitive to signals from
sources in binary systems.

The expectation value of the \fstext (Eq.~\ref{eq:Fstat3}) resolves
into localized spikes at $\Mo$ frequencies separated by $1/\Po$ Hz and
centered on the intrinsic gravitational wave frequency $\fo$. A template
$\stat{T}$ based on this pattern with amplitude defined by $G_n$, takes
the general form
\begin{equation}
  \temp(f)=\sum_{n=-m'}^{m'}G_n|\tilde{W}\left(f-\fn'\right)|^{2}\label{eq:template}\,,
\end{equation}
with $m'=\texttt{ceil}[2\pi f a']$ and $\fn'=\fo'-n/P'$ where we make a distinction between the intrinsic (unknown) values of each parameter (subscript zero) and values selected in the template construction (denoted with a prime). The
window function $\tilde{W}$ is dependent only on the times for which
data is present and is, therefore, also known exactly. 

We define our new detection statistic \ctext as
\begin{eqnarray}
  \mathcal{C}(f)&\equiv&\sum_{j}2\fstat(f_{j})\temp(f_{j}-f)\nonumber\\
  &=&\left(2\fstat\ast\temp\right)(f),
\end{eqnarray}
where the sum over the index $j$ indicates the sum over the discrete
frequency bins $f_{j}$ and $f$ is the search
frequency. Since the template's ``zero frequency'' represents the intrinsic gravitational wave frequency, $f$ corresponds to the intrinsic frequency.  We see that the \cstext is, in fact, the convolution of the
\fstext with our template, assuming the template is constant
with search frequency (an issue we address in the next section).

The benefit of this approach is that the computation of the \fstext for a
known sky position and without accounting for binary effects has
relatively low computational cost.  Similarly, the construction of a
template on the \fstext is independent of the orbital phase parameter
and only weakly dependent upon the orbital semi-major axis and
eccentricity.  The template is highly dependent upon the orbital
period, which, for the sources of interest, is known to high
precision.  Also, since the \cstext is the result of a convolution, we
can make use of the convolution theorem and the speed of the \ac{FFT}. 
Computing $\mathcal{C}$ for all frequencies requires only three applications
of the \ac{FFT}. In practice, the \cstext is computed using
\begin{align}\label{eq:disC}
  &\mathcal{C}(f_{k})=\left(2\fstat\ast\temp\right)(f_{k})\nonumber\\
  &=\sum_{j=0}^{N-1}e^{2\pi
    ijk/N}\left(\sum_{p=0}^{N-1}e^{-2\pi ipj/N}2\fstat(f_{p})\right)\left(\sum_{q=0}^{N-1}e^{-2\pi ijq/N}\temp(f_{q})\right),
\end{align}
which is simply the inverse Fourier transform of the product of the
Fourier transforms of the \fstext and the template.  The \fstext and
the template are both sampled on the same uniform frequency grid
containing $N$ frequency bins. The \cstext is then also output as a
function of the same frequency grid.

\subsection{Choice of \fstext template} \label{subsec:template}

\begin{figure}
\includegraphics[width=\columnwidth]{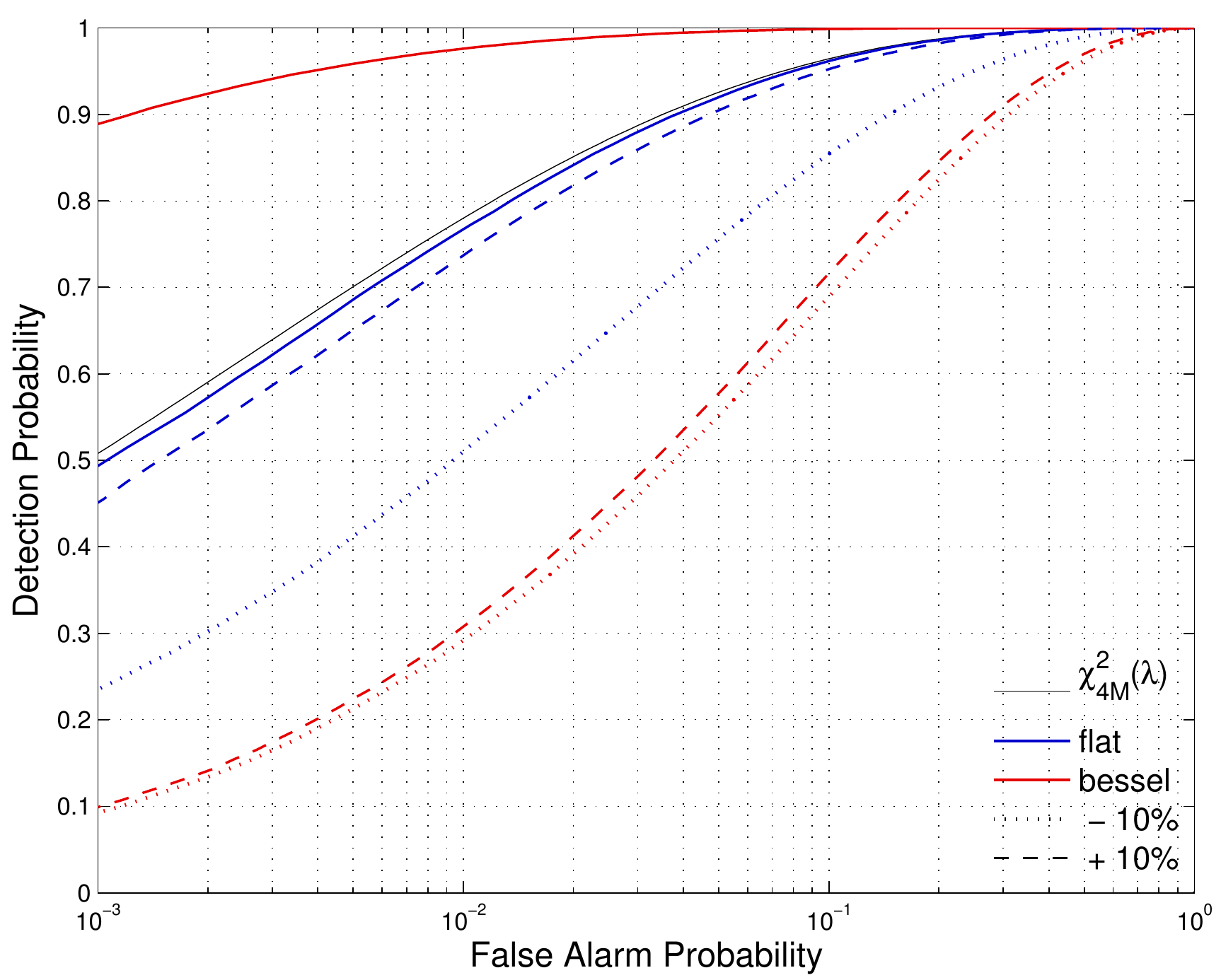}
\caption{\label{fig:ROCtempcompare}\ac{ROC} curves comparing performance of the flat (blue) and Bessel function weighted (red)
  templates, described by Eqs \ref{eq:flattemp} and \ref{eq:besseltemp} respectively. The theoretical (fine black) curve is constructed from a
  non-central $\chi^2_{4M}(\lambda)$ distributed statistic with non-centrality parameter $\lambda=\ro^2=20$ and represents the expected performance of the flat template. Dashed and dotted curves represent a template with a positive and negative $10\%$ error on the semi-major axis respectively. Here the signal parameters were chosen such that the number of sidebands were $M=2001$ and curves were constructed using $10^6$ realizations of noise.}
\end{figure}

Treating the \fstext as the pre-processed input dataset to the \cstext
computation, it might be assumed that the optimal choice of template is
that which exactly matches the expected form of $2\fstat$ in the
presence of a signal.  As shown in Fig.~\ref{fig:ROCtempcompare}, this
approach is highly sensitive to the accuracy with which the projected orbital
semi-major axis is known.

We instead propose the use of a far simpler template: one that
captures the majority of the information contained within the \fstext
and, by design, is relatively insensitive to the orbital semi-major
axis.  We explicitly choose
\begin{equation}\label{eq:flattemp}
  \flattemp(f_k)=\sum_{j=-m}^{m}\delta_{k\,\lj}\,,
\end{equation}
for discrete frequency $f_k$, where $\delta_{ij}$ is the Kronecker delta-function. The frequency
index $l$, defined by,
\begin{equation}\label{eq:findex}
  \lj\equiv\texttt{round}\left[\frac{j}{P'd_f}\right],
\end{equation}
is a function of the best-guess orbital period $P'$ and the frequency resolution $d_f$. The $\texttt{round}[]$ function returns the integer closest to its argument.  The template is therefore composed of the sum of $M=2m+1$ unit amplitude ``spikes''
positioned at discrete frequency bins closest to the predicted locations of the frequency-modulated sidebands (relative to the intrinsic gravitational wave frequency).  The subscript $\text{F}$ refers to the constant amplitude, ``Flat'' template.

If we now convolve this template with the frequency-modulated \fstext we obtain the corresponding \cstext, which reduces to
\begin{equation}
 \cstat(f_{k})=
 \sum_{j=-m}^{m}2\fstat(f_{k-\lj})\,.
\label{eq:cstat_comb}
\end{equation}
Equation \ref{eq:cstat_comb} is simply the sum of $2\fstat$ values taken from discrete
frequency bins positioned at the predicted locations of the
frequency-modulation sidebands. An example is shown in the right hand
panel of Fig. \ref{fig:search}, where the most significant \cstext
value is located at $f=\fo$, and is the point where all sidebands included in
the sum contain some signal.

From Eq. \ref{eq:Fstat3}, the expectation value for the \cstext using the flat-template can be expressed as
\begin{equation}\label{eq:ECstat}
  \text{E}[\mathcal{C}(f_k)]= 4M + 
\ro^{2}\sum_{n=-m}^{m}J^{2}_{n}(2\pi
\fo\atrue)|\tilde{W}(f_{n}-f_{k}+l_{[n]}\fres)|^2,
\end{equation}
where the argument of the window function is the frequency difference between
the location of the $n^\text{th}$ signal sideband and the $n^\text{th}$ template sideband on the discrete frequency grid.  Note that the \cstext is a sum of $M$ statistically independent non-central $\chi^{2}_4$ statistics and hence the result is
itself a non-central $\chi^{2}_{4M}$ statistic, i.e. with $4M$ degrees of freedom, where $M=2\texttt{ceil}[2\pi f \asini] +1$ is the number of sidebands in the expected modulation pattern. The non-centrality parameter is equal to the sum of the
non-centrality parameters from each of the summed $2\fstat$ values.
For a flat template with perfectly matched intrinsic frequency $f=f_0$ and orbital
period $P'=P$, infinite precision $\fres\rightarrow 0$, and where
the number of sidebands included in the analysis matches or exceeds
the true number, the second term in the above equation reduces to
$\ro^{2}$.  In this case we will have recovered all of
the power from the signal but also significantly increased the
contribution from the noise through the incoherent summation of
\fstext from independent frequency bins.  In general, where the
orbital period is known well, but not exactly, and the frequency
resolution is finite, the signal power recovery will be reduced by
imperfect sampling of the window function term in Eq.~\ref{eq:ECstat},
i.e. evaluation at arguments $\neq 0$.

In terms of the generic template defined in equation
Eq.~\ref{eq:template}, the discrete-frequency flat template is
approximately equivalent to the weighting scheme $G_n=1$. A more sensitive approach could use
\begin{equation}
  \temp_{\text{B}}(f_k)=\sum_{j=-m}^{m}J^{2}_{j}(2\pi \fo\, \atrue)\delta_{k\,\lj}\label{eq:besseltemp}
\end{equation}
for the template, following the expected form of the \fstext given in
Eq.~\ref{eq:Fstat3} and using a subscript $\text{B}$ to denote
Bessel function weighting.  Although this would
increase sensitivity for closely matched signal templates (constructed with well constrained signal parameters), this
performance is highly sensitive to the number of sidebands included in the template and therefore sensitive to the semi-major
axis since $M=2\texttt{ceil}[2\pi f \asini] +1$. This is mainly due to the
``double horned'' shape of the expected signal (see the left hand
panel of Fig.~\ref{fig:search}).  A large enough offset between the
true and assumed semi-major axis will significantly change the template's overlap with the sidebands in the \fstext and reduce the significance of the \cstext. Considering the semi-major axis is not well constrained for many \acp{LMXB}, a search over many templates would be necessary,
each with incrementally different semi-major axis values.

The simpler, flat-template (Eq. \ref{eq:flattemp}) has the benefit of being far more
robust against the semi-major axis uncertainty.  In this case the semi-major axis
parameter controls only the number of sidebands to use in the template
and does not control the weighting applied to each sideband. It also
simplifies the statistical properties of the \cstext, making a
Bayesian analysis of the output statistics (as described in Section
\ref{sec:stats}) far easier to apply.

The \ac{ROC} curves shown in Fig.~\ref{fig:ROCtempcompare} compare the
performance of the sideband search with both choices of template (flat
and Bessel function weighted) for the case of signal with optimal
\ac{SNR} $\ro=20$. As seen from the figure, the Bessel function
weighted template for exact number of sidebands provides improved
sensitivity over the flat template. However, when considering the
possible (and highly likely) error on the number of sidebands in the
template, the performance of the Bessel template is already
drastically diminished, even with only a $10\%$ error on the semi-major axis
parameter. It is also interesting to note that for the flat-template
the result of an incorrect semi-major axis is asymmetric with respect
to an under or over-estimate.  The sensitivity degradation is far
less pronounced when the template has over-estimated the semi-major
axis and, therefore, also over-estimated the number of sidebands. This
feature is discussed in more detail in Sec.~\ref{subsec:asini}.  

\subsection{Approximate binary demodulation}\label{subsec:demod}

When a putative source has a highly localized position in the sky, the effect of the Earth's motion with respect to the \ac{SSB} can be accurately removed from the signal during the calculation of the \fstext. This leaves only the Doppler
modulation from the binary orbit. It is also possible to demodulate
the binary orbit (Doppler) modulation in the \fstext calculation provided
the binary orbital parameters $(a,\Porb,\ta)$ are well known. A fully-coherent (sky position- and binary-) demodulated
\fstext search would be very sensitive to any errors in the
sky position of binary orbital parameters. It would therefore be necessary to construct
a bank of templates spanning the parameter space defined by the
uncertainties in these parameters. Adding dimensions to the parameter space increases computational costs and
the search becomes unfeasible considering we are already searching over frequency. A fully coherent search of this type would be possible for
known sources with known emission frequencies, for example pulsing
sources like \acp{msp}.

In this section we show how prior information regarding the binary
orbit of a source can be used to increase the sensitivity of our
semi-coherent approach, without increasing computational costs. By
performing a ``best guess'' binary phase demodulation within the
\fstext, we show that the number of sidebands in the template is
reduced by a factor proportional to the fractional uncertainty in the
orbital semi-major axis. Consequently a reduction in the number of
sidebands increases the sensitivity of the search by reducing the
number of degrees of freedom (see Sec. \ref{sec:stats}).

Expressing our current best estimate for each parameter $\bm{\theta}$ as the sum of
the true value $\bm{\theta_\submath{0}}$ and an error $\Delta\bm{\theta_\submath{0}}$, such that
\begin{equation}
  \bm{\theta} = \bm{\theta_\submath{0}} +\Delta\bm{\theta_\submath{0}},
\end{equation}
we can determine the phase offset of the binary orbit from the
error in the binary orbital parameters. The offset in phase
is the difference between the true and best estimate binary phase and
using Eq.~\ref{eq:phi_bin} can be approximated by
\begin{eqnarray}
 \Delta \phi_\text{bin} &\simeq&-2\pi f\left\{\phantom{\frac{}{}}\left(\atrue\Delta f + f\Delta\atrue\right)\sin\left[\Omega_\submath{0}(t -
        \ta)\right] \right. \nonumber \\
 & +&
 \left. \left[f \atrue\left(\Delta\Omega_\submath{0}\left(t-t_{\text{a}}\right)-\Omega_\submath{0}\Delta
     \ta\right)\right]\cos\left[\Omega_\submath{0}(t -
        \ta)\right]\phantom{\frac{}{}}\right\}+\mathcal{O}(\bm{\Delta\theta}^{2}) \nonumber \\
&\simeq& -2\pi f \Delta \atrue\,\kappa \sin\left[\Omega_\submath{0}(t-\ta) + \gamma\right],
\end{eqnarray}
with
\begin{eqnarray}
  \kappa &=& \sqrt{\left(1+\frac{\atrue}{\Delta \atrue}\frac{\Delta f}{f}\right)^{2}+\left(\frac{\atrue}{\Delta
      \atrue}\left(\Delta\Omega_\submath{0}(t-\ta)-\Omega_\submath{0}\Delta \ta\right)\right)^{2}},  \label{eq:moddepth}\\
\gamma &=& \text{tan}^{-1} \left[
    \frac{\Delta\Omega_\submath{0}(t-\ta) - \Omega_\submath{0}\Delta
      t_{\text{a}}}{(\Delta \atrue/\atrue)-(\Delta f/f)} \right]+\left\{ 
  \begin{array}{l l}
    0 &\,\text{if}\,\left(\frac{\Delta \atrue}{\atrue}\right)-\left(\frac{\Delta f}{f}\right)\ge 0\\
    \pi &\,\text{if}\,\left(\frac{\Delta \atrue}{\atrue}\right)-\left(\frac{\Delta f}{f}\right)< 0.
  \end{array} \right.\nonumber\\
\end{eqnarray}
Here we have expanded the binary phase difference to leading order in
the parameter uncertainties and obtained a phase expression similar in
form to the original binary phase.  In the specific regime where the
fractional uncertainty in the orbital semi-major axis far exceeds the
fractional uncertainty in the intrinsic frequency we see that the
first term in Eq.~\ref{eq:moddepth} becomes $\approx 1$.  Similarly,
if the fractional uncertainty in the orbital semi-major axis also far
exceeds the deviation in orbital angular position
$\Delta\Omega(t-\ta)-\Omega\Delta \ta$ then the
second term $\approx 0$.  This is generally the case for the known
\acp{LMXB} (see Table~\ref{tab:targets}) and in this regime $\kappa$
can be accurately approximated as unity, yielding
\begin{align}
 \Delta \phi_{\text{bin}} &\approx -2\pi f\Delta \atrue\sin[\Omega(t-\ta)+\gamma]. \label{eq:Dphi_bin}
\end{align}
Hence, after approximate binary demodulation, the argument of the
Bessel function and the summation limits in the expected form of the
\fstext (in Eq.~\ref{eq:Fstat3} for example) can be replaced with
$\Delta z_\submath{0}=2\pi f \Delta \atrue$. The number of frequency-modulated
sidebands is now reduced by a factor of $\Delta \atrue/\atrue < 1$. We must
stress that $\Delta a$ is an unknown quantity and is the difference
between the best estimate value of $\asini$ and the true value $\atrue$.  The
\fstext after such a demodulation process will therefore have a
reduced but unknown number of sidebands, although it will still retain
the standard sideband frequency spacing $1/P$.  The sideband phasing
$\phi_{n}$ will also be unknown due to the presence of the phase term
$\gamma$ but is of no consequence to the search since the \fstext is
insensitive to phase.

\section{Parameter space}\label{sec:params}

In this section we will discuss each of the parameters involved in the
search and how the search sensitivity depends upon the uncertainty in
these parameters. Demodulation of the signal phase due to the Earth's
motion requires accurate knowledge of the source sky position. If the
observation time is long enough, we need to consider the sky position
as a search parameter, as discussed in Sec.~\ref{subsec:skypos}. The
gravitational wave frequency is the primary search parameter. In
Sec.~\ref{subsec:spinf} we discuss the limitations on
our search strategy due to its uncertainty. The orbital period
and semi-major axis are discussed in Sections \ref{subsec:period} and
\ref{subsec:asini} respectively. The effects of ignoring the orbital
eccentricity are discussed in Sec. \ref{subsec:ecc}.

\subsection{Sky position and proper motion}\label{subsec:skypos}

In order to quantify the allowable uncertainty in sky position we will
define a simplistic model describing the phase $\Psi(t)$ received at
Earth from a monochromatic source at infinity at sky position
$(\alpha,\delta)$.  If we neglect the detector motion due to the spin of
the Earth and consider only the Earth's orbital motion then we have
\begin{equation}
 \Psi(t) = 2\pi\fo \left[t + R_{\oplus}\cos\delta\cos\left(\Omega_{\oplus} (t-t_{\text{ref}}) + \alpha\right)\right],
\end{equation}
where $\fo$ is the signal frequency, $\alpha$ and $\delta$ are the
true right ascension and declination and $R_{\oplus}$ and
$\Omega_{\oplus}$ are the distance of the Earth from the \ac{SSB} and
the Earth's orbital angular frequency respectively.  We also define a
reference time $t_{\text{ref}}$ that represents the time at which the
detector passes through the vernal equinox. For an observed sky
position $(\alpha',\delta')=(\alpha+\Delta\alpha,\delta+\Delta\delta)$
the corresponding phase offset $\Delta
\Psi(t,\Delta\alpha,\Delta\delta) = \Psi(t,\alpha',\delta') -
\Psi(t,\alpha,\delta)$ amounts to
\begin{eqnarray}
\Delta \Psi &\approx& -2\pi\fo R_{\oplus}
\Big[\Delta\delta\sin\delta\cos\left(\Omega_{\oplus}(t-t_{\text{ref}})+\alpha\right)\nonumber\\
&&+\Delta\alpha\cos\delta\sin\left(\Omega_{\oplus}(t-t_{\text{ref}})+\alpha\right)\Big],
\end{eqnarray}
where we have expanded the expression to leading order in the sky
position errors.  We now make the reasonable assumption that our
analysis would be unable to tolerate a deviation in phase between the
signal and our template of more than $\mathcal{O}(1)$ radian over the
course of an observation on the same timescale of the Earth's orbit.

If we also notice that the worst case scenario (smallest allowable sky
position errors) corresponds to sky positions for which the
trigonometric terms in the previous expression are largest, i.e. of
order unity, then we have
\begin{equation} \label{eq:dbeta}
 |\Delta \alpha,\Delta\delta| \leq (2\pi\fo R_{\oplus})^{-1}.
\end{equation}
If we consider signals of frequency $1$kHz, this gives a maximum
allowable sky position offset of $|\Delta \alpha, \Delta\delta| \simeq
100$ mas. This expression also validates our model
assumption that the sky position sensitivity to the Earth spin would
be dominated by the effect from the Earth orbit for long observation
times.

A similar argument can be made for the proper motion of the source
where we would be safe to model the sky position as fixed if the
change
$(\Delta\alpha,\Delta\delta)=(\mu_{\alpha},\mu_{\delta})\Tspan$, over
the course of the observation also satisfied Eq.~\ref{eq:dbeta}.

\subsection{Spin frequency}\label{subsec:spinf}

The spin frequency $\spinf$ of some \acp{LMXB} can be directly
measured from X-ray pulsations, believed to originate from a hot-spot
on the stellar surface, where accreted material is funneled onto the
magnetic pole with the magnetic axis generally misaligned with the
spin axis. X-ray pulsations have been observed in 13 \ac{LMXB} systems
so far, three of which are intermittent~\cite{LambEA_2011}.

Some \acp{LMXB} exhibit recurrent thermonuclear X-ray bursts. Fourier
spectra reveal oscillations during the rise and tail of many bursts,
which are believed to originate from asymmetric brightness patterns on
the stellar surface. In seven \acp{LMXB} which exhibit both pulsations
and bursts, the asymptotic burst oscillation frequency at late times
matches the pulse frequency. Where there are no pulsations, many
bursts need to be observed to measure the asymptotic burst oscillation
reliably. The spin frequency of an additional ten systems has been
determined from burst oscillations only~\cite{Watts_2012}, but due to
the uncertainties involved, are usually quoted to within uncertainties
of $\pm(5-10)$ Hz.

Another class of \acp{LMXB} exhibit high frequency \acp{QPO} in their
persistent X-ray emission. These kHz \acp{QPO} usually come in pairs,
although singles and triples are occasionally observed and the
\ac{QPO} peak frequencies usually change over time. In some cases the
separation of the \ac{QPO} peaks is roughly constant, but this is not
always the case \cite{vanderKlis_1998, ZhangEA_2012,
  vanderKlisEA_1996}. For the few \ac{QPO} systems where $\spinf$ can
be determined from pulses or burst oscillations there has been no
evidence suggesting consistency with an existing \ac{QPO} model that
links the \ac{QPO} and spin frequencies.  For our purposes, $\spinf$
is considered unknown in sources without pulsations or confirmed
bursts.

In addition to potentially broad uncertainties in $\spinf$, we know
little about how its value may fluctuate over time due to
accretion. Changes in the accretion flow will exert a time varying
torque on the star which will result in a stochastic wandering of the
spin frequency.  In this case the signal can no longer be assumed
monochromatic over a given observation time. To quantify the resulting
phase wandering, we assume that the fluctuating component of the
torque $\delta \Na$ flips sign randomly on the timescale $\tspinw$
consistent with the inferred variation in accretion rate.  If the mean
torque $\Na = \dot{M}(GMR)^{1/2}$ due to steady-state disk-fed
accretion, then the angular spin frequency $\Omega_{s}=2\pi\spinf$
experiences a random walk with step size $(\delta \Na/I)\tspinw$, where
$I$ is the stellar moment of inertia. After time $\Tspan$, the
root-mean-square drift is
\begin{equation}
\langle (\delta \Omega)^2 \rangle ^{1/2} = \left({\Tspan}/{\tspinw}\right)^{1/2} \frac{\delta \Na \tspinw}{I}.
\end{equation}
This frequency drift will wander outside a Fourier frequency bin width if $\langle
(\delta \Omega)^2 \rangle ^{1/2} > 2\pi/\Tspan$. If we choose $\tspinw$ such that the accretion rate can vary up to a factor of two in this time, then the worst case $\delta\Na=\Na$ leads to the restriction
\begin{eqnarray} \label{eq:Ts_spin}
 \Tspansw < \frac{(2\pi)^{2/3}}{(GMR)^{1/3}}
\left(\frac{I}{\dot{M}}\right)^{2/3} \left(\frac{1}{\tspinw}\right)^{1/3}.
\end{eqnarray}
This is the primary reason why an application of the the basic
sideband search, as described here, must be limited in the length of
data it is allowed to analyse.  By exceeding this limit it becomes
increasingly likely that the spin wandering inherent to a true signal
will cause signal power to leak between adjacent frequency bins.
Consequently the assumption that \fstext signal power is localized in
frequency-modulated sidebands will become invalid and the sensitivity
of the \cstext will deteriorate.

\subsection{Orbital Period}\label{subsec:period}

The sideband search relies on relatively precise \ac{EM} measurements
of the orbital period in order to construct a search template. The
duration of the orbit defines the minimum observation time, since $T
\gtrsim 3\Porb /2$ is required before sidebands
become clearly resolved in the spectrum~\cite{RansomEA_2003}. The
uncertainty in the orbital period will determine the number of
templates required to fully sample the search space, or equivalently,
the maximum observation time allowed for a single value of $\Porb$.

We will now provide an indication of the sensitivity of the search to
errors in the orbital period. If our estimate (observation) of the orbital
period $\Pmeas$ is offset from the true value $P_\submath{0}$ by an amount
$\Delta P$, we would expect the error to seriously affect the \cstext recovered
from the search once it is large enough to shift the outermost ``tooth'' in the sideband template by one canonical frequency bin away from the true sideband location. In this case, the offset between the template and true sideband frequency is
proportional to the number of sidebands from the central spike. There
will be low mismatch at the center of the template extending to
$\mathcal{O}(100\%)$ mismatch at the edges.  It follows that the
average signal power recovered from such a mismatched template will
be $\mathcal{O}(50\%)$ and therefore serves as a useful threshold by
which to determine the maximum allowed $\Delta\Porb$.

If we use the measured value of $\Porb$ as our template parameter, the template
centered at frequency $f$ then consists of $\simeq 4\pi
f\asini$ unit spikes (or teeth) separated by $1/\Porb$. Assuming that the
central spike is exactly equal to the true intrinsic gravitational wave frequency, any
errors in the orbital period will be propagated along the comb, causing the
offset between the true and template frequency of any particular sideband to
grow progressively larger.  The frequency difference $\Delta f_P$
between the outermost template sideband, at frequency $f + 2\pi f
\asini/\Porb'$, and the outermost signal sideband at $f + 2\pi f
\asini/\Porb$, is given by
\begin{equation} \label{eq:deltaf}
  \Delta f_P \approx 2\pi f \asini \left(\frac{|\Delta \Porb|}{\Porb^{2}}\right),
\end{equation}
for $\Delta \Porb \ll \Porb$.  To satisfy the condition described
above we now require that this frequency shift should be less than the size of
one frequency bin. The true frequency bin size $\fres$ is determined by
the observation time span and is given by 
\begin{equation}\label{eq:fres}
\fres=\frac{1}{r\Tspan},
\end{equation}
where $r$ is the resolution used in the \fstext calculation.\footnote{The default resolution factor is $r=2$} Using Eqs. \ref{eq:deltaf} and \ref{eq:fres} and imposing the condition that $\Delta f_P < \fres$ provides an estimate for the maximum allowable (orbital period limited) coherent
observation timespan, 
\begin{eqnarray}\label{eq:Ts_Porb}
\Tspan^{\Porb} \approx \frac{\Porb^{2}}{2\pi f\asini |\Delta \Porb|}.
\end{eqnarray}
Given a relatively poorly constrained orbital period uncertainty, this
restriction may provide too short a duration.  This could be because
it is then in conflict with the requirement that $T>3\Porb/2$ or
simply because more \ac{SNR} is desired from the signal.  In either
case, the orbital period space must then be divided into templates
with spacing $\delta \Porb$ derived from simply rearranging
Eq.~\ref{eq:Ts_Porb} to solve for $\Delta \Porb$. In relative terms the sideband search places very strong constraints
on the prior knowledge of the orbital period compared to the other
search parameters.

\subsection{Semi-major axis}\label{subsec:asini}

An error in the value of the orbital semi-major axis results in an
incorrect choice for the number of sidebands in the template.  As can
be seen in Eq.~\ref{eq:ECstat}, an underestimate results in the
summation of a fraction of the total power in the signal whereas an
overestimate results in a dilution of the total power by summing
additional noise from sideband frequencies containing no signal
contribution. 

If we define the true semi-major axis parameter $\atrue$ as the measured value
$\ameas$ and some (unknown) fraction $\xi$ (where $\xi\in \mathbb{R}$) of the
measurement error $\Delta \ameas$ (i.e. $\atrue = \ameas +\xi\Delta \ameas$), we can investigate
the effects of errors on the semi-major axis parameter in terms of this offset
parameter $\xi$. We consider the advantage of using a deliberately offset value
$\atemp$ instead of the observed value $\ameas$ in order to minimize losses in
recovered \ac{SNR}.

\begin{figure}
\includegraphics[width=\columnwidth]{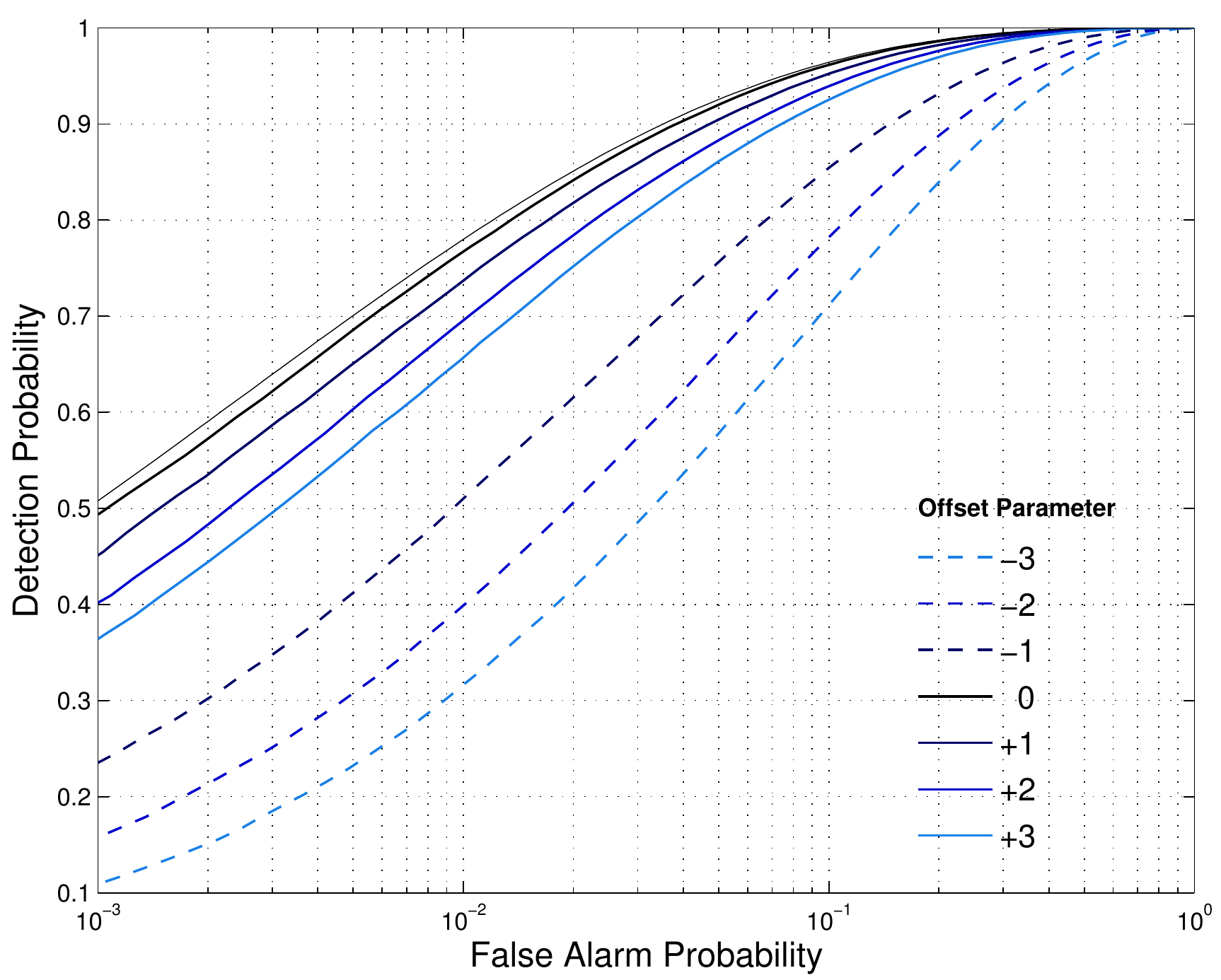}
\caption{\label{fig:asiROCflat}\ac{ROC} curves showing how the
  performance of the flat template \cstext is affected by an offset in
  the orbital semi-major axis assuming it is measured exactly (i.e. $\ameas=\atrue$).
The thick black curve represents a zero offset ($\xi=0$). Thick colored curves
represent a positive offset in the semi-major axis  ($\xi>0$). Dashed colored
curves represent negative offsets ($\xi<0$). The fainter black curve is
constructed from a statistic governed by a non-central $\chi^2_{4M}(\lambda)$
  distribution with a non-centrality parameter $\lambda=\ro^2$ and represents
the theoretical expected behavior of a perfectly matched template. Signal
  parameters are the same as described in Fig. \ref{fig:ROCtempcompare}, with
$\ro=20$, $M=2001$ and using $10^6$ realizations of noise.}
\end{figure}

\begin{figure}
\includegraphics[width=\columnwidth]{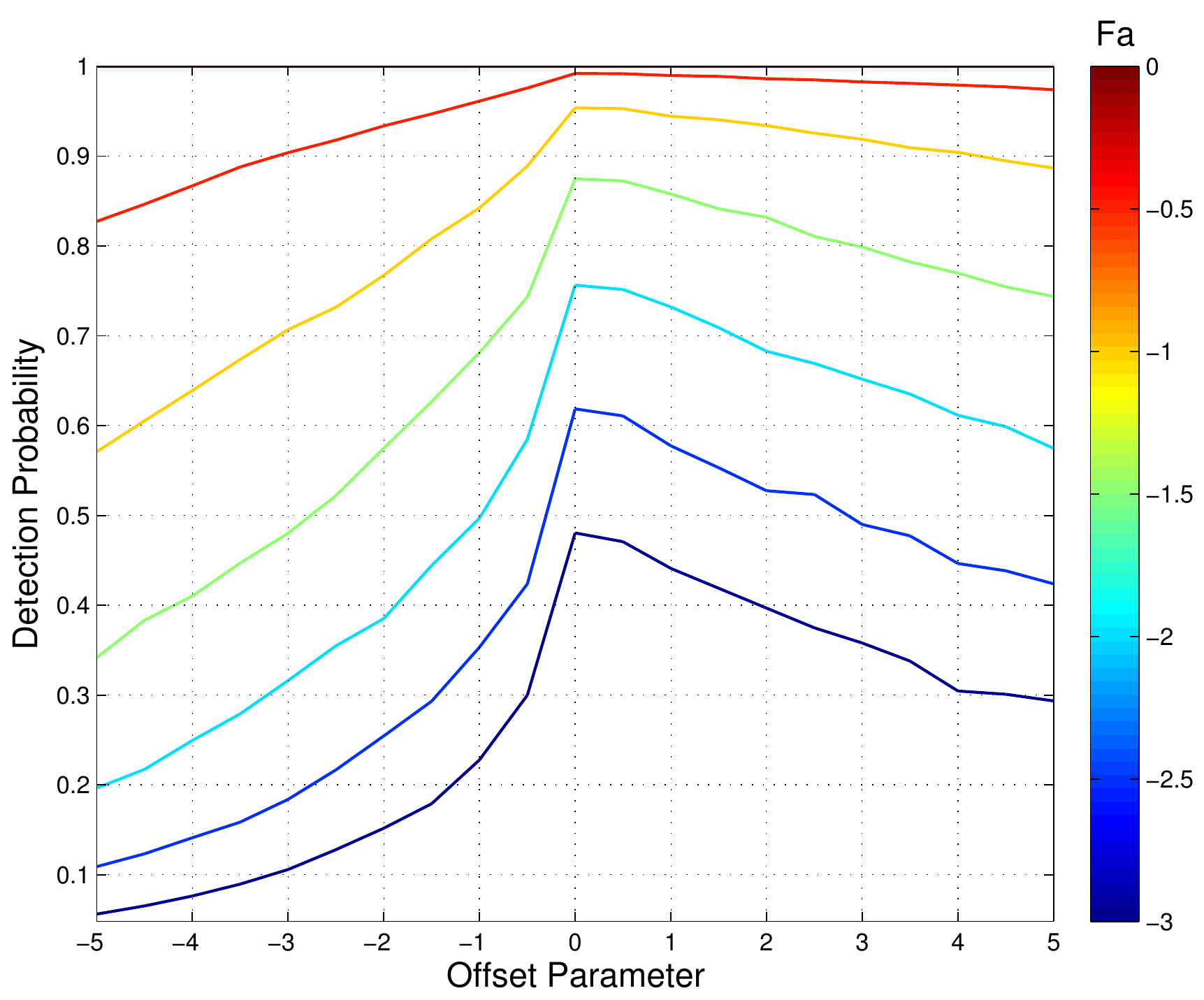}
\caption{\label{fig:asiROCperformance}Performance of the \cstext with respect to offsets in the semi-major axis at given false alarm probability $F_a$. The
offset parameter $\xi$ quantifies the semi-major axis error in terms of known parameters $\ameas$ and $\Delta \ameas$. The color-bar represents the log of the false alarm probability, ranging from $10^{-3}$ (bottom, blue) to 0 (top, red). Signal parameters are the same as Figs. \ref{fig:ROCtempcompare} and \ref{fig:asiROCflat}, with $\ro=20$, $M=2001$ and a $\Delta \ameas/\ameas=10\%$ fractional uncertainty on the semi-major axis, using $10^6$ realizations of noise.}
\end{figure}

The \ac{ROC} curves shown in Fig. \ref{fig:asiROCflat} show the
effects of these offsets, and clearly illustrate degradation in the
performance of the \cstext as $|\Delta \ameas|$ ($|\xi|$) increases. The
reduction in detection confidence at a given false alarm probability
is much faster for $\atemp < \atrue$ ($\xi<0$), when the template underestimates
the width of the sideband structure, than for $\atemp>\atrue$ ($\xi>0$). This is
natural considering the ``horned'' shape of the signal (see the left
hand panel of Fig.~\ref{fig:search} and
Sec.~\ref{subsec:template}). Although it is already clear from this
figure that the performance of the search is not symmetric about
$\atemp=\atrue$, this asymmetry is much better illustrated in
Fig. \ref{fig:asiROCperformance} where for different values of the
false alarm rate we show the detection probability plotted against the
offset parameter $\xi$.

This plot provides us with a rough scheme by which to improve the
search performance by exploiting the asymmetry in search sensitivity
with respect to $\xi$. In general, we are keen to probe the
low false alarm and high detection probability regime in which it is
clear that using a template based on an orbital semi-major axis value
$>$ the best estimate reduces the possibility that the bulk of the
signal power (in the horns) will be missed.  Based on
Fig.~\ref{fig:asiROCperformance} we choose
\begin{equation}
  \atemp = \ameas + \Delta \ameas
\end{equation}
as our choice of semi-major axis with which to generate the search template.
%

\subsection{Orbital eccentricity}\label{subsec:ecc}

The orbital eccentricity $e$ of the \ac{LMXB} sources is expected to be
highly circularized ($e<10^{-3}$) by the time mass transfer occurs
within the system.  In Eq.~\ref{eq:bintime2} we give the first-order
correction (proportional to $e$) of the retarded time at the
\ac{SSB}. If we include higher order terms in the expansion, the phase
(Eq.~\ref{eq:binphase}) can be written as
\begin{align}
  \Phi(t) \simeq
  2\pi\fo\atrue\left\{\sum_{k=1}^{\infty}c_{k}\sin\omega\cos\left[k\Omega\left(t-\tp\right)\right]\right.\nonumber\\
  +d_{k}\cos\omega\sin\left[k\Omega\left(t-\tp\right)\right]\Bigg\},
\end{align}
where the first 4 coefficients (expanded to $\mathcal{O}(e^4)$) in the
sum are given by
\begin{subequations}
\begin{eqnarray}
  c_{1} &=& 1-\frac{3}{8}e^{2}+\frac{5}{192}e^{4} +
    \mathcal{O}(e^{6})\\
  d_{1} &=& 1-\frac{5}{8}e^{2}-\frac{11}{192}e^{4} +
    \mathcal{O}(e^{6})\\
    c_{2} &=& \frac{1}{2}e-\frac{1}{3}e^{3} + \mathcal{O}(e^{5})\\
    d_{2} &=& \frac{1}{2}e-\frac{5}{12}e^{3} + \mathcal{O}(e^{5}).
\end{eqnarray}
\end{subequations}
Hence the phase for $e \neq 0$ is a sum of harmonics of the orbital
frequency. When including additional eccentric phase components in
this way, the sum inside the exponential can be expressed as a product of sums such that the Jacobi-Anger expansion (Eq.~\ref{eq:eiz}) can be
modified such that
\begin{equation}
  \exp\left[i\sum_{k}z_{k}\sin k\theta\right]=\prod_{k=1}^{\infty}\sum_{n=-\infty}^{\infty}J_{n}(z_{k})e^{ink\theta}, \label{eq:eizP}
\end{equation}
where $z_k$ corresponds to the $k^\text{th}$ amplitude term (on the left hand side) that defines the argument of the Bessel function for each $k$ in the product (on the right hand side). Equation \ref{eq:eizP} tells us that eccentric signals can be thought of in a similar
way to circular orbit cases.  The signal can be modeled as being
composed of many harmonics all separated by some integer number of the
inverse of the orbital period.  In the eccentric case $k$ is allowed
to be $>1$ and power can be spread over a far greater range of
harmonics.  What is important to note however, is that the signal
power remains restricted to those discrete harmonics.

If we consider only leading order terms in the eccentricity expansion
(as in Eq.~\ref{eq:bintime2}) the form of the Jacobi-Anger expression given
above becomes the product of 2 sums where we consider only $k=1,2$.
The $k=1$ terms are simply the circular orbit terms and describe how
the signal power is distributed amongst $\approx 2z_{1}$
sidebands at frequencies offset from the intrinsic source frequency by
integer multiples of $\theta$.
 
In our low eccentricity case we notice that the next to leading order
term in the expansion, $k=2$, has a corresponding Bessel function
argument of $\mathcal{O}(z_{1}e)$ and will therefore have far fewer,
$\sim 2z_{1}e$, non-negligible terms in sum over $n$.  Taking the
product between the $k=1$ and $k=2$ sums will then produce a
redistribution of the signal power amongst a slightly expanded range
of harmonic frequencies.  For the circular orbit case we expect power
to be spread amongst $\approx 2z_{1}$ sidebands whereas we now expect
the same power to be divided amongst $\approx 2z_{1}(1+2e)$ sidebands.

In general, orbital eccentricity causes a redistribution of signal
power amongst the existing circular orbit sidebands and will cause
negligible leakage of signal power into additional sidebands at the
boundaries of the sideband structure. Orbital eccentricity also has
the effect of modifying the phase of each sideband.  However, as shown
in Section~\ref{subsec:fmod}, the standard sideband search is
insensitive to the phase of individual sidebands.

\section{Primary sources}\label{sec:sources}

The benefit of the sideband search is that it is robust and
computationally cheap enough to be run over a wide frequency
band~\cite{MW2007}. The most suitable targets are those with
well-measured sky position and orbital periods, reasonably well
constrained semi-major axes, and poorly or unconstrained spin
frequency. 

The most suitable candidates in terms of these criteria are \acp{LMXB} due to their high accretion rate (directly related to gravitational wave amplitude) and their visibility in the electromagnetic regime (predominantly X-ray, but optical and radio observations also provide accurate sky position, ephemeris and sometimes orbital information). They are classified into three main types depending on the behavior of their X-ray emissions: pulsing, bursting or \ac{QPO} sources. Pulsating and frequently bursting \acp{LMXB} usually have a well determined spin frequency and are better suited to the more sensitive, narrow-band
techniques, such as \ac{LIGO}'s known pulsar pipeline \cite{LIGO_S5_Crab_2008,LIGO_S5_KnownPulsars_2010} including corrections for the binary motion. Non-pulsing burst sources with irregular or infrequent bursts still have a fairly wide ($\mathcal{O}$ a few Hz) range around the suspected spin frequency. A convincing relationship between \acp{QPO} and the spin frequency of the neutron star has not yet been determined, so the spin frequency of purely \ac{QPO} sources is considered unknown for our applications. 

The gravitational wave strain amplitude is directly
proportional to the square root of the x-ray flux $\ho \propto
\Fx^{1/2}$ (Eq. \ref{eq:hc_Fx_300}), so the most luminous sources,
which are usually the \ac{qpo} sources, will be the most detectable. In addition, the
(already weak) gravitational wave strain amplitude is proportional to the inverse of the distance to the source, so closer (i.e. galactic) sources
are also favorable.

In this section we present possible sources to which the sideband
search can be applied. We start with galactic \acp{LMXB} and consider
the most detectable sources in terms of their parameter
constraints. We exclusively consider sources requiring wide frequency
search bands ($\gtrsim 5$ Hz), and so neglect the accreting
millisecond pulsars. The detectability of a wider range of accreting
sources, with some measurement or estimate of spin frequency, in terms
of general gravitational wave searches was reviewed in \cite{WattsEA_2008}.

\subsection{Galactic LMXBs}\label{subsec:population}

\begin{table*}
\caption{Target sources for the sideband method. The different columns list the X-ray flux $F_X$ (in units of $10^{-8} \text{ erg cm}^{-2}\text{s}^{-1}$), distance $d$, sky position uncertainty $\Delta\beta$, fractional error on the semi-major axis $\Delta a/a$ and orbital period $\Delta P/P$, and the orbital period limited observation time at a frequency of 1 kHz $\Tspan^\Porb|_\text{1 kHz}$. The horizontal line separates \ac{qpo} (top) and burst (bottom) sources.}\label{tab:targets} 
\begin{ruledtabular}
\begin{tabular}{lcccccc}
Source & $\Fx \;(F_*)$ \footnote{$F_*= 10^{-8} \text{ erg cm}^{-2}\text{s}^{-1}$} & d (kpc) &
$\Delta\beta$ (arc sec) &  $\Delta \ameas/\ameas$ & $(\Delta \Porb/\Porb)
(\times 10^{-7})$ & $\Tspan^\Porb|_\text{1 kHz}$ \\ \hline 

Sco X-1	      & 40     & 2.8   & $3\times 10^{-4}$ 	& 0.13 & 9	& 50 days	\\
4U 1820-30    & 2.1    & 7.4   & 0.15 			& 0.48 & 2	& 300 days	\\
Cyg X-2	      & 1.1    & 10.55 & 0.5    		& 0.12 & 0.004	& 400 years	\\
J 2123-058    & 0.21   & 9.6   & 0.6 			& 0.19 & 0.9	& 3.5 years	\\ \hline

4U 1636-536   & 0.84   & 6     & $< 60$			& 0.11 & 10	& 17 days 	\\
X 1658-298    & 0.67   & 12    & 0.1			& 0.82 & 0.1	& 5 years 	\\
XB 1254-690   & 0.09   & 13    & 0.6			& 0.12 & 500	& 6 hours 	\\
EXO 0748-676  & 0.036  & 7.4   & 0.7			& 0.77 & 6 	& 40 days 	\\
4U 1916-053   & 0.027  & 8     & 0.6			& 0.72 & 3	& 2 years

\end{tabular}
\end{ruledtabular} 
\end{table*}

The sideband search is best suited to \acp{LMXB} with a relatively
large uncertainty in the spin frequency ($\gtrsim$ a few Hz), so
\ac{qpo} and poorly constrained burst sources are the best targets. The requirement of a
relatively well defined sky position and orbital period excludes many
sources including those that are considered to be X-ray bright.
Table~\ref{tab:targets} lists some of the galactic \acp{LMXB}, and
their limiting parameters, for which the sideband search is most
applicable. The parameters displayed in the table allow us to
determine the most suitable targets for the search.

For each source the table lists the bolometric X-ray flux $\Fx$, the
distance to the source $\dist$, the error in the sky position $\Delta\beta=(\Delta
\alpha,\Delta\delta)$, the fractional error in the semi-major axis
$\Delta \ameas/\ameas$ and the orbital period limited observation time
$\Tspan^\Porb|_\text{1 kHz}$ calculated using
Eq.~\ref{eq:Ts_Porb} at a frequency of 1 kHz. Although we could expect gravitational wave emission up to $\sim1500 Hz$ (from the currently measured spin distributions of \acp{LMXB} maxing out at $\sim720 Hz$), 1 kHz is chosen as an upper bound on the search frequency since the sensitivity of \ac{LIGO} detectors is limited at high frequencies and the amplitudes of these systems are not expected to be very strong. Sources with poorly constrained ($\Delta \asini/\asini > 0.9$) semi-major axis and sky position ($\Delta\beta > 60''$) have not been included. The sources are
listed in order of their bolometric X-ray flux within each source
group with \ac{QPO} sources in the top and burst sources in the bottom
half of the table. The distance is included as a reference but is
already taken into account in calculation of $\Fx$. From these factors
alone, \ac{ScoX1} is already the leading candidate source.  The sky
position error $\Delta\beta$ should be less than 100
mas for a source with $\fo=1$ kHz (see
Sec. \ref{subsec:skypos}). \ac{ScoX1} is the only candidate that falls
easily within this basic limit, although a few other sources are
borderline cases. The fractional error in semi-major axis is included
also as a guide. Although a smaller error on this parameter improves
our sensitivity, as shown in Sec.~\ref{subsec:asini} we are relatively
insensitive to $\asini$ uncertainties on the scale of $10$'s of
percent. The final column lists the orbital period limited
observation timespan $\Tspan^\Porb$ at a frequency of 1 kHz. Although the spin frequencies of the burst sources are better constrained than \ac{QPO} sources, the comparison of $\Tspan^\Porb$ is still made at 1 kHz so that a direct comparison on the source parameters (rather than search performance) can be made. This column is included for reference as the orbital
period may not be the tightest constraint on the observation time
(c.f. Sec. \ref{subsec:spinf}). It does, however, give an indication of
how well the orbital period of the source is constrained and specifically how it affects the search performance. 

\subsection{Sco X-1}\label{subsec:ScoX1}

\begin{table*}
\caption{Sco X-1 system parameters required for the sideband search. Directly observable parameters are presented in the top half of the table. The bottom half, separated by the horizontal line, displays search limits and constraints derived from these.}\label{tab:ScoX1} 
\begin{ruledtabular}
\begin{tabular}{lccrllcr}
Parameter 			& Symbol 	  && Value   & Units  & Uncertainty  && References	\\ \hline 
Right Ascension 		& $\alpha$ 	  && $16^h19^m55^s.0850$ & mas & $\pm0.3$ && \cite{BradshawEA_1999} \\
Declination 			& $\delta$ 	  && $-15\dg 38' 24.9" $ & mas & $\pm0.3$ && \cite{BradshawEA_1999}	\\
Proper motion			& $\mu$		  && $14.1$ & mas yr${}^{-1}$	&	  && \cite{BradshawEA_1999}	\\
Parallax			& $\pi_\beta$	  && $0.36 $ & mas 	&$\pm0.04$ 	  && \cite{BradshawEA_1999}	\\
Moment of inertia		& $I$		  && $10^{38}$ & $\text{kg m}^2$ &	  && \cite{BradshawEA_1999}	\\
Accretion rate			& $\dot{M}$	  && $1.23 \times 10^{15}$ & kg s${}^{-1}$& && \cite{BradshawEA_1999}	\\
Bolometric X-ray flux		& $\Fx$		  && $40 \times 10^{-8}$ & $\text{erg cm}^{-2}\text{s}^{-1}$ &	&& \cite{WattsEA_2008}  \\
Projected semi-major axis light travel time & $\ameas$   && 1.44 & s & $\pm 0.18$   		 && \cite{Steeghs_Casares_2002}	\\
Orbital Period 			& $\Pmeas$	  && 68023.82 & s & $\pm 0.06048$   	 && \cite{GallowayEA_2012} 	\\
NS spin inclination angle 	& $\iota$ 	  && $44 $ & deg & $\pm 6$		 && \cite{FomalontEA_2001b}	\\
GW polarization angle 		& $\psi$ 	  && $234$ & deg & $\pm 3$ 		 && \cite{FomalontEA_2001b} 	\\
Time of periapse passage (SSB) 	& $\tp$ 	  && 614638484 & s & $\pm$ 400  	 && \cite{Steeghs_Casares_2002, Messenger2005} \\ \hline \\

Strain amplitude (at $\spinf = 300$ Hz)	& $\ho^{300}$   && $3.5 \times 10^{-26}$ && 	 && Eq. \ref{eq:hc_Fx_300} 	\\
Spin limited observation timespan 		& $\Tspansw$ 	&& 13 &days 	&	 && Eq. \ref{eq:Ts_spin} 		

\end{tabular}
\end{ruledtabular}
\end{table*}

\ac{ScoX1}, the first \ac{LMXB} to be discovered, is also the
brightest extra-solar x-ray source in the sky. The direct relation
between gravitational wave strain and x-ray flux given by Eq.~\ref{eq:hc_Fx_300}
makes it also the most likely to be a strong gravitational wave emitter. This, as
well as the parameter constraints displayed in
Table~\ref{tab:targets}, makes it an ideal candidate for the sideband
search.  Table~\ref{tab:ScoX1} provides a list of \ac{ScoX1}
parameters determined from various electromagnetic observations. The
table includes the parameters required to run the sideband search
together with some values used for calculating limits and constraints
on the performance and sensitivity of the search. The bottom section
of the table lists some of the limits and constraints derived using
the above mentioned parameters.

Running the standard version of the sideband search requires accurate
knowledge of the sky position and orbital period and approximate
knowledge of the semi-major axis. The sky position $\beta= (\alpha,\delta)$ listed for \ac{ScoX1} is accurate to within 0.3 mas. This error is well within the 100 mas limit defined in Sec.~\ref{subsec:skypos}, justifying the assumption of a fixed sky
position. The accuracy of measurements of the orbital period require
only a single sideband template if the observation timespan is within
$\Tspan < \Tspan^\text{obs} \approx 49$ days (for \ac{ScoX1} at 1
kHz). The semi-major axis and its measurement error are also required
for construction of the sideband template.

Estimates of the primary (accreting neutron star) and secondary (donor
star) masses, as well measurements of the bolometric X-ray flux
($\Fx$) are required to estimate the indirect, torque balance, gravitational wave
strain upper limit $\htorq$ using Eq.~\ref{eq:hc_Fx_300}, displayed in
the bottom section of the table. The spin frequency limited
observation timespan is also listed here and requires values for the
accretion rate $\dot{M}$ and moment of inertia $I$ to calculate this
value for \ac{ScoX1} using Eq.~\ref{eq:Ts_spin} assuming a
spin-wandering timescale $\tspinw = 1$ day.\footnote{Assuming the instantaneous accretion rate does not vary more than the x-ray flux, observations of the x-ray variability of \ac{ScoX1} show that the accretion rate can vary by roughly a factor of two over a timescale $\tspinw =1$ day. \cite{BradtEA_1975, HertzEA_1992}} The corresponding value of
$\Tspansw\approx 13$ days displayed in the table is more
restrictive than the orbital period limited timespan and is our
limiting time constraint in the search.

\section{Statistical analysis}\label{sec:stats}

Let us first assume that our analysis has yielded no significant
candidate signal given a designated significance threshold.  In this
case, with no evidence for detection, we place an upper limit on the
possible strength of an underlying signal. In the literature on
continuous-wave gravitational signals, it is common to determine these
upper limits
numerically~\cite{LIGO_S2_ScoX1_2007,LIGO_2008_PRD77,LIGO_S5_2012arXiv}
or semi-analytically~\cite{LIGO_CasA_2010, Wette2012} using
frequentist Monte Carlo methods. In these cases simulated signals are
repeatedly added to data over a range of frequencies and recovered
using a localized, computationally cheap, search around the point of
injection.

The sideband algorithm combines signal from many (typically $\sim
10^3$) correlated \ftext-statistic frequency bins which must be
computed over a relatively wide frequency band for each simulated
signal. Such computations represent a computational cost far in excess
of existing methods and are only manageable for a small parameter
space, e.g. injection studies where the signal frequency is known and
$\mathcal{O}(10^2)$ realizations are feasible. The computations become
daunting for a wide-band search covering more than a few Hz.

We choose to optimize the process by calculating upper limits within a Bayesian
framework. This is an especially appealing alternative since the \ac{pdf} of the
\ctext-statistic takes a relatively simple, closed, analytic form. Bayesian
upper limits have been computed in time-domain gravitational wave
searches targeting known sources (pulsars)
\cite{DupuisWoan_2005,LIGO_S5_KnownPulsars_2010,LIGO_Vela_2011}, and
cross-correlation searches for the stochastic background
\cite{LIGO_S4_ScoX1_2007,LIGO_2007_PRD76,LIGO_S5stoch_ScoX1_2011}. Comparisons
on specific data sets have shown that Bayesian and frequentist upper limits are
consistent \cite{LIGO_2004_PRD69,RMP2011,LIGO_Vela_2011}.

\subsection{Bayes Theorem}\label{subsec:Bayes}

In the Bayesian framework, the posterior probability density of the hypothesis
$H$ given the data $D$ and our background information $I$ is defined as
\begin{equation}\label{eq:Bayes}
 p(\bm{\theta}|D,H,I) = p(\bm{\theta}|H,I)\frac{p(D|\bm{\theta},H,I)}{p(D|H,I)},
\end{equation}
where $p(\bm{\theta}|H,I)$ denotes the prior probability distribution
of our model parameters $\bm{\theta}$ given a model $H$ assuming the
background information $I$.  The quantity $p(D|\bm{\theta},H,I)$ is
the direct probability density (or likelihood function) of the data
given the parameters, model and background information. The term
$p(D|H,I)$ is known as the evidence of $D$ given our model and acts as
a normalisation constant and does not affect the shape of the
posterior distribution
$p(\bm{\theta}|D,H,I)$~\cite{Bretthorst1988}. The background
information $I$ (which represents our signal model, assumptions on
Gaussian noise, physicality of parameters etc.) remains constant
throughout our analysis and will not be mentioned hereafter.

\subsection{Likelihood}\label{subsec:likelihood}

When there is no signal in the data, we will say the null hypothesis
$\nullH$, that the data contains any Gaussian noise, is true. Under
these conditions, each $\cstat$ value is drawn from a central
$\chi^2_{4M}$ distribution. Hence the $\nullH$ model is parametrized
entirely by $M=2m+1$, the number of sidebands in the template, where
$m=\texttt{ceil}[2\pi f a]$ and depends on the search
frequency $f$ and semi-major axis $a$.

The signal hypothesis $\sigH$ is true if the data contains Gaussian
noise plus a signal. The signal is defined by the set of parameters $\bm{\theta}=\{\ho, \cos\iota, \psi, \phio, a, P\}$. In the case of a signal present in the data, each \ctext-statistic is drawn from
a non-central $\chi^2_{4M}[\lambda(\bm{\theta})]$ distribution. The non-centrality
parameter $\lambda(\bm{\theta})$ is defined by the signal parameters $\bm{\theta}$ and is given by 
\begin{equation}
 \lambda(\bm{\theta}) = \ro^{2}\sum_{n=-m}^{m}J^{2}_{n}(2\pi
\fo\atrue)|\tilde{W}(f_{n}-f_{k}+l(n)\Delta f)|^2. \label{eq:lambda}
\end{equation}
It represents the total recovered optimal \ac{SNR} contained within
the sidebands. The likelihood function (the probability of our measured $\cstat$
value given a parameter set $\bm{\theta}$) is then given by
\begin{equation}\label{eq:likelihood}
 p(\cstat|\bm{\theta}) = \frac{1}{2}
\exp\left(-\frac{1}{2}\left[\cstat+\lambda\left(\bm{\theta}\right)\right]\right)
\left(\frac{\cstat}{\lambda\left(\bm{\theta}\right)}\right)^{M-\frac{1}{2}}
\textrm{I}_{2M-1}\Bigg(
\sqrt{\cstat \lambda\left(\bm{\theta}\right)}\Bigg).
\end{equation}
It should be noted that although the quantity $M$ is a function of the
semi-major axis and intrinsic gravitational wave frequency, it has been
fixed according to the predefined number of teeth used in the sideband
template. It is therefore not a function of $\bm{\theta}$.

\subsection{Priors}\label{subsec:priors}

When searching for weak signals, an overly prescriptive prior is
undesirable because it may dominate the posterior. Hence, to be
conservative, we adopt a uniform prior on $\ho\geq 0$; the possibility
of $\ho=0$ excludes the use of a fully scale-invariant Jeffreys prior
$\propto 1/\ho$ \cite{DupuisWoan_2005}.  The upper limit thus derived
is consistent with the data, not just a re-iteration of the prior. The
same $\ho$ prior has been adopted in previous searches
\cite{LIGO_2004_PRD69, LIGO_S2_ScoX1_2007, LIGO_2008_PRD77,
  PrixKrishnan_2009}; the motivation is discussed in more detail in
\cite{DupuisWoan_2005}.

Electromagnetic measurements of the orbital period $\Pmeas$ and
semi-major axis $\ameas$ are assumed to carry normally distributed
random errors. Hence we adopt Gaussian priors on the actual values
$\Po$ and $\atrue$. Specifically we take
$p(\Po)=\mathcal{N}(\Pmeas,\Delta\Pmeas)$ and $p(\atrue)=
\mathcal{N}(\ameas,\Delta\ameas)$, where $\mathcal{N}(\mu,\sigma)$
denotes a Gaussian (normal) distribution with mean $\mu$ given by the
electromagnetic observation and standard
deviation $\sigma$ taken as the error in that observation.

The reference phase $\phi_{0}$ is automatically maximized over
within the \ftext stage of the analysis and therefore does not
directly affect our (semi-coherent) analysis. The remaining amplitude
parameters serve only to influence the optimal \ac{SNR}, and therefore
also the \ctext.  Without prior information from electromagnetic
observations, we select the least informative (ignorant)
\emph{physical} priors such that $p(\cos\iota)=1/2$ and
$p(\psi)=1/2\pi$ on the domains $(-1,1)$ and $(0,2\pi)$ respectively.

Any prior informative measurements (e.g. electromagnetic) on the
amplitude parameters can be incorporated into the analysis, and serve
to narrow the prior probability distributions.  For the Sco X-1
search, we can deduce measurements for $\cos\iota$ and $\psi$ from
observations if we assume the rotation axes of the neutron star and
accretion disk are aligned. This implies the neutron star inclination
$\iota$ is equal to the orbital inclination. We can then set $\iota =
44\dg \pm 6\dg$ from the inclination of the orbital plane suggested
from observations of the radio components of Sco
X-1~\cite{FomalontEA_2001b}. The same observations measure a position
angle of these radio jets of $54\pm 3\dg$. Under the alignment
assumption, the position angle is directly related to the gravitational wave
polarization angle $\psi$, but with a phase shift of $180\dg$,
i.e. $\psi = 234\pm 3\dg$.
The above assumes the usual mass-quadrupole emission; for current-quadrupole
emission from $r$-modes the results are the same with $\psi \rightarrow \psi + 45\dg$
\cite{Owen2010}.

\subsection{Posteriors}\label{subsec:posts}

The \ac{PDF} on our search parameters given a single \ctext-statistic
value is
\begin{equation}
 p({\bm{\theta}}|\cstat) \propto p(\cstat|\bm{\theta})p(\bm{\theta}),
\end{equation}
and assuming that the prior \acp{PDF} on our parameters are
independent, we can express the posterior \ac{PDF} as
\begin{eqnarray}
 p(\ho,\cos\iota,\psi, \Porb, \asini| \cstat)\propto p(\cstat|\bm{\theta})p(\ho)p(\cos\iota)p(\psi)p(\Porb)p(\asini). \nonumber \\
\end{eqnarray}
To perform inference on the gravitational wave strain $\ho$, we can marginalize
this joint distribution over the other parameters leaving us with
\begin{eqnarray}\label{eq:h0_post}
  p(\ho|\cstat)\!&\propto&\!\int\limits_{-\infty}^{\infty}\!d\asini
\!\int\limits_{\infty}^{\infty}\!d\Porb\!\int\limits_{0}^{2\pi}\!d\psi
\!\int\limits_{-1}^{1}\!d\cos\iota~p(\cstat|\bm{\theta})\mathcal{N}(\Porb,\Delta{P})\mathcal{N}(\asini,\Delta{\asini}), \nonumber \\
\end{eqnarray}
where the flat priors on $\ho$, $\cos\iota$ and $\psi$ are absorbed into the
proportionality.  Note that the amplitude parameters act through the
non-centrality parameter $\lambda(\bm{\theta})$ (Eq.~\ref{eq:lambda})
via the optimal \ac{SNR} term (Eq.~\ref{eq:rho2}), in the likelihood.
The orbital parameters $\asini,P$ dictate the fraction of recovered
\ac{SNR} based on the mismatch in the predicted quantity and location
of frequency-modulated sidebands (Eq.~\ref{eq:lambda}).

\subsection{Detection criteria and upper limits}\label{subsed:detection}

To determine whether or not a signal is present in the data, we
compute a threshold value of the \cstext such that the probability of
achieving such a value or greater due to noise alone is $\Pa$, the
false alarm probability.  For a single measurement of the \cstext
this threshold is computed via 
\begin{eqnarray}
 \Pa &=& \int\limits_{\cstar}^{\infty}p(\cstat|\ho=0) \nonumber \\
     &=& 1 - \mathcal{P}\left(2M,\cstar/2\right),\label{eq:Pa}
\end{eqnarray}
where the likelihood on $\cstat$ in the noise only case becomes a
central $\chi^2$ distribution and $\mathcal{P}(k/2,x/2)$ is the regularized Gamma function with $k$ degrees of freedom (the cumulative distribution function of a central $\chi^2_k$ distribution) defined at $x$. 

In the case of $N$ measurements of the \cstext, assuming statistically
independent trials, the false alarm probability is given by
\begin{eqnarray}
 P_{a|N} &=& 1-(1-\Pa)^N \nonumber \\
         &=& 1-\left[\mathcal{P}\left(2M, \cstar/2\right)\right]^N.
\end{eqnarray}
The corresponding threshold $\cstar_{N}$ such that the probability that one or more of these values
exceeds that threshold is obtained by solving
\begin{equation}
\mathcal{P}(2M,\cstar_{N}/2)= \left(1-P_{a|N}\right)^{1/N}. \label{eq:P_cstar}
\end{equation}
This solution is obtained numerically but can be represented notationally by
\begin{equation}\label{eq:cstar}
 \cstar_N = 2 \mathcal{P}^{-1}\left(2M,\left[1-P_{a|N}\right]^{1/N}\right),
\end{equation}
where $\mathcal{P}^{-1}$ represents the inverse function of $\mathcal{P}$.

In practice the \cstext values will not be statistically
independent as assumed above.  The level of independence between
adjacent frequency bins will be reduced (i.e. values will be
become increasingly correlated) as the frequency resolution of the
\cstext is made finer.  Additionally, due to the comb structure of the
signal and template we find that results at frequencies separated by
an integer number of frequency-modulated sideband spacings $j/P$ Hz for $j<m$ are
highly correlated.  This is due to the fact that these results will
have been constructed from sums of \fstext values containing many
common values.  This latter effect is dominant over the former and as
an approximation it can be assumed that within the frequency span of a
single comb template there are $rT/P$ independent \cstext results.\footnote{This comes from the number of bins in between each sideband, given by the sideband separation $1/P$ divided by the bin size $\fres = (r\Tspan)^{-1}$ (Eq. \ref{eq:fres}).} The number of templates spans per unit search frequency is
$\sim P/rM$ which leaves us with $\sim T/M$ independent \cstext values per unit Hz.
This is a reduction by a factor of $M$ in the number of statistically
independent results expected.

In the event of there being no candidate \cstext values, the search
allows us to compute upper-limits on the amplitude of gravitational waves from
our target source.  We define the upper limit on the wave strain
$\ho$ as the value $\hUL$ that bounds the fraction UL of the area of
the marginalized posterior distribution $p(\ho|\cstat)$. This value is
obtained numerically by solving
\begin{equation}
 \text{UL} = \int\limits_{0}^{\hUL}p(\ho|\cstat) \,\, d\ho. \label{eq:h0_UL}
\end{equation}
We note that this procedure allows us to compute an upper-limit for
each \cstext value output from a search.  The standard practice in
continuous gravitational wave data analysis is to perform a frequentist
upper-limit using computationally expensive Monte-Carlo simulations
involving repeated signal injections.  The results of these injections
are then compared to loudest detection statistic recovered from the
actual search~\cite{LIGO_S2_ScoX1_2007}.  In our approach, by virtue
of the fact that we are able to compute upper-limits very efficiently
for each \cstext value and the upper-limit value is a monotonic
function of $\cstat$ we naturally also include the worst case (loudest
event) result.  The difference in the upper-limits obtained from both
strategies then becomes an issue of Bayesian versus Frequentist
interpretation.  However, as shown in~\cite{RMP2011}, in the limit of
large \ac{SNR} these upper-limit results become indistinguishable.
When searching wide parameter spaces with large numbers of templates,
as is the case for the sideband search, the most likely largest
detection statistic value will be consistent with large \ac{SNR}.

\section{Sensitivity}\label{sec:sensitivity}

\begin{figure}
\includegraphics[width=\columnwidth]{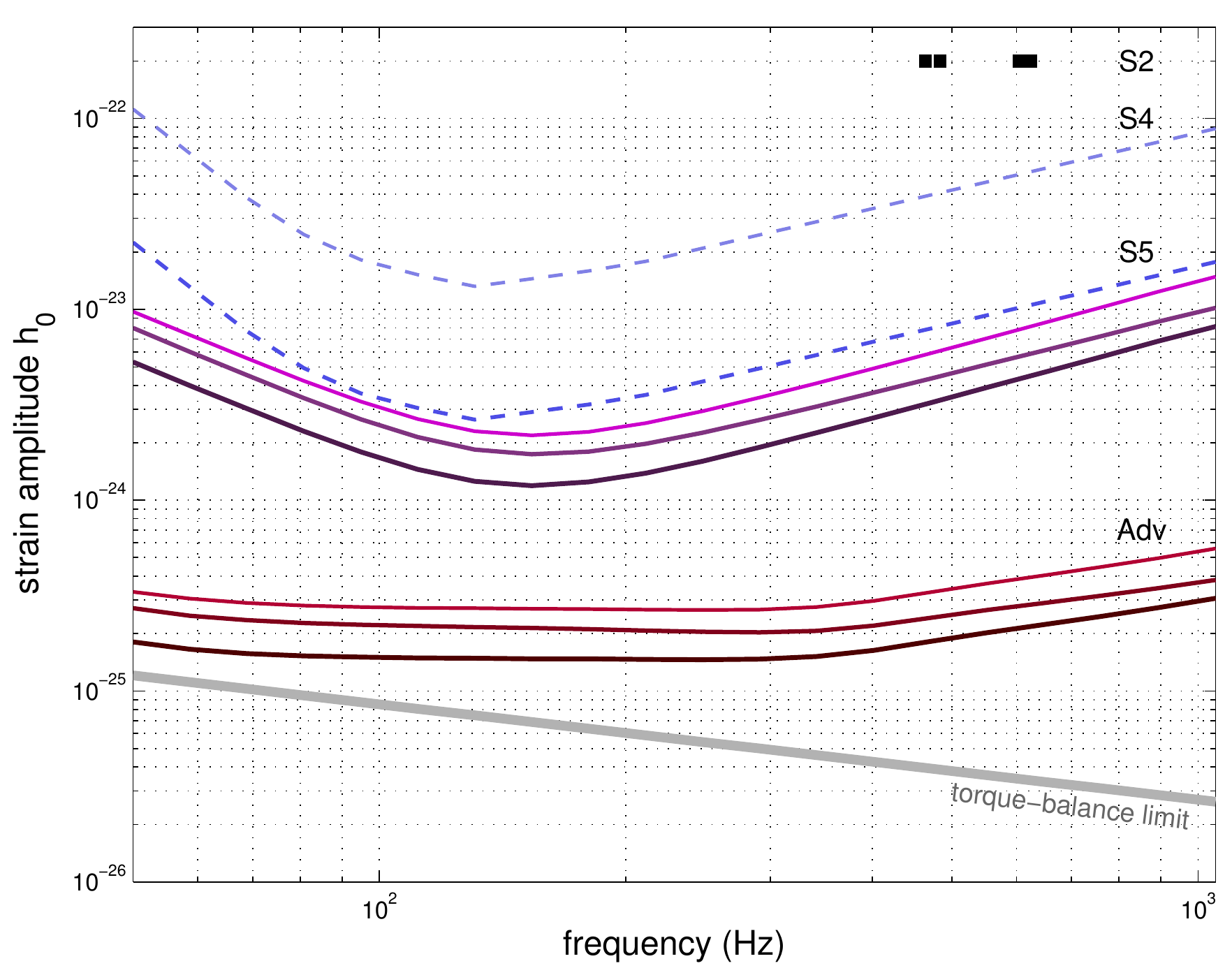}
\caption{\label{fig:Bayes_sens} 
  Sensitivity estimate for a 10 day standard, approximate demodulated and approximate demodulated with known priors sideband search (fine, medium and bold solid curves respectively) using \ac{LIGO} (H1L1) S5  data (upper, purple group), and using the 3-detector (H1L1V) advanced \ac{LIGO} configuration (lower, red group). Also shown are results from the the previous coherent search for \ac{ScoX1} in S2 data (solid black dashes) \cite{LIGO_S2_ScoX1_2007} and the maximum upper limits for each Hz band of the directed stochastic (radiometer) search in S4 and S5 data (light and dark blue dashed curves, respectively) \cite{LIGO_S4_ScoX1_2007,LIGO_S5stoch_ScoX1_2011}. The theoretical torque-balance gravitational wave strain upper
  limit ($\htorq$ from Eq. \ref{eq:hc_Fx_300}) for Sco X-1 is indicated by the thick gray straight line.}
\end{figure}

The sensitivity of a future search can be predicted in a variety of ways.  We choose to estimate the expected gravitational wave strain upper-limits
for Initial \ac{LIGO} data in order to compare against previous results. We also compare this to the expected sensitivity of the search with Advanced \ac{LIGO}.  

If the search is conducted such that the frequency space is split into small sub-bands, the sensitivity can be estimated by computing upper limits on the expected maximum from each of the sub-bands in Gaussian noise. This is equivalent to assigning a false alarm probability $\Pa=50\%$ for $N=\Tspan/M$ trials for
each, say one Hz frequency sub-band, and using Eq.~\ref{eq:cstar} as the expected \cstext. We can then calculate the posterior distribution of $\ho$ from Eqs.~\ref{eq:likelihood} and Eq ~\ref{eq:h0_post}. 

Figure~\ref{fig:Bayes_sens} shows the sensitivity estimate of the
$90\%$ upper limit (UL=0.9) for the sideband search in different
modes: standard (described in Section \ref{sec:sideband}, represented by the thin solid curves), binary demodulated
(described in Section \ref{subsec:demod}, represented by the medium solid curves), and binary demodulated with known priors
on $\cos\iota$ and $\psi$ (described in Section \ref{subsec:priors}, represented by the bold solid curves). It compares
the sensitivity of the search in two-detector (H1L1) LIGO S5 data (upper, purple group) and three-detector (H1L1V) Advanced LIGO data (red group) with previous searches for \ac{ScoX1} in LIGO S2 (black dashes) \cite{LIGO_S2_ScoX1_2007}, S4 and S5 data (light and dark blue dashed curves, respectively) \cite{LIGO_S4_ScoX1_2007,LIGO_S5stoch_ScoX1_2011}. The $h_{\text{rms}}$ upper limit quoted in the latter two (radiometer) searches is optimized for the special case of a circularly polarized signal and hence less conservative than the angle averaged $\ho$ quoted in \cite{LIGO_S2_ScoX1_2007} and commonly used when quoting upper limits for continuous gravitational wave searches. Converting detector-strain rms upper limits $h_\text{rms}$ to source-strain amplitude upper limit $\ho$ requires $\ho\sim 2.43 h_{\text{rms}}$ (see \cite{Messenger2010}). The different confidence on the coherent S2 analysis and S4 ans S5 radiometer analyses (90 and 95$\%$ respectively) also complexify any direct comparisons. The theoretical indirect wave strain limit $\htorq$ for gravitational waves from \acp{LMXB} represented by the thick gray line comes from Eq. \ref{eq:hc_Fx_300}.

The sensitivity curves in Fig. \ref{fig:Bayes_sens} show that the standard sideband search should improve current upper limits on gravitational waves from \ac{ScoX1}, even though it is limited to only 10 days of consecutive data. Running a demodulated search with known $\cos \iota$ and $\psi$ comes close to setting constraints on the indirect (torque-balance) upper limits in the advanced detector era.

\section{Discussion}\label{sec:discussion}

We have described the sideband algorithm and shown that it provides a
computationally efficient method to search for gravitational waves from sources
in binary systems. It requires accurate knowledge of the sky position
of the source and the orbital period of the binary, and less accurate
knowledge on the semi-major axis. Effects of spin wandering can be
ignored over a short enough coherent integration time. 

The tolerance on the errors of relevant search parameters was computed, defining the range over which they can be assumed constant (Section \ref{sec:params}). In light of these limits, electromagnetic observations suggest several candidates (Section \ref{sec:sources}). Of these sources, \ac{ScoX1} is
identified as the strongest candidate based on the gravitational wave strain
recovered from the torque-balance argument (Eq.
\ref{eq:hc_Fx_300}). In future, the search can also be directed at
several other of the suitable \ac{LMXB} candidates presented in
Section \ref{sec:sources}.

A Bayesian upper limit strategy was presented in Section \ref{sec:stats}, rather than the frequentist methods commonly employed in
frequency-based (\ac{LIGO}) searches. Knowing the likelihood function in closed
analytic form makes the Bayesian approach computationally more
feasible than Monte Carlo simulations (see
\ref{subsec:likelihood}). Knowing the gravitational wave polarization angle and
inclination leads to additional sensitivity improvements using this
framework (see \ref{subsec:priors}).

The sensitivity of the search, described in Section
\ref{sec:sensitivity}, is estimated by performing the Bayesian
analysis on the design curves of the S5 and Advanced \ac{LIGO} noise
floors. The sensitivity of a 10 day limited \ac{ScoX1} directed sideband search compared to previous \ac{LIGO} searches is shown in Fig. \ref{fig:Bayes_sens}. 
It shows that measurements of an orbital reference time and phase (the time and argument of periapse)
can be employed to improve search sensitivity by a factor of $~1.5$ in the approximate
demodulated version of the search. Also, prior information on the polarization and inclination of the gravitational wave signal constrains the upper limit calculation improving the sensitivity by another factor of $1.5$. In its most sensitive configuration (approximated binary demodulated assuming known $\cos\iota$ and $\psi$ in the Advanced detector era), the sideband search brings us closer to testing the theoretical indirect torque-balance limit.

The studies presented here assume pure Gaussian noise. The performance
for realistic \ac{LIGO}-like noise will be presented elsewhere, in a report on the results from the \ac{ScoX1} directed search performed
on \ac{LIGO} (S5) data. The search could also look forward to running on
the next-generation Advanced \ac{LIGO} data.

\section*{ACKNOWLEDGEMENTS}
%
This work has benefited from many useful discussions and helpful comments. In particular we would like to thank Karl Wette, Reinhard Prix, Evan Goetz, Holger Pletsch, Miroslav Shaltev, Badri Krishnan, Paul Lasky, Mark Bennett, Sterl Phinney and the LIGO Scientific Collaboration continuous waves working group. LS is indebted to Bruce Allen for his support and gratefully acknowledges the Max-Planck Society (Albert-Einstein Institut) for support and hospitality during periods when significant progress on this work was made. LS and CM acknowledge the Gravitational wave group at Cardiff University for their support. LS was supported by an Australian Postgraduate Award and a Melbourne University Overseas Research Experience Scholarship. AM and BJO acknowledge ARC grant DP-110103347. BJO was supported by NSF Grants PHY-0855589 and PHY-1206027.
%

\bibliography{references}

\begin{thebibliography}{86}
\expandafter\ifx\csname natexlab\endcsname\relax\def\natexlab#1{#1}\fi
\expandafter\ifx\csname bibnamefont\endcsname\relax
  \def\bibnamefont#1{#1}\fi
\expandafter\ifx\csname bibfnamefont\endcsname\relax
  \def\bibfnamefont#1{#1}\fi
\expandafter\ifx\csname citenamefont\endcsname\relax
  \def\citenamefont#1{#1}\fi
\expandafter\ifx\csname url\endcsname\relax
  \def\url#1{\texttt{#1}}\fi
\expandafter\ifx\csname urlprefix\endcsname\relax\def\urlprefix{URL }\fi
\providecommand{\bibinfo}[2]{#2}
\providecommand{\eprint}[2][]{\url{#2}}

\bibitem[{\citenamefont{Bildsten}(1998)}]{Bildsten_1998}
\bibinfo{author}{\bibfnamefont{L.}~\bibnamefont{Bildsten}},
  \bibinfo{journal}{Astrophys. J. Lett.} \textbf{\bibinfo{volume}{501}},
  \bibinfo{pages}{L89} (\bibinfo{year}{1998}),
  \urlprefix\url{http://stacks.iop.org/1538-4357/501/i=1/a=L89}.

\bibitem[{\citenamefont{{Ushomirsky}
  et~al.}(2000{\natexlab{a}})\citenamefont{{Ushomirsky}, {Cutler}, and
  {Bildsten}}}]{UCB_2000}
\bibinfo{author}{\bibfnamefont{G.}~\bibnamefont{{Ushomirsky}}},
  \bibinfo{author}{\bibfnamefont{C.}~\bibnamefont{{Cutler}}}, \bibnamefont{and}
  \bibinfo{author}{\bibfnamefont{L.}~\bibnamefont{{Bildsten}}},
  \bibinfo{journal}{Mon. Not. R. Astron. Soc.} \textbf{\bibinfo{volume}{319}},
  \bibinfo{pages}{902} (\bibinfo{year}{2000}{\natexlab{a}}),
  \urlprefix\url{http://dx.doi.org/10.1046/j.1365-8711.2000.03938.x}.

\bibitem[{\citenamefont{{Owen}}(2005)}]{Owen_2005}
\bibinfo{author}{\bibfnamefont{B.~J.} \bibnamefont{{Owen}}},
  \bibinfo{journal}{Phys. Rev. Lett.} \textbf{\bibinfo{volume}{95}},
  \bibinfo{eid}{211101} (\bibinfo{year}{2005}),
  \eprint{arXiv:astro-ph/0503399}.

\bibitem[{\citenamefont{Haskell et~al.}(2007)\citenamefont{Haskell, Andersson,
  Jones, and Samuelsson}}]{HaskellEA_2007}
\bibinfo{author}{\bibfnamefont{B.}~\bibnamefont{Haskell}},
  \bibinfo{author}{\bibfnamefont{N.}~\bibnamefont{Andersson}},
  \bibinfo{author}{\bibfnamefont{D.~I.} \bibnamefont{Jones}}, \bibnamefont{and}
  \bibinfo{author}{\bibfnamefont{L.}~\bibnamefont{Samuelsson}},
  \bibinfo{journal}{Phys. Rev. Lett.} \textbf{\bibinfo{volume}{99}},
  \bibinfo{pages}{231101} (\bibinfo{year}{2007}),
  \urlprefix\url{http://link.aps.org/doi/10.1103/PhysRevLett.99.231101}.

\bibitem[{\citenamefont{Lin}(2007)}]{Lin2007}
\bibinfo{author}{\bibfnamefont{L.-M.} \bibnamefont{Lin}},
  \bibinfo{journal}{Phys. Rev.} \textbf{\bibinfo{volume}{D76}},
  \bibinfo{pages}{081502} (\bibinfo{year}{2007}), \eprint{0708.2965}.

\bibitem[{\citenamefont{Johnson-McDaniel and Owen}(2013)}]{JM_Owen_2013}
\bibinfo{author}{\bibfnamefont{N.~K.} \bibnamefont{Johnson-McDaniel}}
  \bibnamefont{and} \bibinfo{author}{\bibfnamefont{B.~J.} \bibnamefont{Owen}},
  \bibinfo{journal}{\prd} \textbf{\bibinfo{volume}{88}},
  \bibinfo{pages}{044004} (\bibinfo{year}{2013}), \eprint{1208.5227}.

\bibitem[{\citenamefont{Bonazzola and Gourgoulhon}(1996)}]{B_G_1996}
\bibinfo{author}{\bibfnamefont{S.}~\bibnamefont{Bonazzola}} \bibnamefont{and}
  \bibinfo{author}{\bibfnamefont{E.}~\bibnamefont{Gourgoulhon}},
  \bibinfo{journal}{Astron. Astrophys.} \textbf{\bibinfo{volume}{312}},
  \bibinfo{pages}{675} (\bibinfo{year}{1996}), \eprint{astro-ph/9602107}.

\bibitem[{\citenamefont{{Cutler}}(2002)}]{Cutler_2002}
\bibinfo{author}{\bibfnamefont{C.}~\bibnamefont{{Cutler}}},
  \bibinfo{journal}{\prd} \textbf{\bibinfo{volume}{66}}, \bibinfo{eid}{084025}
  (\bibinfo{year}{2002}), \eprint{arXiv:gr-qc/0206051}.

\bibitem[{\citenamefont{Melatos and Payne}(2005)}]{Melatos_Payne_2005}
\bibinfo{author}{\bibfnamefont{A.}~\bibnamefont{Melatos}} \bibnamefont{and}
  \bibinfo{author}{\bibfnamefont{D.~J.~B.} \bibnamefont{Payne}},
  \bibinfo{journal}{\apj} \textbf{\bibinfo{volume}{623}}, \bibinfo{pages}{1044}
  (\bibinfo{year}{2005}),
  \urlprefix\url{http://stacks.iop.org/0004-637X/623/i=2/a=1044}.

\bibitem[{\citenamefont{{Vigelius} and
  {Melatos}}(2008)}]{Vigelius_Melatos_2008}
\bibinfo{author}{\bibfnamefont{M.}~\bibnamefont{{Vigelius}}} \bibnamefont{and}
  \bibinfo{author}{\bibfnamefont{A.}~\bibnamefont{{Melatos}}},
  \bibinfo{journal}{Mon. Not. R. Astron. Soc.} \textbf{\bibinfo{volume}{386}},
  \bibinfo{pages}{1294} (\bibinfo{year}{2008}), \eprint{0802.3238}.

\bibitem[{\citenamefont{{Haskell} et~al.}(2008)\citenamefont{{Haskell},
  {Samuelsson}, {Glampedakis}, and {Andersson}}}]{HaskellEA_2008}
\bibinfo{author}{\bibfnamefont{B.}~\bibnamefont{{Haskell}}},
  \bibinfo{author}{\bibfnamefont{L.}~\bibnamefont{{Samuelsson}}},
  \bibinfo{author}{\bibfnamefont{K.}~\bibnamefont{{Glampedakis}}},
  \bibnamefont{and}
  \bibinfo{author}{\bibfnamefont{N.}~\bibnamefont{{Andersson}}},
  \bibinfo{journal}{Mon. Not. R. Astron. Soc.} \textbf{\bibinfo{volume}{385}},
  \bibinfo{pages}{531} (\bibinfo{year}{2008}), \eprint{0705.1780}.

\bibitem[{\citenamefont{{Mastrano} and
  {Melatos}}(2012)}]{Mastrano_Melatos_2012}
\bibinfo{author}{\bibfnamefont{A.}~\bibnamefont{{Mastrano}}} \bibnamefont{and}
  \bibinfo{author}{\bibfnamefont{A.}~\bibnamefont{{Melatos}}},
  \bibinfo{journal}{Mon. Not. R. Astron. Soc.} \textbf{\bibinfo{volume}{421}},
  \bibinfo{pages}{760} (\bibinfo{year}{2012}), \eprint{1112.1542}.

\bibitem[{\citenamefont{{Jones} and {Andersson}}(2002)}]{Jones_Andersson_2002}
\bibinfo{author}{\bibfnamefont{D.~I.} \bibnamefont{{Jones}}} \bibnamefont{and}
  \bibinfo{author}{\bibfnamefont{N.}~\bibnamefont{{Andersson}}},
  \bibinfo{journal}{Mon. Not. R. Astron. Soc.} \textbf{\bibinfo{volume}{331}},
  \bibinfo{pages}{203} (\bibinfo{year}{2002}), \eprint{arXiv:gr-qc/0106094}.

\bibitem[{\citenamefont{{Jones}}(2012)}]{Jones_2012}
\bibinfo{author}{\bibfnamefont{D.~I.} \bibnamefont{{Jones}}},
  \bibinfo{journal}{Mon. Not. R. Astron. Soc.} \textbf{\bibinfo{volume}{420}},
  \bibinfo{pages}{2325} (\bibinfo{year}{2012}), \eprint{1107.3503}.

\bibitem[{\citenamefont{Owen et~al.}(1998)\citenamefont{Owen, Lindblom, Cutler,
  Schutz, Vecchio et~al.}}]{OwenEA_1998}
\bibinfo{author}{\bibfnamefont{B.~J.} \bibnamefont{Owen}},
  \bibinfo{author}{\bibfnamefont{L.}~\bibnamefont{Lindblom}},
  \bibinfo{author}{\bibfnamefont{C.}~\bibnamefont{Cutler}},
  \bibinfo{author}{\bibfnamefont{B.~F.} \bibnamefont{Schutz}},
  \bibinfo{author}{\bibfnamefont{A.}~\bibnamefont{Vecchio}},
  \bibnamefont{et~al.}, \bibinfo{journal}{Phys. Rev.}
  \textbf{\bibinfo{volume}{D58}}, \bibinfo{pages}{084020}
  (\bibinfo{year}{1998}), \eprint{gr-qc/9804044}.

\bibitem[{\citenamefont{{Andersson} et~al.}(1999)\citenamefont{{Andersson},
  {Kokkotas}, and {Stergioulas}}}]{AKS_1999}
\bibinfo{author}{\bibfnamefont{N.}~\bibnamefont{{Andersson}}},
  \bibinfo{author}{\bibfnamefont{K.~D.} \bibnamefont{{Kokkotas}}},
  \bibnamefont{and}
  \bibinfo{author}{\bibfnamefont{N.}~\bibnamefont{{Stergioulas}}},
  \bibinfo{journal}{\apj} \textbf{\bibinfo{volume}{516}}, \bibinfo{pages}{307}
  (\bibinfo{year}{1999}), \eprint{arXiv:astro-ph/9806089}.

\bibitem[{\citenamefont{Bondarescu et~al.}(2007)\citenamefont{Bondarescu,
  Teukolsky, and Wasserman}}]{Bondarescu2007}
\bibinfo{author}{\bibfnamefont{R.}~\bibnamefont{Bondarescu}},
  \bibinfo{author}{\bibfnamefont{S.~A.} \bibnamefont{Teukolsky}},
  \bibnamefont{and}
  \bibinfo{author}{\bibfnamefont{I.}~\bibnamefont{Wasserman}},
  \bibinfo{journal}{Phys. Rev.} \textbf{\bibinfo{volume}{D76}},
  \bibinfo{pages}{064019} (\bibinfo{year}{2007}), \eprint{0704.0799}.

\bibitem[{\citenamefont{{Haskell} et~al.}(2012)\citenamefont{{Haskell},
  {Degenaar}, and {Ho}}}]{Haskell2012}
\bibinfo{author}{\bibfnamefont{B.}~\bibnamefont{{Haskell}}},
  \bibinfo{author}{\bibfnamefont{N.}~\bibnamefont{{Degenaar}}},
  \bibnamefont{and} \bibinfo{author}{\bibfnamefont{W.~C.~G.}
  \bibnamefont{{Ho}}}, \bibinfo{journal}{Mon. Not. R. Astron. Soc.}
  \textbf{\bibinfo{volume}{424}}, \bibinfo{pages}{93} (\bibinfo{year}{2012}),
  \eprint{1201.2101}.

\bibitem[{\citenamefont{Bondarescu and Wasserman}(2013)}]{Bondarescu2013}
\bibinfo{author}{\bibfnamefont{R.}~\bibnamefont{Bondarescu}} \bibnamefont{and}
  \bibinfo{author}{\bibfnamefont{I.}~\bibnamefont{Wasserman}}
  (\bibinfo{year}{2013}), \eprint{1305.2335}.

\bibitem[{\citenamefont{Nayyar and Owen}(2006)}]{Nayyar_Owen_2006}
\bibinfo{author}{\bibfnamefont{M.}~\bibnamefont{Nayyar}} \bibnamefont{and}
  \bibinfo{author}{\bibfnamefont{B.~J.} \bibnamefont{Owen}},
  \bibinfo{journal}{\prd} \textbf{\bibinfo{volume}{73}},
  \bibinfo{pages}{084001} (\bibinfo{year}{2006}),
  \urlprefix\url{http://link.aps.org/doi/10.1103/PhysRevD.73.084001}.

\bibitem[{\citenamefont{Melatos}(2007)}]{Melatos2007}
\bibinfo{author}{\bibfnamefont{A.}~\bibnamefont{Melatos}},
  \bibinfo{journal}{Advances in Space Research} \textbf{\bibinfo{volume}{40}},
  \bibinfo{pages}{1472 } (\bibinfo{year}{2007}), ISSN
  \bibinfo{issn}{0273-1177},
  \urlprefix\url{http://www.sciencedirect.com/science/article/pii/S02731177070%
03511}.

\bibitem[{\citenamefont{van Eysden and Melatos}(2008)}]{vEysden_Melatos_2008}
\bibinfo{author}{\bibfnamefont{C.~A.} \bibnamefont{van Eysden}}
  \bibnamefont{and} \bibinfo{author}{\bibfnamefont{A.}~\bibnamefont{Melatos}},
  \bibinfo{journal}{Class. Quant. Grav.} \textbf{\bibinfo{volume}{25}},
  \bibinfo{pages}{225020} (\bibinfo{year}{2008}),
  \urlprefix\url{http://stacks.iop.org/0264-9381/25/i=22/a=225020}.

\bibitem[{\citenamefont{{Vigelius} and
  {Melatos}}(2009)}]{Vigelius_Melatos_2009}
\bibinfo{author}{\bibfnamefont{M.}~\bibnamefont{{Vigelius}}} \bibnamefont{and}
  \bibinfo{author}{\bibfnamefont{A.}~\bibnamefont{{Melatos}}},
  \bibinfo{journal}{Mon. Not. R. Astron. Soc.} \textbf{\bibinfo{volume}{395}},
  \bibinfo{pages}{1972} (\bibinfo{year}{2009}), \eprint{0902.4264}.

\bibitem[{\citenamefont{{Andersson} et~al.}(2011)\citenamefont{{Andersson},
  {Ferrari}, {Jones}, {Kokkotas}, {Krishnan}, {Read}, {Rezzolla}, and
  {Zink}}}]{AnderssonEA_2011}
\bibinfo{author}{\bibfnamefont{N.}~\bibnamefont{{Andersson}}},
  \bibinfo{author}{\bibfnamefont{V.}~\bibnamefont{{Ferrari}}},
  \bibinfo{author}{\bibfnamefont{D.~I.} \bibnamefont{{Jones}}},
  \bibinfo{author}{\bibfnamefont{K.~D.} \bibnamefont{{Kokkotas}}},
  \bibinfo{author}{\bibfnamefont{B.}~\bibnamefont{{Krishnan}}},
  \bibinfo{author}{\bibfnamefont{J.~S.} \bibnamefont{{Read}}},
  \bibinfo{author}{\bibfnamefont{L.}~\bibnamefont{{Rezzolla}}},
  \bibnamefont{and} \bibinfo{author}{\bibfnamefont{B.}~\bibnamefont{{Zink}}},
  \bibinfo{journal}{General Relativity and Gravitation}
  \textbf{\bibinfo{volume}{43}}, \bibinfo{pages}{409} (\bibinfo{year}{2011}),
  \eprint{0912.0384}.

\bibitem[{\citenamefont{{Papaloizou} and
  {Pringle}}(1978)}]{Papaloizou_Pringle_1978}
\bibinfo{author}{\bibfnamefont{J.}~\bibnamefont{{Papaloizou}}}
  \bibnamefont{and} \bibinfo{author}{\bibfnamefont{J.~E.}
  \bibnamefont{{Pringle}}}, \bibinfo{journal}{Mon. Not. R. Astron. Soc.}
  \textbf{\bibinfo{volume}{184}}, \bibinfo{pages}{501} (\bibinfo{year}{1978}).

\bibitem[{\citenamefont{{Wagoner}}(1984)}]{Wagoner_1984}
\bibinfo{author}{\bibfnamefont{R.~V.} \bibnamefont{{Wagoner}}},
  \bibinfo{journal}{\apj} \textbf{\bibinfo{volume}{278}}, \bibinfo{pages}{345}
  (\bibinfo{year}{1984}).

\bibitem[{\citenamefont{{Verbunt}}(1993)}]{Verbunt_1993}
\bibinfo{author}{\bibfnamefont{F.}~\bibnamefont{{Verbunt}}},
  \bibinfo{journal}{Annu. Rev. Astron. Astrophys.}
  \textbf{\bibinfo{volume}{31}}, \bibinfo{pages}{93} (\bibinfo{year}{1993}).

\bibitem[{\citenamefont{{Watts} et~al.}(2008)\citenamefont{{Watts}, {Krishnan},
  {Bildsten}, and {Schutz}}}]{WattsEA_2008}
\bibinfo{author}{\bibfnamefont{A.~L.} \bibnamefont{{Watts}}},
  \bibinfo{author}{\bibfnamefont{B.}~\bibnamefont{{Krishnan}}},
  \bibinfo{author}{\bibfnamefont{L.}~\bibnamefont{{Bildsten}}},
  \bibnamefont{and} \bibinfo{author}{\bibfnamefont{B.~F.}
  \bibnamefont{{Schutz}}}, \bibinfo{journal}{Mon. Not. R. Astron. Soc.}
  \textbf{\bibinfo{volume}{389}}, \bibinfo{pages}{839} (\bibinfo{year}{2008}).

\bibitem[{\citenamefont{{Cook} et~al.}(1994)\citenamefont{{Cook}, {Shapiro},
  and {Teukolsky}}}]{CST_1994}
\bibinfo{author}{\bibfnamefont{G.~B.} \bibnamefont{{Cook}}},
  \bibinfo{author}{\bibfnamefont{S.~L.} \bibnamefont{{Shapiro}}},
  \bibnamefont{and} \bibinfo{author}{\bibfnamefont{S.~A.}
  \bibnamefont{{Teukolsky}}}, \bibinfo{journal}{\apj}
  \textbf{\bibinfo{volume}{424}}, \bibinfo{pages}{823} (\bibinfo{year}{1994}).

\bibitem[{\citenamefont{{Ushomirsky}
  et~al.}(2000{\natexlab{b}})\citenamefont{{Ushomirsky}, {Bildsten}, and
  {Cutler}}}]{UBC_2000}
\bibinfo{author}{\bibfnamefont{G.}~\bibnamefont{{Ushomirsky}}},
  \bibinfo{author}{\bibfnamefont{L.}~\bibnamefont{{Bildsten}}},
  \bibnamefont{and} \bibinfo{author}{\bibfnamefont{C.}~\bibnamefont{{Cutler}}},
  in \emph{\bibinfo{booktitle}{American Institute of Physics Conference
  Series}}, edited by \bibinfo{editor}{\bibnamefont{{S.~Meshkov}}}
  (\bibinfo{year}{2000}{\natexlab{b}}), vol. \bibinfo{volume}{523} of
  \emph{\bibinfo{series}{American Institute of Physics Conference Series}}, pp.
  \bibinfo{pages}{65--74}, \eprint{arXiv:astro-ph/0001129}.

\bibitem[{\citenamefont{{Chakrabarty} et~al.}(2003)\citenamefont{{Chakrabarty},
  {Morgan}, {Muno}, {Galloway}, {Wijnands}, {van der Klis}, and
  {Markwardt}}}]{ChakrabartyEA_Nature_2003}
\bibinfo{author}{\bibfnamefont{D.}~\bibnamefont{{Chakrabarty}}},
  \bibinfo{author}{\bibfnamefont{E.~H.} \bibnamefont{{Morgan}}},
  \bibinfo{author}{\bibfnamefont{M.~P.} \bibnamefont{{Muno}}},
  \bibinfo{author}{\bibfnamefont{D.~K.} \bibnamefont{{Galloway}}},
  \bibinfo{author}{\bibfnamefont{R.}~\bibnamefont{{Wijnands}}},
  \bibinfo{author}{\bibfnamefont{M.}~\bibnamefont{{van der Klis}}},
  \bibnamefont{and} \bibinfo{author}{\bibfnamefont{C.~B.}
  \bibnamefont{{Markwardt}}}, \bibinfo{journal}{Nature}
  \textbf{\bibinfo{volume}{424}}, \bibinfo{pages}{42} (\bibinfo{year}{2003}),
  \eprint{arXiv:astro-ph/0307029}.

\bibitem[{\citenamefont{{Bhattacharyya}}(2007)}]{Bhattacharyya2007}
\bibinfo{author}{\bibfnamefont{S.}~\bibnamefont{{Bhattacharyya}}},
  \bibinfo{journal}{Mon. Not. R. Astron. Soc.} \textbf{\bibinfo{volume}{377}},
  \bibinfo{pages}{198} (\bibinfo{year}{2007}), \eprint{arXiv:astro-ph/0605510}.

\bibitem[{\citenamefont{{Galloway} et~al.}(2010)\citenamefont{{Galloway},
  {Lin}, {Chakrabarty}, and {Hartman}}}]{GallowayEA_2010}
\bibinfo{author}{\bibfnamefont{D.~K.} \bibnamefont{{Galloway}}},
  \bibinfo{author}{\bibfnamefont{J.}~\bibnamefont{{Lin}}},
  \bibinfo{author}{\bibfnamefont{D.}~\bibnamefont{{Chakrabarty}}},
  \bibnamefont{and} \bibinfo{author}{\bibfnamefont{J.~M.}
  \bibnamefont{{Hartman}}}, \bibinfo{journal}{Astrophys. J. Lett.}
  \textbf{\bibinfo{volume}{711}}, \bibinfo{pages}{L148} (\bibinfo{year}{2010}),
  \eprint{0910.5546}.

\bibitem[{\citenamefont{{Abbott}
  et~al.}(2009{\natexlab{a}})\citenamefont{{Abbott}, {Abbott}, {Adhikari},
  {Ajith}, {Allen}, {Allen}, {Amin}, {Anderson}, {Anderson}, {Arain}
  et~al.}}]{LIGO_2009}
\bibinfo{author}{\bibfnamefont{B.~P.} \bibnamefont{{Abbott}}},
  \bibinfo{author}{\bibfnamefont{R.}~\bibnamefont{{Abbott}}},
  \bibinfo{author}{\bibfnamefont{R.}~\bibnamefont{{Adhikari}}},
  \bibinfo{author}{\bibfnamefont{P.}~\bibnamefont{{Ajith}}},
  \bibinfo{author}{\bibfnamefont{B.}~\bibnamefont{{Allen}}},
  \bibinfo{author}{\bibfnamefont{G.}~\bibnamefont{{Allen}}},
  \bibinfo{author}{\bibfnamefont{R.~S.} \bibnamefont{{Amin}}},
  \bibinfo{author}{\bibfnamefont{S.~B.} \bibnamefont{{Anderson}}},
  \bibinfo{author}{\bibfnamefont{W.~G.} \bibnamefont{{Anderson}}},
  \bibinfo{author}{\bibfnamefont{M.~A.} \bibnamefont{{Arain}}},
  \bibnamefont{et~al.}, \bibinfo{journal}{Reports on Progress in Physics}
  \textbf{\bibinfo{volume}{72}}, \bibinfo{pages}{076901}
  (\bibinfo{year}{2009}{\natexlab{a}}), \eprint{0711.3041}.

\bibitem[{\citenamefont{{Abadie}
  et~al.}(2010{\natexlab{a}})\citenamefont{{Abadie}, {Abbott}, {Abbott},
  {Abernathy}, {Adams}, {Adhikari}, {Ajith}, {Allen}, {Allen}, {Amador Ceron}
  et~al.}}]{LIGO_S5cal_Abadie2010}
\bibinfo{author}{\bibfnamefont{J.}~\bibnamefont{{Abadie}}},
  \bibinfo{author}{\bibfnamefont{B.~P.} \bibnamefont{{Abbott}}},
  \bibinfo{author}{\bibfnamefont{R.}~\bibnamefont{{Abbott}}},
  \bibinfo{author}{\bibfnamefont{M.}~\bibnamefont{{Abernathy}}},
  \bibinfo{author}{\bibfnamefont{C.}~\bibnamefont{{Adams}}},
  \bibinfo{author}{\bibfnamefont{R.}~\bibnamefont{{Adhikari}}},
  \bibinfo{author}{\bibfnamefont{P.}~\bibnamefont{{Ajith}}},
  \bibinfo{author}{\bibfnamefont{B.}~\bibnamefont{{Allen}}},
  \bibinfo{author}{\bibfnamefont{G.}~\bibnamefont{{Allen}}},
  \bibinfo{author}{\bibfnamefont{E.}~\bibnamefont{{Amador Ceron}}},
  \bibnamefont{et~al.}, \bibinfo{journal}{Nuclear Instruments and Methods in
  Physics Research A} \textbf{\bibinfo{volume}{624}}, \bibinfo{pages}{223}
  (\bibinfo{year}{2010}{\natexlab{a}}), \eprint{1007.3973}.

\bibitem[{LIG()}]{LIGO}
\urlprefix\url{http://www.ligo.caltech.edu/}.

\bibitem[{VIR()}]{VIRGO}
\urlprefix\url{https://wwwcascina.virgo.infn.it/}.

\bibitem[{\citenamefont{Accadia et~al.}(2012)}]{Virgo2}
\bibinfo{author}{\bibfnamefont{T.}~\bibnamefont{Accadia}} \bibnamefont{et~al.}
  (\bibinfo{collaboration}{VIRGO Collaboration}), \bibinfo{journal}{JINST}
  \textbf{\bibinfo{volume}{7}}, \bibinfo{pages}{P03012} (\bibinfo{year}{2012}).

\bibitem[{\citenamefont{Grote}(2010)}]{GEO2}
\bibinfo{author}{\bibfnamefont{H.}~\bibnamefont{Grote}}
  (\bibinfo{collaboration}{LIGO Scientific Collaboration}),
  \bibinfo{journal}{Class. Quant. Grav.} \textbf{\bibinfo{volume}{27}},
  \bibinfo{pages}{084003} (\bibinfo{year}{2010}).

\bibitem[{GEO()}]{GEO600}
\urlprefix\url{http://www.geo600.org/}.

\bibitem[{\citenamefont{Brady et~al.}(1998)\citenamefont{Brady, Creighton,
  Cutler, and Schutz}}]{BradyEA_1998}
\bibinfo{author}{\bibfnamefont{P.~R.} \bibnamefont{Brady}},
  \bibinfo{author}{\bibfnamefont{T.}~\bibnamefont{Creighton}},
  \bibinfo{author}{\bibfnamefont{C.}~\bibnamefont{Cutler}}, \bibnamefont{and}
  \bibinfo{author}{\bibfnamefont{B.~F.} \bibnamefont{Schutz}},
  \bibinfo{journal}{\prd} \textbf{\bibinfo{volume}{57}}, \bibinfo{pages}{2101}
  (\bibinfo{year}{1998}),
  \urlprefix\url{http://link.aps.org/doi/10.1103/PhysRevD.57.2101}.

\bibitem[{\citenamefont{{Abbott} et~al.}(2004)\citenamefont{{Abbott}, {Abbott},
  {Adhikari}, {Ageev}, {Allen}, {Amin}, {Anderson}, {Anderson}, {Araya},
  {Armandula} et~al.}}]{LIGO_2004_PRD69}
\bibinfo{author}{\bibfnamefont{B.}~\bibnamefont{{Abbott}}},
  \bibinfo{author}{\bibfnamefont{R.}~\bibnamefont{{Abbott}}},
  \bibinfo{author}{\bibfnamefont{R.}~\bibnamefont{{Adhikari}}},
  \bibinfo{author}{\bibfnamefont{A.}~\bibnamefont{{Ageev}}},
  \bibinfo{author}{\bibfnamefont{B.}~\bibnamefont{{Allen}}},
  \bibinfo{author}{\bibfnamefont{R.}~\bibnamefont{{Amin}}},
  \bibinfo{author}{\bibfnamefont{S.~B.} \bibnamefont{{Anderson}}},
  \bibinfo{author}{\bibfnamefont{W.~G.} \bibnamefont{{Anderson}}},
  \bibinfo{author}{\bibfnamefont{M.}~\bibnamefont{{Araya}}},
  \bibinfo{author}{\bibfnamefont{H.}~\bibnamefont{{Armandula}}},
  \bibnamefont{et~al.} (\bibinfo{collaboration}{The LIGO Scientific
  Collaboration}), \bibinfo{journal}{\prd} \textbf{\bibinfo{volume}{69}},
  \bibinfo{eid}{082004} (\bibinfo{year}{2004}), \eprint{arXiv:gr-qc/0308050}.

\bibitem[{\citenamefont{{Abbott} et~al.}(2010)\citenamefont{{Abbott}, {Abbott},
  {Acernese}, {Adhikari}, {Ajith}, {Allen}, {Allen}, {Alshourbagy}, {Amin},
  {Anderson} et~al.}}]{LIGO_S5_KnownPulsars_2010}
\bibinfo{author}{\bibfnamefont{B.~P.} \bibnamefont{{Abbott}}},
  \bibinfo{author}{\bibfnamefont{R.}~\bibnamefont{{Abbott}}},
  \bibinfo{author}{\bibfnamefont{F.}~\bibnamefont{{Acernese}}},
  \bibinfo{author}{\bibfnamefont{R.}~\bibnamefont{{Adhikari}}},
  \bibinfo{author}{\bibfnamefont{P.}~\bibnamefont{{Ajith}}},
  \bibinfo{author}{\bibfnamefont{B.}~\bibnamefont{{Allen}}},
  \bibinfo{author}{\bibfnamefont{G.}~\bibnamefont{{Allen}}},
  \bibinfo{author}{\bibfnamefont{M.}~\bibnamefont{{Alshourbagy}}},
  \bibinfo{author}{\bibfnamefont{R.~S.} \bibnamefont{{Amin}}},
  \bibinfo{author}{\bibfnamefont{S.~B.} \bibnamefont{{Anderson}}},
  \bibnamefont{et~al.} (\bibinfo{collaboration}{The LIGO Scientific
  Collaboration and the Virgo Collaboration}), \bibinfo{journal}{\apj}
  \textbf{\bibinfo{volume}{713}}, \bibinfo{pages}{671} (\bibinfo{year}{2010}),
  \eprint{0909.3583}.

\bibitem[{\citenamefont{{Abbott}
  et~al.}(2008{\natexlab{a}})\citenamefont{{Abbott}, {Abbott}, {Adhikari},
  {Ajith}, {Allen}, {Allen}, {Amin}, {Anderson}, {Anderson}, {Arain}
  et~al.}}]{LIGO_S5_Crab_2008}
\bibinfo{author}{\bibfnamefont{B.}~\bibnamefont{{Abbott}}},
  \bibinfo{author}{\bibfnamefont{R.}~\bibnamefont{{Abbott}}},
  \bibinfo{author}{\bibfnamefont{R.}~\bibnamefont{{Adhikari}}},
  \bibinfo{author}{\bibfnamefont{P.}~\bibnamefont{{Ajith}}},
  \bibinfo{author}{\bibfnamefont{B.}~\bibnamefont{{Allen}}},
  \bibinfo{author}{\bibfnamefont{G.}~\bibnamefont{{Allen}}},
  \bibinfo{author}{\bibfnamefont{R.}~\bibnamefont{{Amin}}},
  \bibinfo{author}{\bibfnamefont{S.~B.} \bibnamefont{{Anderson}}},
  \bibinfo{author}{\bibfnamefont{W.~G.} \bibnamefont{{Anderson}}},
  \bibinfo{author}{\bibfnamefont{M.~A.} \bibnamefont{{Arain}}},
  \bibnamefont{et~al.} (\bibinfo{collaboration}{The LIGO Scientific
  Collaboration}), \bibinfo{journal}{Astrophys. J. Lett.}
  \textbf{\bibinfo{volume}{683}}, \bibinfo{pages}{L45}
  (\bibinfo{year}{2008}{\natexlab{a}}), \eprint{0805.4758}.

\bibitem[{\citenamefont{{Abadie}
  et~al.}(2011{\natexlab{a}})\citenamefont{{Abadie}, {Abbott}, {Abbott},
  {Abernathy}, {Accadia}, {Acernese}, {Adams}, {Adhikari}, {Affeldt}, {Allen}
  et~al.}}]{LIGO_Vela_2011}
\bibinfo{author}{\bibfnamefont{J.}~\bibnamefont{{Abadie}}},
  \bibinfo{author}{\bibfnamefont{B.~P.} \bibnamefont{{Abbott}}},
  \bibinfo{author}{\bibfnamefont{R.}~\bibnamefont{{Abbott}}},
  \bibinfo{author}{\bibfnamefont{M.}~\bibnamefont{{Abernathy}}},
  \bibinfo{author}{\bibfnamefont{T.}~\bibnamefont{{Accadia}}},
  \bibinfo{author}{\bibfnamefont{F.}~\bibnamefont{{Acernese}}},
  \bibinfo{author}{\bibfnamefont{C.}~\bibnamefont{{Adams}}},
  \bibinfo{author}{\bibfnamefont{R.}~\bibnamefont{{Adhikari}}},
  \bibinfo{author}{\bibfnamefont{C.}~\bibnamefont{{Affeldt}}},
  \bibinfo{author}{\bibfnamefont{B.}~\bibnamefont{{Allen}}},
  \bibnamefont{et~al.} (\bibinfo{collaboration}{The LIGO Scientific
  Collaboration and the Virgo Collaboration}), \bibinfo{journal}{\apj}
  \textbf{\bibinfo{volume}{737}}, \bibinfo{eid}{93}
  (\bibinfo{year}{2011}{\natexlab{a}}), \eprint{1104.2712}.

\bibitem[{\citenamefont{{Abadie}
  et~al.}(2010{\natexlab{b}})\citenamefont{{Abadie}, {Abbott}, {Abbott},
  {Abernathy}, {Adams}, {Adhikari}, {Ajith}, {Allen}, {Allen}, {Amador Ceron}
  et~al.}}]{LIGO_CasA_2010}
\bibinfo{author}{\bibfnamefont{J.}~\bibnamefont{{Abadie}}},
  \bibinfo{author}{\bibfnamefont{B.~P.} \bibnamefont{{Abbott}}},
  \bibinfo{author}{\bibfnamefont{R.}~\bibnamefont{{Abbott}}},
  \bibinfo{author}{\bibfnamefont{M.}~\bibnamefont{{Abernathy}}},
  \bibinfo{author}{\bibfnamefont{C.}~\bibnamefont{{Adams}}},
  \bibinfo{author}{\bibfnamefont{R.}~\bibnamefont{{Adhikari}}},
  \bibinfo{author}{\bibfnamefont{P.}~\bibnamefont{{Ajith}}},
  \bibinfo{author}{\bibfnamefont{B.}~\bibnamefont{{Allen}}},
  \bibinfo{author}{\bibfnamefont{G.}~\bibnamefont{{Allen}}},
  \bibinfo{author}{\bibfnamefont{E.}~\bibnamefont{{Amador Ceron}}},
  \bibnamefont{et~al.} (\bibinfo{collaboration}{LIGO Scientific
  Collaboration}), \bibinfo{journal}{\apj} \textbf{\bibinfo{volume}{722}},
  \bibinfo{pages}{1504} (\bibinfo{year}{2010}{\natexlab{b}}),
  \eprint{1006.2535}.

\bibitem[{\citenamefont{{Abbott}
  et~al.}(2008{\natexlab{b}})\citenamefont{{Abbott}, {Abbott}, {Adhikari},
  {Agresti}, {Ajith}, {Allen}, {Amin}, {Anderson}, {Anderson}, {Arain}
  et~al.}}]{LIGO_S4_AllSKy_2008}
\bibinfo{author}{\bibfnamefont{B.}~\bibnamefont{{Abbott}}},
  \bibinfo{author}{\bibfnamefont{R.}~\bibnamefont{{Abbott}}},
  \bibinfo{author}{\bibfnamefont{R.}~\bibnamefont{{Adhikari}}},
  \bibinfo{author}{\bibfnamefont{J.}~\bibnamefont{{Agresti}}},
  \bibinfo{author}{\bibfnamefont{P.}~\bibnamefont{{Ajith}}},
  \bibinfo{author}{\bibfnamefont{B.}~\bibnamefont{{Allen}}},
  \bibinfo{author}{\bibfnamefont{R.}~\bibnamefont{{Amin}}},
  \bibinfo{author}{\bibfnamefont{S.~B.} \bibnamefont{{Anderson}}},
  \bibinfo{author}{\bibfnamefont{W.~G.} \bibnamefont{{Anderson}}},
  \bibinfo{author}{\bibfnamefont{M.}~\bibnamefont{{Arain}}},
  \bibnamefont{et~al.} (\bibinfo{collaboration}{LIGO Scientific
  Collaboration}), \bibinfo{journal}{\prd} \textbf{\bibinfo{volume}{77}},
  \bibinfo{eid}{022001} (\bibinfo{year}{2008}{\natexlab{b}}),
  \eprint{0708.3818}.

\bibitem[{\citenamefont{{Abbott}
  et~al.}(2009{\natexlab{b}})\citenamefont{{Abbott}, {Abbott}, {Adhikari},
  {Ajith}, {Allen}, {Allen}, {Amin}, {Anderson}, {Anderson}, {Anderson}
  et~al.}}]{LIGO_S4_EaH_2009}
\bibinfo{author}{\bibfnamefont{B.}~\bibnamefont{{Abbott}}},
  \bibinfo{author}{\bibfnamefont{R.}~\bibnamefont{{Abbott}}},
  \bibinfo{author}{\bibfnamefont{R.}~\bibnamefont{{Adhikari}}},
  \bibinfo{author}{\bibfnamefont{P.}~\bibnamefont{{Ajith}}},
  \bibinfo{author}{\bibfnamefont{B.}~\bibnamefont{{Allen}}},
  \bibinfo{author}{\bibfnamefont{G.}~\bibnamefont{{Allen}}},
  \bibinfo{author}{\bibfnamefont{R.}~\bibnamefont{{Amin}}},
  \bibinfo{author}{\bibfnamefont{D.~P.} \bibnamefont{{Anderson}}},
  \bibinfo{author}{\bibfnamefont{S.~B.} \bibnamefont{{Anderson}}},
  \bibinfo{author}{\bibfnamefont{W.~G.} \bibnamefont{{Anderson}}},
  \bibnamefont{et~al.} (\bibinfo{collaboration}{The LIGO Scientific
  Collaboration}), \bibinfo{journal}{\prd} \textbf{\bibinfo{volume}{79}},
  \bibinfo{eid}{022001} (\bibinfo{year}{2009}{\natexlab{b}}),
  \eprint{0804.1747}.

\bibitem[{\citenamefont{{Abbott}
  et~al.}(2009{\natexlab{c}})\citenamefont{{Abbott}, {Abbott}, {Adhikari},
  {Ajith}, {Allen}, {Allen}, {Amin}, {Anderson}, {Anderson}, {Arain}
  et~al.}}]{LIGO_S5i_Powerflux_2009}
\bibinfo{author}{\bibfnamefont{B.~P.} \bibnamefont{{Abbott}}},
  \bibinfo{author}{\bibfnamefont{R.}~\bibnamefont{{Abbott}}},
  \bibinfo{author}{\bibfnamefont{R.}~\bibnamefont{{Adhikari}}},
  \bibinfo{author}{\bibfnamefont{P.}~\bibnamefont{{Ajith}}},
  \bibinfo{author}{\bibfnamefont{B.}~\bibnamefont{{Allen}}},
  \bibinfo{author}{\bibfnamefont{G.}~\bibnamefont{{Allen}}},
  \bibinfo{author}{\bibfnamefont{R.~S.} \bibnamefont{{Amin}}},
  \bibinfo{author}{\bibfnamefont{S.~B.} \bibnamefont{{Anderson}}},
  \bibinfo{author}{\bibfnamefont{W.~G.} \bibnamefont{{Anderson}}},
  \bibinfo{author}{\bibfnamefont{M.~A.} \bibnamefont{{Arain}}},
  \bibnamefont{et~al.} (\bibinfo{collaboration}{The LIGO Scientific
  Collaboration}), \bibinfo{journal}{Phys. Rev. Lett.}
  \textbf{\bibinfo{volume}{102}}, \bibinfo{eid}{111102}
  (\bibinfo{year}{2009}{\natexlab{c}}), \eprint{0810.0283}.

\bibitem[{\citenamefont{{Abbott}
  et~al.}(2009{\natexlab{d}})\citenamefont{{Abbott}, {Abbott}, {Adhikari},
  {Ajith}, {Allen}, {Allen}, {Amin}, {Anderson}, {Anderson}, {Arain}
  et~al.}}]{LIGO_S5_EaH_2009}
\bibinfo{author}{\bibfnamefont{B.~P.} \bibnamefont{{Abbott}}},
  \bibinfo{author}{\bibfnamefont{R.}~\bibnamefont{{Abbott}}},
  \bibinfo{author}{\bibfnamefont{R.}~\bibnamefont{{Adhikari}}},
  \bibinfo{author}{\bibfnamefont{P.}~\bibnamefont{{Ajith}}},
  \bibinfo{author}{\bibfnamefont{B.}~\bibnamefont{{Allen}}},
  \bibinfo{author}{\bibfnamefont{G.}~\bibnamefont{{Allen}}},
  \bibinfo{author}{\bibfnamefont{R.~S.} \bibnamefont{{Amin}}},
  \bibinfo{author}{\bibfnamefont{S.~B.} \bibnamefont{{Anderson}}},
  \bibinfo{author}{\bibfnamefont{W.~G.} \bibnamefont{{Anderson}}},
  \bibinfo{author}{\bibfnamefont{M.~A.} \bibnamefont{{Arain}}},
  \bibnamefont{et~al.} (\bibinfo{collaboration}{The LIGO Scientific
  Collaboration and the Virgo Collaboration}), \bibinfo{journal}{\prd}
  \textbf{\bibinfo{volume}{80}}, \bibinfo{eid}{042003}
  (\bibinfo{year}{2009}{\natexlab{d}}), \eprint{0905.1705}.

\bibitem[{\citenamefont{{Abadie} et~al.}(2012)\citenamefont{{Abadie}, {Abbott},
  {Abbott}, {Abbott}, {Abernathy}, {Accadia}, {Acernese}, {Adams}, {Adhikari},
  {Affeldt} et~al.}}]{LIGO_S5full_Powerflux_2012}
\bibinfo{author}{\bibfnamefont{J.}~\bibnamefont{{Abadie}}},
  \bibinfo{author}{\bibfnamefont{B.~P.} \bibnamefont{{Abbott}}},
  \bibinfo{author}{\bibfnamefont{R.}~\bibnamefont{{Abbott}}},
  \bibinfo{author}{\bibfnamefont{T.~D.} \bibnamefont{{Abbott}}},
  \bibinfo{author}{\bibfnamefont{M.}~\bibnamefont{{Abernathy}}},
  \bibinfo{author}{\bibfnamefont{T.}~\bibnamefont{{Accadia}}},
  \bibinfo{author}{\bibfnamefont{F.}~\bibnamefont{{Acernese}}},
  \bibinfo{author}{\bibfnamefont{C.}~\bibnamefont{{Adams}}},
  \bibinfo{author}{\bibfnamefont{R.}~\bibnamefont{{Adhikari}}},
  \bibinfo{author}{\bibfnamefont{C.}~\bibnamefont{{Affeldt}}},
  \bibnamefont{et~al.} (\bibinfo{collaboration}{The LIGO Scientific
  Collaboration}), \bibinfo{journal}{\prd} \textbf{\bibinfo{volume}{85}},
  \bibinfo{eid}{022001} (\bibinfo{year}{2012}), \eprint{1110.0208}.

\bibitem[{\citenamefont{Goetz and Riles}(2011)}]{Goetz_Riles_TwoSpect_2011}
\bibinfo{author}{\bibfnamefont{E.}~\bibnamefont{Goetz}} \bibnamefont{and}
  \bibinfo{author}{\bibfnamefont{K.}~\bibnamefont{Riles}},
  \bibinfo{journal}{Class. Quant. Grav.} \textbf{\bibinfo{volume}{28}},
  \bibinfo{pages}{215006} (\bibinfo{year}{2011}),
  \urlprefix\url{http://stacks.iop.org/0264-9381/28/i=21/a=215006}.

\bibitem[{\citenamefont{Abbott et~al.}(2007{\natexlab{a}})\citenamefont{Abbott,
  Abbott, Adhikari, Agresti, Ajith, Allen, Amin, Anderson, Anderson, Arain
  et~al.}}]{LIGO_S2_ScoX1_2007}
\bibinfo{author}{\bibfnamefont{B.}~\bibnamefont{Abbott}},
  \bibinfo{author}{\bibfnamefont{R.}~\bibnamefont{Abbott}},
  \bibinfo{author}{\bibfnamefont{R.}~\bibnamefont{Adhikari}},
  \bibinfo{author}{\bibfnamefont{J.}~\bibnamefont{Agresti}},
  \bibinfo{author}{\bibfnamefont{P.}~\bibnamefont{Ajith}},
  \bibinfo{author}{\bibfnamefont{B.}~\bibnamefont{Allen}},
  \bibinfo{author}{\bibfnamefont{R.}~\bibnamefont{Amin}},
  \bibinfo{author}{\bibfnamefont{S.~B.} \bibnamefont{Anderson}},
  \bibinfo{author}{\bibfnamefont{W.~G.} \bibnamefont{Anderson}},
  \bibinfo{author}{\bibfnamefont{M.}~\bibnamefont{Arain}}, \bibnamefont{et~al.}
  (\bibinfo{collaboration}{LIGO Scientific Collaboration}),
  \bibinfo{journal}{\prd} \textbf{\bibinfo{volume}{76}},
  \bibinfo{pages}{082001} (\bibinfo{year}{2007}{\natexlab{a}}).

\bibitem[{\citenamefont{Jaranowski et~al.}(1998)\citenamefont{Jaranowski,
  Kr\'olak, and Schutz}}]{JKS1998}
\bibinfo{author}{\bibfnamefont{P.}~\bibnamefont{Jaranowski}},
  \bibinfo{author}{\bibfnamefont{A.}~\bibnamefont{Kr\'olak}}, \bibnamefont{and}
  \bibinfo{author}{\bibfnamefont{B.~F.} \bibnamefont{Schutz}},
  \bibinfo{journal}{\prd} \textbf{\bibinfo{volume}{58}},
  \bibinfo{pages}{063001} (\bibinfo{year}{1998}).

\bibitem[{\citenamefont{Abbott et~al.}(2007{\natexlab{b}})\citenamefont{Abbott,
  Abbott, Adhikari, Agresti, Ajith, Allen, Amin, Anderson, Anderson, Arain
  et~al.}}]{LIGO_S4_ScoX1_2007}
\bibinfo{author}{\bibfnamefont{B.}~\bibnamefont{Abbott}},
  \bibinfo{author}{\bibfnamefont{R.}~\bibnamefont{Abbott}},
  \bibinfo{author}{\bibfnamefont{R.}~\bibnamefont{Adhikari}},
  \bibinfo{author}{\bibfnamefont{J.}~\bibnamefont{Agresti}},
  \bibinfo{author}{\bibfnamefont{P.}~\bibnamefont{Ajith}},
  \bibinfo{author}{\bibfnamefont{B.}~\bibnamefont{Allen}},
  \bibinfo{author}{\bibfnamefont{R.}~\bibnamefont{Amin}},
  \bibinfo{author}{\bibfnamefont{S.~B.} \bibnamefont{Anderson}},
  \bibinfo{author}{\bibfnamefont{W.~G.} \bibnamefont{Anderson}},
  \bibinfo{author}{\bibfnamefont{M.}~\bibnamefont{Arain}}, \bibnamefont{et~al.}
  (\bibinfo{collaboration}{LIGO Scientific Collaboration}),
  \bibinfo{journal}{\prd} \textbf{\bibinfo{volume}{76}},
  \bibinfo{pages}{082003} (\bibinfo{year}{2007}{\natexlab{b}}).

\bibitem[{\citenamefont{{Abadie}
  et~al.}(2011{\natexlab{b}})\citenamefont{{Abadie}, {Abbott}, {Abbott},
  {Abernathy}, {Accadia}, {Acernese}, {Adams}, {Adhikari}, {Ajith}, {Allen}
  et~al.}}]{LIGO_S5stoch_ScoX1_2011}
\bibinfo{author}{\bibfnamefont{J.}~\bibnamefont{{Abadie}}},
  \bibinfo{author}{\bibfnamefont{B.~P.} \bibnamefont{{Abbott}}},
  \bibinfo{author}{\bibfnamefont{R.}~\bibnamefont{{Abbott}}},
  \bibinfo{author}{\bibfnamefont{M.}~\bibnamefont{{Abernathy}}},
  \bibinfo{author}{\bibfnamefont{T.}~\bibnamefont{{Accadia}}},
  \bibinfo{author}{\bibfnamefont{F.}~\bibnamefont{{Acernese}}},
  \bibinfo{author}{\bibfnamefont{C.}~\bibnamefont{{Adams}}},
  \bibinfo{author}{\bibfnamefont{R.}~\bibnamefont{{Adhikari}}},
  \bibinfo{author}{\bibfnamefont{P.}~\bibnamefont{{Ajith}}},
  \bibinfo{author}{\bibfnamefont{B.}~\bibnamefont{{Allen}}},
  \bibnamefont{et~al.} (\bibinfo{collaboration}{LIGO Scientific Collaboration,
  Virgo Collaboration}), \bibinfo{journal}{Phys. Rev. Lett.}
  \textbf{\bibinfo{volume}{107}}, \bibinfo{eid}{271102}
  (\bibinfo{year}{2011}{\natexlab{b}}), \eprint{1109.1809}.

\bibitem[{\citenamefont{{Brady} and {Creighton}}(2000)}]{Brady_Creighton_2000}
\bibinfo{author}{\bibfnamefont{P.~R.} \bibnamefont{{Brady}}} \bibnamefont{and}
  \bibinfo{author}{\bibfnamefont{T.}~\bibnamefont{{Creighton}}},
  \bibinfo{journal}{\prd} \textbf{\bibinfo{volume}{61}}, \bibinfo{eid}{082001}
  (\bibinfo{year}{2000}), \eprint{arXiv:gr-qc/9812014}.

\bibitem[{\citenamefont{{Prix} and {Shaltev}}(2012)}]{Prix_Shaltev_2012}
\bibinfo{author}{\bibfnamefont{R.}~\bibnamefont{{Prix}}} \bibnamefont{and}
  \bibinfo{author}{\bibfnamefont{M.}~\bibnamefont{{Shaltev}}},
  \bibinfo{journal}{\prd} \textbf{\bibinfo{volume}{85}}, \bibinfo{eid}{084010}
  (\bibinfo{year}{2012}), \eprint{1201.4321}.

\bibitem[{\citenamefont{{Messenger} and {Woan}}(2007)}]{MW2007}
\bibinfo{author}{\bibfnamefont{C.}~\bibnamefont{{Messenger}}} \bibnamefont{and}
  \bibinfo{author}{\bibfnamefont{G.}~\bibnamefont{{Woan}}},
  \bibinfo{journal}{Class. Quant. Grav.} \textbf{\bibinfo{volume}{24}},
  \bibinfo{pages}{469} (\bibinfo{year}{2007}), \eprint{arXiv:gr-qc/0703155}.

\bibitem[{\citenamefont{Ransom et~al.}(2003)\citenamefont{Ransom, Cordes, and
  Eikenberry}}]{RansomEA_2003}
\bibinfo{author}{\bibfnamefont{S.~M.} \bibnamefont{Ransom}},
  \bibinfo{author}{\bibfnamefont{J.~M.} \bibnamefont{Cordes}},
  \bibnamefont{and} \bibinfo{author}{\bibfnamefont{S.~S.}
  \bibnamefont{Eikenberry}}, \bibinfo{journal}{\apj}
  \textbf{\bibinfo{volume}{589}}, \bibinfo{pages}{911} (\bibinfo{year}{2003}),
  \urlprefix\url{http://stacks.iop.org/0004-637X/589/i=2/a=911}.

\bibitem[{\citenamefont{Tauris and van~den
  Heuvel}(2006)}]{Tauris_vandenHeuvel_ch16}
\bibinfo{author}{\bibfnamefont{T.~M.} \bibnamefont{Tauris}} \bibnamefont{and}
  \bibinfo{author}{\bibfnamefont{E.~P.~J.} \bibnamefont{van~den Heuvel}}
  (\bibinfo{publisher}{Cambridge University Press}, \bibinfo{year}{2006}),
  chap.~\bibinfo{chapter}{16},
  \urlprefix\url{http://dx.doi.org/10.1017/CBO9780511536281}.

\bibitem[{\citenamefont{{van Paradijs} and
  {White}}(1995)}]{vanParadijs_White_1995}
\bibinfo{author}{\bibfnamefont{J.}~\bibnamefont{{van Paradijs}}}
  \bibnamefont{and} \bibinfo{author}{\bibfnamefont{N.}~\bibnamefont{{White}}},
  \bibinfo{journal}{"Astrophys. J. Lett."} \textbf{\bibinfo{volume}{447}},
  \bibinfo{pages}{L33+} (\bibinfo{year}{1995}).

\bibitem[{\citenamefont{Owen}(2010)}]{Owen2010}
\bibinfo{author}{\bibfnamefont{B.~J.} \bibnamefont{Owen}},
  \bibinfo{journal}{Phys. Rev.} \textbf{\bibinfo{volume}{D82}},
  \bibinfo{pages}{104002} (\bibinfo{year}{2010}), \eprint{1006.1994}.

\bibitem[{\citenamefont{{Taylor} and {Weisberg}}(1989)}]{Taylor_Weisberg_1989}
\bibinfo{author}{\bibfnamefont{J.~H.} \bibnamefont{{Taylor}}} \bibnamefont{and}
  \bibinfo{author}{\bibfnamefont{J.~M.} \bibnamefont{{Weisberg}}},
  \bibinfo{journal}{\apj} \textbf{\bibinfo{volume}{345}}, \bibinfo{pages}{434}
  (\bibinfo{year}{1989}).

\bibitem[{\citenamefont{Prix}(2007)}]{Prix_2007}
\bibinfo{author}{\bibfnamefont{R.}~\bibnamefont{Prix}}, \bibinfo{journal}{\prd}
  \textbf{\bibinfo{volume}{75}}, \bibinfo{pages}{023004}
  (\bibinfo{year}{2007}).

\bibitem[{\citenamefont{{Lo} et~al.}(2011)\citenamefont{{Lo}, {Lamb},
  {Boutloukos}, and {Miller}}}]{LambEA_2011}
\bibinfo{author}{\bibfnamefont{K.-H.} \bibnamefont{{Lo}}},
  \bibinfo{author}{\bibfnamefont{F.~K.} \bibnamefont{{Lamb}}},
  \bibinfo{author}{\bibfnamefont{S.}~\bibnamefont{{Boutloukos}}},
  \bibnamefont{and} \bibinfo{author}{\bibfnamefont{M.~C.}
  \bibnamefont{{Miller}}}, in \emph{\bibinfo{booktitle}{AAS/High Energy
  Astrophysics Division}} (\bibinfo{year}{2011}), vol.~\bibinfo{volume}{12} of
  \emph{\bibinfo{series}{AAS/High Energy Astrophysics Division}}, p.
  \bibinfo{pages}{44.05}.

\bibitem[{\citenamefont{{Watts}}(2012)}]{Watts_2012}
\bibinfo{author}{\bibfnamefont{A.~L.} \bibnamefont{{Watts}}},
  \bibinfo{journal}{"Ann. Rev. Astron. Astrophys."}
  \textbf{\bibinfo{volume}{50}}, \bibinfo{pages}{609} (\bibinfo{year}{2012}),
  \eprint{1203.2065}.

\bibitem[{\citenamefont{van~der Klis}(1998)}]{vanderKlis_1998}
\bibinfo{author}{\bibfnamefont{M.}~\bibnamefont{van~der Klis}},
  \bibinfo{journal}{Advances in Space Research} \textbf{\bibinfo{volume}{22}},
  \bibinfo{pages}{925 } (\bibinfo{year}{1998}), ISSN \bibinfo{issn}{0273-1177},
  \bibinfo{note}{x-Ray Timing and Cosmic Gamma Ray Bursts},
  \urlprefix\url{http://www.sciencedirect.com/science/article/B6V3S-3X8G9H0-3/%
2/f18d9fad8f3d9f9d 15b6c69105c749b1}.

\bibitem[{\citenamefont{{Zhang} et~al.}(2012)\citenamefont{{Zhang}, {Pan},
  {Wang}, {Taani}, and {Zhao}}}]{ZhangEA_2012}
\bibinfo{author}{\bibfnamefont{C.~M.} \bibnamefont{{Zhang}}},
  \bibinfo{author}{\bibfnamefont{Y.~Y.} \bibnamefont{{Pan}}},
  \bibinfo{author}{\bibfnamefont{J.}~\bibnamefont{{Wang}}},
  \bibinfo{author}{\bibfnamefont{A.}~\bibnamefont{{Taani}}}, \bibnamefont{and}
  \bibinfo{author}{\bibfnamefont{Y.~H.} \bibnamefont{{Zhao}}},
  \bibinfo{journal}{Int. J. Mod. Phys. Conference Series}
  \textbf{\bibinfo{volume}{12}}, \bibinfo{pages}{414} (\bibinfo{year}{2012}).

\bibitem[{\citenamefont{van~der Klis et~al.}(1996)\citenamefont{van~der Klis,
  Swank, Zhang, Jahoda, Morgan, Lewin, Vaughan, and van
  Paradijs}}]{vanderKlisEA_1996}
\bibinfo{author}{\bibfnamefont{M.}~\bibnamefont{van~der Klis}},
  \bibinfo{author}{\bibfnamefont{J.~H.} \bibnamefont{Swank}},
  \bibinfo{author}{\bibfnamefont{W.}~\bibnamefont{Zhang}},
  \bibinfo{author}{\bibfnamefont{K.}~\bibnamefont{Jahoda}},
  \bibinfo{author}{\bibfnamefont{E.~H.} \bibnamefont{Morgan}},
  \bibinfo{author}{\bibfnamefont{W.~H.~G.} \bibnamefont{Lewin}},
  \bibinfo{author}{\bibfnamefont{B.}~\bibnamefont{Vaughan}}, \bibnamefont{and}
  \bibinfo{author}{\bibfnamefont{J.}~\bibnamefont{van Paradijs}},
  \bibinfo{journal}{Astrophys. J. Lett.} \textbf{\bibinfo{volume}{469}},
  \bibinfo{pages}{L1} (\bibinfo{year}{1996}),
  \urlprefix\url{http://stacks.iop.org/1538-4357/469/i=1/a=L1}.

\bibitem[{\citenamefont{Bradshaw et~al.}(1999)\citenamefont{Bradshaw, Fomalont,
  and Geldzahler}}]{BradshawEA_1999}
\bibinfo{author}{\bibfnamefont{C.~F.} \bibnamefont{Bradshaw}},
  \bibinfo{author}{\bibfnamefont{E.~B.} \bibnamefont{Fomalont}},
  \bibnamefont{and} \bibinfo{author}{\bibfnamefont{B.~J.}
  \bibnamefont{Geldzahler}}, \bibinfo{journal}{Astrophys. J. Lett.}
  \textbf{\bibinfo{volume}{512}}, \bibinfo{pages}{L121} (\bibinfo{year}{1999}),
  \urlprefix\url{http://stacks.iop.org/1538-4357/512/i=2/a=L121}.

\bibitem[{\citenamefont{Steeghs and Casares}(2002)}]{Steeghs_Casares_2002}
\bibinfo{author}{\bibfnamefont{D.}~\bibnamefont{Steeghs}} \bibnamefont{and}
  \bibinfo{author}{\bibfnamefont{J.}~\bibnamefont{Casares}},
  \bibinfo{journal}{The Astrophysical Journal} \textbf{\bibinfo{volume}{568}},
  \bibinfo{pages}{273} (\bibinfo{year}{2002}),
  \urlprefix\url{http://stacks.iop.org/0004-637X/568/i=1/a=273}.

\bibitem[{\citenamefont{{Galloway} et~al.}()\citenamefont{{Galloway},
  {Premachandra}, {Steeghs}, {Casares}, and {Messenger}}}]{GallowayEA_2012}
\bibinfo{author}{\bibfnamefont{D.}~\bibnamefont{{Galloway}}},
  \bibinfo{author}{\bibfnamefont{S.}~\bibnamefont{{Premachandra}}},
  \bibinfo{author}{\bibfnamefont{D.}~\bibnamefont{{Steeghs}}},
  \bibinfo{author}{\bibfnamefont{J.}~\bibnamefont{{Casares}}},
  \bibnamefont{and}
  \bibinfo{author}{\bibfnamefont{C.}~\bibnamefont{{Messenger}}},
  \bibinfo{note}{{GWPAW Hannover}}.

\bibitem[{\citenamefont{Fomalont et~al.}(2001)\citenamefont{Fomalont,
  Geldzahler, and Bradshaw}}]{FomalontEA_2001b}
\bibinfo{author}{\bibfnamefont{E.~B.} \bibnamefont{Fomalont}},
  \bibinfo{author}{\bibfnamefont{B.~J.} \bibnamefont{Geldzahler}},
  \bibnamefont{and} \bibinfo{author}{\bibfnamefont{C.~F.}
  \bibnamefont{Bradshaw}}, \bibinfo{journal}{The Astrophysical Journal}
  \textbf{\bibinfo{volume}{558}}, \bibinfo{pages}{283} (\bibinfo{year}{2001}),
  \urlprefix\url{http://stacks.iop.org/0004-637X/558/i=1/a=283}.

\bibitem[{\citenamefont{Messenger}(2006)}]{Messenger2005}
\bibinfo{author}{\bibfnamefont{C.~J.} \bibnamefont{Messenger}}, Ph.D. thesis,
  \bibinfo{school}{The University of Birmingham} (\bibinfo{year}{2006}).

\bibitem[{\citenamefont{{Bradt} et~al.}(1975)\citenamefont{{Bradt}, {Moore},
  {Braes}, {Miley}, {Forman}, {Kellogg}, {Hesser}, {Kunkel}, {Hiltner},
  {Hjellming} et~al.}}]{BradtEA_1975}
\bibinfo{author}{\bibfnamefont{H.~V.} \bibnamefont{{Bradt}}},
  \bibinfo{author}{\bibfnamefont{G.}~\bibnamefont{{Moore}}},
  \bibinfo{author}{\bibfnamefont{L.~L.~E.} \bibnamefont{{Braes}}},
  \bibinfo{author}{\bibfnamefont{G.~K.} \bibnamefont{{Miley}}},
  \bibinfo{author}{\bibfnamefont{W.}~\bibnamefont{{Forman}}},
  \bibinfo{author}{\bibfnamefont{E.}~\bibnamefont{{Kellogg}}},
  \bibinfo{author}{\bibfnamefont{J.~E.} \bibnamefont{{Hesser}}},
  \bibinfo{author}{\bibfnamefont{W.~E.} \bibnamefont{{Kunkel}}},
  \bibinfo{author}{\bibfnamefont{W.~A.} \bibnamefont{{Hiltner}}},
  \bibinfo{author}{\bibfnamefont{R.}~\bibnamefont{{Hjellming}}},
  \bibnamefont{et~al.}, \bibinfo{journal}{\apj} \textbf{\bibinfo{volume}{197}},
  \bibinfo{pages}{443} (\bibinfo{year}{1975}).

\bibitem[{\citenamefont{{Hertz} et~al.}(1992)\citenamefont{{Hertz}, {Vaughan},
  {Wood}, {Norris}, {Mitsuda}, {Michelson}, and {Dotani}}}]{HertzEA_1992}
\bibinfo{author}{\bibfnamefont{P.}~\bibnamefont{{Hertz}}},
  \bibinfo{author}{\bibfnamefont{B.}~\bibnamefont{{Vaughan}}},
  \bibinfo{author}{\bibfnamefont{K.~S.} \bibnamefont{{Wood}}},
  \bibinfo{author}{\bibfnamefont{J.~P.} \bibnamefont{{Norris}}},
  \bibinfo{author}{\bibfnamefont{K.}~\bibnamefont{{Mitsuda}}},
  \bibinfo{author}{\bibfnamefont{P.~F.} \bibnamefont{{Michelson}}},
  \bibnamefont{and} \bibinfo{author}{\bibfnamefont{T.}~\bibnamefont{{Dotani}}},
  \bibinfo{journal}{\apj} \textbf{\bibinfo{volume}{396}}, \bibinfo{pages}{201}
  (\bibinfo{year}{1992}).

\bibitem[{\citenamefont{{Abbott}
  et~al.}(2008{\natexlab{c}})\citenamefont{{Abbott}, {Abbott}, {Adhikari},
  {Agresti}, {Ajith}, {Allen}, {Amin}, {Anderson}, {Anderson}, {Arain}
  et~al.}}]{LIGO_2008_PRD77}
\bibinfo{author}{\bibfnamefont{B.}~\bibnamefont{{Abbott}}},
  \bibinfo{author}{\bibfnamefont{R.}~\bibnamefont{{Abbott}}},
  \bibinfo{author}{\bibfnamefont{R.}~\bibnamefont{{Adhikari}}},
  \bibinfo{author}{\bibfnamefont{J.}~\bibnamefont{{Agresti}}},
  \bibinfo{author}{\bibfnamefont{P.}~\bibnamefont{{Ajith}}},
  \bibinfo{author}{\bibfnamefont{B.}~\bibnamefont{{Allen}}},
  \bibinfo{author}{\bibfnamefont{R.}~\bibnamefont{{Amin}}},
  \bibinfo{author}{\bibfnamefont{S.~B.} \bibnamefont{{Anderson}}},
  \bibinfo{author}{\bibfnamefont{W.~G.} \bibnamefont{{Anderson}}},
  \bibinfo{author}{\bibfnamefont{M.}~\bibnamefont{{Arain}}},
  \bibnamefont{et~al.} (\bibinfo{collaboration}{The LIGO Scientific
  Collaboration}), \bibinfo{journal}{\prd} \textbf{\bibinfo{volume}{77}},
  \bibinfo{eid}{022001} (\bibinfo{year}{2008}{\natexlab{c}}),
  \eprint{0708.3818}.

\bibitem[{\citenamefont{{Aasi} et~al.}(2012)\citenamefont{{Aasi}, {Abadie},
  {Abbott}, {Abbott}, {Abbott}, {Abernathy}, {Accadia}, {Acernese}, {Adams},
  {Adams} et~al.}}]{LIGO_S5_2012arXiv}
\bibinfo{author}{\bibfnamefont{J.}~\bibnamefont{{Aasi}}},
  \bibinfo{author}{\bibfnamefont{J.}~\bibnamefont{{Abadie}}},
  \bibinfo{author}{\bibfnamefont{B.~P.} \bibnamefont{{Abbott}}},
  \bibinfo{author}{\bibfnamefont{R.}~\bibnamefont{{Abbott}}},
  \bibinfo{author}{\bibfnamefont{T.~D.} \bibnamefont{{Abbott}}},
  \bibinfo{author}{\bibfnamefont{M.}~\bibnamefont{{Abernathy}}},
  \bibinfo{author}{\bibfnamefont{T.}~\bibnamefont{{Accadia}}},
  \bibinfo{author}{\bibfnamefont{F.}~\bibnamefont{{Acernese}}},
  \bibinfo{author}{\bibfnamefont{C.}~\bibnamefont{{Adams}}},
  \bibinfo{author}{\bibfnamefont{T.}~\bibnamefont{{Adams}}},
  \bibnamefont{et~al.} (\bibinfo{collaboration}{The LIGO Scientific
  Collaboration and the Virgo Collaboration}), \bibinfo{journal}{ArXiv
  e-prints}  (\bibinfo{year}{2012}), \eprint{1207.7176}.

\bibitem[{\citenamefont{Wette}(2012)}]{Wette2012}
\bibinfo{author}{\bibfnamefont{K.}~\bibnamefont{Wette}},
  \bibinfo{journal}{\prd} \textbf{\bibinfo{volume}{85}},
  \bibinfo{pages}{042003} (\bibinfo{year}{2012}),
  \urlprefix\url{http://link.aps.org/doi/10.1103/PhysRevD.85.042003}.

\bibitem[{\citenamefont{{Dupuis} and {Woan}}(2005)}]{DupuisWoan_2005}
\bibinfo{author}{\bibfnamefont{R.~J.} \bibnamefont{{Dupuis}}} \bibnamefont{and}
  \bibinfo{author}{\bibfnamefont{G.}~\bibnamefont{{Woan}}},
  \bibinfo{journal}{\prd} \textbf{\bibinfo{volume}{72}}, \bibinfo{eid}{102002}
  (\bibinfo{year}{2005}), \eprint{arXiv:gr-qc/0508096}.

\bibitem[{\citenamefont{{Abbott} et~al.}(2007)\citenamefont{{Abbott}, {Abbott},
  {Adhikari}, {Agresti}, {Ajith}, {Allen}, {Amin}, {Anderson}, {Anderson},
  {Arain} et~al.}}]{LIGO_2007_PRD76}
\bibinfo{author}{\bibfnamefont{B.}~\bibnamefont{{Abbott}}},
  \bibinfo{author}{\bibfnamefont{R.}~\bibnamefont{{Abbott}}},
  \bibinfo{author}{\bibfnamefont{R.}~\bibnamefont{{Adhikari}}},
  \bibinfo{author}{\bibfnamefont{J.}~\bibnamefont{{Agresti}}},
  \bibinfo{author}{\bibfnamefont{P.}~\bibnamefont{{Ajith}}},
  \bibinfo{author}{\bibfnamefont{B.}~\bibnamefont{{Allen}}},
  \bibinfo{author}{\bibfnamefont{R.}~\bibnamefont{{Amin}}},
  \bibinfo{author}{\bibfnamefont{S.~B.} \bibnamefont{{Anderson}}},
  \bibinfo{author}{\bibfnamefont{W.~G.} \bibnamefont{{Anderson}}},
  \bibinfo{author}{\bibfnamefont{M.}~\bibnamefont{{Arain}}},
  \bibnamefont{et~al.} (\bibinfo{collaboration}{LIGO Scientific Collaboration
  and ALLEGRO Collaboration}), \bibinfo{journal}{\prd}
  \textbf{\bibinfo{volume}{76}}, \bibinfo{eid}{022001} (\bibinfo{year}{2007}),
  \eprint{arXiv:gr-qc/0703068}.

\bibitem[{\citenamefont{{R{\"o}ver} et~al.}(2011)\citenamefont{{R{\"o}ver},
  {Messenger}, and {Prix}}}]{RMP2011}
\bibinfo{author}{\bibfnamefont{C.}~\bibnamefont{{R{\"o}ver}}},
  \bibinfo{author}{\bibfnamefont{C.}~\bibnamefont{{Messenger}}},
  \bibnamefont{and} \bibinfo{author}{\bibfnamefont{R.}~\bibnamefont{{Prix}}},
  \bibinfo{journal}{ArXiv e-prints}  (\bibinfo{year}{2011}),
  \eprint{1103.2987}.

\bibitem[{\citenamefont{Bretthorst}(1988)}]{Bretthorst1988}
\bibinfo{author}{\bibfnamefont{G.~L.} \bibnamefont{Bretthorst}},
  \emph{\bibinfo{title}{Bayesian Spectrum Analysis and Parameter Estimation}},
  vol.~\bibinfo{volume}{48} of \emph{\bibinfo{series}{Lecture Notes in
  Statistics}} (\bibinfo{publisher}{Springer-Verlag}, \bibinfo{year}{1988}).

\bibitem[{\citenamefont{Prix and Krishnan}(2009)}]{PrixKrishnan_2009}
\bibinfo{author}{\bibfnamefont{R.}~\bibnamefont{Prix}} \bibnamefont{and}
  \bibinfo{author}{\bibfnamefont{B.}~\bibnamefont{Krishnan}},
  \bibinfo{journal}{Class. Quant. Grav.} \textbf{\bibinfo{volume}{26}},
  \bibinfo{pages}{204013} (\bibinfo{year}{2009}),
  \urlprefix\url{http://stacks.iop.org/0264-9381/26/i=20/a=204013}.

\bibitem[{\citenamefont{Messenger}(2010)}]{Messenger2010}
\bibinfo{author}{\bibfnamefont{C.}~\bibnamefont{Messenger}},
  \emph{\bibinfo{title}{Understanding the sensitivity of the stochastic
  radiometer analysis in terms of the strain tensor amplitude}},
  \bibinfo{howpublished}{LIGO Document T1000195} (\bibinfo{year}{2010}).

\end{thebibliography}


\end{document}